\documentclass[a4paper,flushrt,twocolumn]{aastex62}
\usepackage{amsmath,amsfonts,amssymb,graphicx,rotating,natbib,enumitem}
\usepackage{indentfirst}    %zastosowanie akapitu bezposrednio po rozpoczeciu nowego rozdzialu
\usepackage{epstopdf}
\DeclareGraphicsExtensions{.png,.pdf}
\bibpunct{(}{)}{;}{a}{}{,}
\newcommand{\cmD}{cm$^{-3}$}
\newcommand{\fluxU}{$\left[ \mathrm{cm}^{-2}\mathrm{s}^{-1}\mathrm{sr}^{-1} \right]$}
\newcommand{\elong}{$\varepsilon_{\mathrm{FOV}}$}
\newcommand{\elongF}{$\varepsilon_{\mathrm{FOV,1}}$}
\newcommand{\elongT}{$\varepsilon_{\mathrm{FOV,2}}$}
\newcommand{\priHe}{He$_{\mathrm{pri}}$}
\newcommand{\secHe}{He$_{\mathrm{sec}}$}
\newcommand{\isnHe}{He$_{\mathrm{ISN}}$}
\newcommand{\priH}{H$_{\mathrm{pri}}$}
\newcommand{\secH}{H$_{\mathrm{sec}}$}
\newcommand{\isnH}{H$_{\mathrm{ISN}}$}
\newcommand{\priD}{D$_{\mathrm{pri}}$}
\newcommand{\secD}{D$_{\mathrm{sec}}$}
\newcommand{\isnD}{D$_{\mathrm{ISN}}$}
\newcommand{\ionD}{D$^{-}$}

\shortauthors{Sok{\'o\l} et al.}
\begin{document}
\title{\textbf{\large{Science Opportunities from Observations of the Interstellar Neutral Gas with Adjustable Boresight Direction}}}
\correspondingauthor{Justyna M. Sok{\'o\l}}
\email{jsokol@cbk.waw.pl, jsokol@princeton.edu}
\author{Justyna M. Sok{\'o\l}}
\altaffiliation{NAWA Bekker Program Visiting Fellow \\ in the Department of Astrophysical Sciences at Princeton University}
\affiliation{Space Research Centre, Polish Academy of Sciences, (CBK PAN), Warsaw, Poland}
\affiliation{Department of Astrophysical Sciences, Princeton University, Princeton, NJ 08544, USA}
\author{Marzena A. Kubiak}
\affiliation{Space Research Centre, Polish Academy of Sciences, (CBK PAN), Warsaw, Poland}
\author{Maciej Bzowski}
\affiliation{Space Research Centre, Polish Academy of Sciences, (CBK PAN), Warsaw, Poland}
\author{Eberhard M\"obius}
\affiliation{University of New Hampshire, Durham, NH 03824, USA}
\author{Nathan A. Schwadron}
\affiliation{University of New Hampshire, Durham, NH 03824, USA}
\affiliation{Department of Astrophysical Sciences, Princeton University, Princeton, NJ 08544, USA}
%
%\date{}
%
\begin{abstract}
The interstellar neutral (ISN) gas enters the heliosphere and is detected at a few au from the Sun, as demonstrated by \textit{Ulysses} and the \textit{Interstellar Boundary Explorer} (\textit{IBEX}) missions. \textit{Ulysses} observed ISN gas from different vantage points in a polar orbit from 1994 to 2007, while \textit{IBEX} has been observing in an Earth orbit in a fixed direction relative to the Sun from 2009.  \citet{mccomas_etal:18b} reported about an \textit{IMAP}-Lo detector on board the \textit{Interstellar Mapping and Acceleration Probe} (\textit{IMAP}), with an ability to track the ISN flux in the sky. We present observation geometries for ISN gas for a detector with the capability to adjust the boresight direction along the Earth orbit over a year within a multichoice ISN observation scheme. We study science opportunities from the observations as a function of time during a year and the phase of solar activity. We identify observation geometries and determine the observation seasons separately for various ISN species and populations. We find that using an adjustable viewing direction allows for ISN gas observations in the upwind hemisphere, where the signal is not distorted by gravitational focusing, in addition to the viewing of ISN species throughout the entire year. Moreover, we demonstrate that with appropriately adjusted observation geometries, primary and secondary populations can be fully separated. Additionally, we show that atoms of ISN gas on indirect trajectories are accessible for detection, and we present their impact on the study of the ionization rates for ISN species.
\end{abstract}

%% Keywords of a maximum number of 6

\section{Introduction \label{sec:intro}}
The neutral component of the very local interstellar matter (VLISM) enters the heliosphere unaffected by the local magnetic field and reaches close distances to the Sun being focused by the solar gravitational force and attenuated by the solar environment \citep[e.g.,][]{axford:72, fahr:78, thomas:78}. Thus, the interstellar neutral (ISN) gas atoms can be used to probe the Sun's motion direction relative to the VLISM, as well as the physical state and composition of the VLISM \citep[e.g.,][]{mobius_etal:04a, bzowski_etal:19a}. ISN gas has been studied directly by in-situ measurements by the GAS experiment on board \textit{Ulysses} \citep{wenzel_etal:89a, witte_etal:92a} and the \textit{IBEX}-Lo experiment on board the \textit{Interstellar Boundary Explorer} (\textit{IBEX}, \citet{mccomas_etal:09a, fuselier_etal:09b, mobius_etal:09b}). Also, indirect methods have been developed to study the ISN gas via observations of the resonant backscatter glow \citep{bertaux_blamont:71, ajello_etal:87, vallerga_etal:04a, quemerais_etal:14a} and pickup ions \citep{mobius_etal:85a, gloeckler_etal:93a, drews_etal:12a}.

\textit{Ulysses} had the capability to track the ISN beam along the polar orbit during ISN observation periods from 1994 to 2007 at distances from the Sun of $1 - 2.5$~au. \textit{IBEX}-Lo operates in Earth's orbit from 2009 and has a field of view (FOV) fixed perpendicularly to the Sun-pointing spin axis of the spacecraft. The fixed observation geometry of \textit{IBEX}-Lo limits the observation time during a year. The measurements by GAS/\textit{Ulysses} and \textit{IBEX}-Lo have provided the ISN gas flow velocity and temperature \citep{witte:04, bzowski_etal:12a, mobius_etal:12a, bzowski_etal:14a, bzowski_etal:15a, schwadron_etal:15a, wood_etal:15a}, the VLISM composition by determination of Ne/O and D/H ratios \citep{bochsler_etal:12a, park_etal:14a, rodriguez_etal:14a, wood_etal:15c}, and the first direct sampling of the secondary population of ISN gas created in the outer heliosheath (OHS), the so-called Warm Breeze of He \citep{kubiak_etal:14a, kubiak_etal:16a, bzowski_etal:17a, wood_etal:17a, kubiak_etal:19a} and the secondary population of O \citep{park_etal:16a, baliukin_etal:17a, park_etal:19a}. The analysis of the first two years of \textit{IBEX}-Lo measurements of ISN~He opened a discussion about the ISN gas flow vector and temperature \citep{frisch_etal:13a, lallement_etal:14a, frisch_etal:15a}. Reanalysis of \textit{Ulysses} measurements \citep{bzowski_etal:14a, katushkina_etal:14c, wood_etal:15a} and analysis of both the primary He gas and the Warm Breeze observations by \textit{IBEX}-Lo \citep{bzowski_etal:15a, swaczyna_etal:15a, mobius_etal:15a, schwadron_etal:15a} led to the conclusion that the ISN flow vector agrees between \textit{Ulysses} and \textit{IBEX}-Lo measurements; however, the temperature is higher than the one determined in the original analysis of the \textit{Ulysses} data \citep{witte:04, witte_etal:04a, mccomas_etal:15a}.

\textit{Ulysses} and \textit{IBEX} measurements revealed opportunities to the study the VLISM and OHS through observations at close distances to the Sun with different observation geometries, but many questions about the ISN gas flow, VLISM, and OHS remain open (Section~\ref{sec:sciOpp}). On 2018 June 1 NASA selected a next heliospheric mission, the Interstellar Mapping and Acceleration Probe (\textit{IMAP}; \citet{mccomas_etal:18b}). The mission is dedicated to studying the acceleration of energetic particles and interaction of the solar wind with the interstellar medium. \textit{IMAP} will be on a halo orbit around the Sun--Earth L1 point. One of the instruments for the planned mission is a single-pixel neutral atom imager, \textit{IMAP}-Lo  mounted on a pivot platform (see Table~5 in \citet{mccomas_etal:18b}). This instrument takes it heritage from \textit{IBEX}-Lo with a better collection power and the capability to change the inclination of the viewing direction to the spacecraft rotation axis (see Section~4.1 in \citet{mccomas_etal:18b} and Section~\ref{sec:ibexImap} here). The goals for \textit{IMAP}-Lo are to deliver energy and angle-resolved measurements of ISN atoms for H, He, Ne, O, and D tracked over many months during a year and energy-resolved global maps of energetic neutral atoms (ENAs). 

In this paper, we present observational opportunities for an ISN gas instrument with an adjustable boresight direction. We study the ISN gas of He, H, Ne, O, and D as seen by an instrument located in Earth's orbit. We discuss how to optimize the observation geometry to maximize the scientific effectiveness of the measurements during a year. We illustrate our findings by simulations, adopting observation geometries of the planned \textit{IMAP}-Lo detector. However, our conclusions can be applied to any instrument located in the ecliptic plane close to Earth's orbit. The key feature of the instrument is the ability to change the boresight inclination relative to the Sun-pointing spin axis of the spacecraft.

We briefly discuss the state-of-the-art knowledge and open questions in the study of ISN gas flow and VLISM and OHS via ISN gas in Section~\ref{sec:sciOpp}. The technology for ISN observation on an example of \textit{IBEX}-Lo is presented in Section~\ref{sec:ibexImap}. The methodology used to calculate the ISN gas flux is described in Section~\ref{sec:methods}. The opportunities for ISN observation with adjustable boresight direction are discussed in Sections~\ref{sec:peaks} and \ref{sec:discussion}. The resulting locations of the maximum flux are discussed in Section~\ref{sec:peaksImap} and a comparison to the fixed geometry of observation is presented in Section~\ref{sec:compIBEX}. The observation geometry to resolve the secondary populations of ISN~He and ISN~H is discussed in Section~\ref{sec:secondaries}. The observation geometry for ISN~Ne and O is presented in Section~\ref{sec:NeO}. Possibilities for detection of ISN~D are discussed in  Section~\ref{sec:D2H}. The indirect beam of ISN~He is presented in Section~\ref{sec:twoBeams}. At the end, we present a potential observation plan for the ISN gas during solar maximum in Section~\ref{sec:OperationPlan}. We summarize the study in Section~\ref{sec:summary}.

\section{From the Earth to the VLISM \label{sec:sciOpp}}
%% Figure with SO and ST in a graph
\begin{figure*}
\includegraphics[scale=0.5]{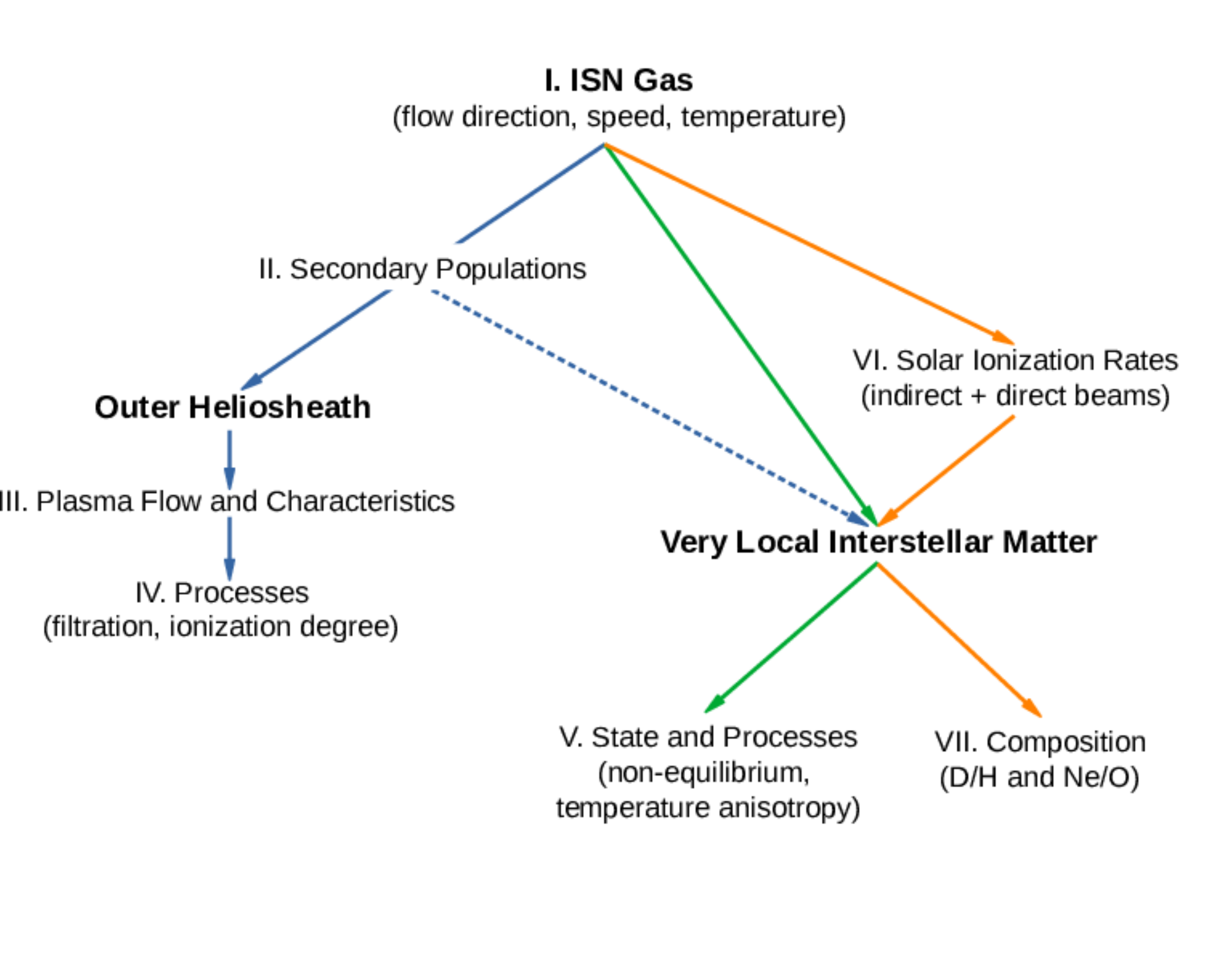}
\centering
\caption{Graphic representation of SOs and relations between them for the study of local interstellar medium accessible by ISN gas observations from the vicinity of Earth. \label{fig:graph}}
\end{figure*}
Observations of ISN gas from inside the heliosphere allow for the following:\\
\begin{tabular}{lp{7cm}}
I. & Precise determination of the ISN flow direction, speed, and temperature. \\
II. & Resolving the secondary population from the primary population flow (for ISN H, He, and O). \\
III. & Investigation of the plasma flow and its characteristics in the OHS (by analysis of the secondary populations). \\
IV. & Analysis of filtration in the OHS and ionization degree of interstellar species at the heliospheric boundary. \\
V. & Study of hypothetical departures of ISN~He in the VLISM from the Maxwell--Boltzmann distribution (nonequilibrium  processes and temperature anisotropy).\\
VI. & Study of the solar ionization rates (by joint analysis of the direct and indirect beams of the ISN He gas flow). \\
VII. & Study of the VLISM composition by determination of the D/H and Ne/O ratios. \\
\end{tabular}

The order of the science opportunities (SOs) listed above is organized so that each following topic requires results of the previous one, as illustrated in Figure~\ref{fig:graph}.  In this section we present relations between these SOs, and later in the paper we discuss how these SOs can be achieved with the capability to adjust the boresight direction of the detector. A potential observation plan to fulfill these SOs during a one year is presented in Section~\ref{sec:OperationPlan}.
 
The fundamental science goal is the precise determination of the ISN gas flow parameters (SO~I). The knowledge of the flow details is a prerequisite for all further investigations and is fundamental to our knowledge of the VLISM. This includes determination of the outer boundary condition for understanding the solar--interstellar interaction. A fixed observation geometry of \textit{IBEX}-Lo led to a parameter correlation tube (see, e.g., Figures~22 and 23 in \citet{bzowski_etal:12a}, Figure~8 in \citet{mobius_etal:12a}, Figure~4 in \citet{swaczyna_etal:15a}, Figure~7 in \citet{schwadron_etal:15a}), which is not present in the analysis of the \textit{Ulysses} measurements from different vantage points (see, e.g., Figures~11 and 12 in \citet{bzowski_etal:14a}). Moreover, as mentioned in Section~\ref{sec:intro}, cross-analysis of \textit{Ulysses} and \textit{IBEX}-Lo data led to a new estimate of a higher VLISM temperature. An opportunity to follow the ISN flow by continuous measurements of ISN gas distribution along the orbit around the Sun should allow for removing of the correlation of the flow parameters (direction, speed, and temperature) known as the parameter tube. Additionally, with continuous observations of the ISN flux throughout the year, observations in the upwind direction are accessible. In this region the atom trajectories are only negligibly deflected by the gravitation force, and the flow is less attenuated by the solar environment (see Figure~\ref{fig:obsScheme}). This allows for a more precise determination of the ISN flow direction. Moreover, the primary He, H, Ne, and O flows are co-aligned in this portion of the heliosphere. Observations in the upwind direction, together with increased measurement statistics and detection efficiency, should enable a detailed study of how closely aligned the flows of these species are. This has not been possible with \textit{Ulysses} and \textit{IBEX}-Lo, due to the measurement locations and low statistics of heavy atoms \citep{mobius_etal:09b, wood_etal:15c}.

Only with precise determination of the primary flow is an accurate analysis of the secondary populations is possible (SO~II, see also discussion in \citet{bzowski_etal:19a} and \citet{kubiak_etal:19a}). The primary population is a parent population for the secondary population \citep{baranov_malama:93, bzowski_etal:17a}, which is a result of the charge exchange process between ISN gas and plasma in the OHS. Thus, the primary flow parameters and a prerequisite knowledge about the plasma flow in the OHS are necessary for the study of the production of the secondary populations. Next, a precise study of the secondary population flow gives an opportunity for a remote study of the details of the plasma flow and characteristics in the OHS (SO~III), including deflection and heating. As presented by \citet{kubiak_etal:14a} (their Figure~12) and by \citet{bzowski_etal:17a} (their Figure~7), secondary population atoms enter the heliosphere from a wide portion of the sky. They represent a probe for a  spatial distribution of the plasma in the OHS, the region of their origin. However, a detailed analysis of the OHS through the Warm Breeze observations is hampered by the fixed observation geometry of \textit{IBEX}-Lo. The observation season is limited in time and covers a low count rate portion of the signal \citep[e.g.,][]{kubiak_etal:19a}. The secondary populations of ISN~He and O have been detected by \textit{IBEX}-Lo \citep{kubiak_etal:14a, park_etal:16a}, but limitations of the \textit{IBEX} observation geometry (more in Section~\ref{sec:ibexImap}) do not allow for a more thorough study, nor for a separate study of secondary ISN~H. Additionally, as pointed out by \citet{park_etal:19a}, the abundance ratio of secondary to primary O, as determined from observations, stands in disagreement with model predictions, raising questions about the plasma characteristics in the OHS. In Section~\ref{sec:secondaries}, we present how to observe the secondaries  without a significant contribution from primary populations and thus fully resolve these two populations.

With the knowledge of the ISN flow details and plasma interaction in the OHS and with better statistics of observations of ISN gas of various species, which can be achieved with an appropriate observation geometry, the study of how OHS filtrates ISN species in the OHS \citep{schwadron_etal:16a} and the ionization degree of the VLISM \citep{bzowski_etal:19a} is feasible (SO~IV). However, as the analysis of \textit{IBEX}-Lo measurements demonstrated, an extended observation period and an increased collection power are necessary to fulfill those goals. Therefore, we present an observation plan for ISN gas with the adjustable boresight direction instrument to achieve adequate statistics to study SO~IV.

The question of a possible departure of the ISN gas from thermal equilibrium has been raised in the past, and more recently in connection with the \textit{IBEX}-Lo measurements of ISN~He gas \citep{bzowski_etal:12a}. A thorough study of the anticipated signature of nonequilibrium processes in the measured ISN gas flow (SO~V) will be facilitated by the precise determination of the flow parameters (SOs~I and II) and the plasma properties in the boundary regions of the heliosphere (SOs~III and IV). But, as shown by, e.g., \citet{sokol_etal:15a, swaczyna_etal:19a, wood_etal:19a}, thus far the search for possible kappa-like signatures or asymmetries in the distribution function has not been conclusive with \textit{IBEX}-Lo and \textit{Ulysses}. A longer observation season at a location free from terrestrial contamination with enhanced sensitivity should resolve this question.

The ISN gas flow is attenuated inside the heliosphere by the solar environment, the ionization rates, and the radiation pressure, which act differently on different species \citep[see, e.g.,][]{sokol_etal:19a}. To study the processes in front of the heliosphere, a correction for the solar modulation of the ISN flux measured at $\sim1$~au along its path through the heliosphere is necessary \citep{bzowski:08a}. The ionization processes acting on the ISN atoms can be assessed from the solar wind and the solar EUV radiation measurements. However, an insight into the effective ionization rates for the ISN gas, in particular, inside 1~au, where electron impact ionization becomes important, will become accessible through a study of the ISN indirect beam (SO~VI). This remote investigation of the solar ionization rates is possible if only the indirect beam is detectable. We show that the indirect beam atoms are accessible for detection with properly directed boresight during the second half of a year.

The ISN flow parameters (SOs~I and II), together with the plasma characteristics in the OHS (SOs~III and IV) and heliospheric ionization losses, obtained in a consistent way for various species (SO~VI), enable the study of the VLISM composition (SO~VII). \textit{IBEX}-Lo has demonstrated a possibility for such a study \citep{bochsler_etal:12a, park_etal:14a}. However, better statistics of ISN Ne and O, as well as D and H, are required to improve the precision of the Ne/O and D/H ratios. We discuss the preferable viewing directions for these species during a year.

\section{Measurements of ISN Gas \label{sec:ibexImap}}
We discuss the measurement technique based on the \textit{IBEX}-Lo instrument, which is the heritage for \textit{IMAP}-Lo.  \textit{IBEX}-Lo collects atoms in the energy range from 0.01 to 2~keV, which allows us to observe the ISN gas of He \citep{mobius_etal:09b}, H \citep{saul_etal:12a}, D \citep{rodriguez_etal:13a}, Ne and O \citep{bochsler_etal:12a}, and H~ENAs \citep{fuselier_etal:12a, galli_etal:14a}. The instrument registers negative ions released from a specially designed conversion surface subjected to impact of various species of ISN atoms  \citep{fuselier_etal:09b, mobius_etal:09a}. The measurement method depends on the species and the ability of the conversion surface of the detector to produce stable negative ions. H and O atoms are observed directly via conversion to negative ions on the carbon conversion surface. However, noble gases do not produce stable negative ions \citep{wurz_etal:08a, mobius_etal:09a}, and thus detection of ISN He and Ne atoms is via analysis of the ratio of H$^-$, C$^-$, and O$^-$ ions sputtered from the conversion surface. The energy distribution of the sputtered negative ions is distinctly different from that of the converted ions. Whereas the latter still peak in energy only with a slightly reduced energy over the incoming neutral atoms, the energy distribution of the sputtered ions has a cutoff at an energy below the incoming energy, rises to a maximum near half the incoming atom energy, and continues at almost the same level to very low energies, as indicated in Figure~1 of \citet{mobius_etal:12a} for ISN He. Additionally, the H$^-$ and O$^-$ ions stem from a permanent monolayer of terrestrial water from outgassing of the sensor. As a consequence, terrestrial \ionD\, ions sputtered off the conversion surface by incoming ISN atoms, mostly He \citep{rodriguez_etal:14a}. Due to their expected low counting statistics, careful differentiation between a foreground of the terrestrial \ionD\, ions sputtered from the conversion surface and \ionD\, ions from the much less abundant ISN~D \citep{kubiak_etal:13a} is needed. Thus, the ISN~D detection is based on a careful statistical analysis of the registered signal, as presented by \citet{rodriguez_etal:14a}.

The \textit{IBEX}-Lo ISN observation season is limited to an interval from mid-November (when the Warm Breeze starts to be detectable) to April (when the ISN~H signal fades out) each year. In the remaining months the ISN gas is not observed owing to magnetospheric contamination and insufficient collection statistics outside the peak of the flow due to the low flux and low impact energy of atoms in the spacecraft frame, as well as the existence of a detector energy threshold \citep{galli_etal:15a, sokol_etal:15a}. As discussed by \citet{sokol_etal:15a} and \citet{swaczyna_etal:19a}, the observation time and energy limitations preclude any study of departures from the Maxwell--Boltzmann distribution function of the ISN He flow \citep[see also][]{wood_etal:19a}. The signatures of kappa-like distribution functions in the source region for ISN~He are expected either in orbits contaminated by the magnetospheric signal or in orbits where the predicted signal is below the instrument energy threshold. Moreover, even when the distribution function of the ISN~He gas in the VLISM is Maxwell--Boltzmann, \citet{kubiak_etal:19a} reported that departures of the observed signal from the two-Maxwellian approximation (including secondary population) observed in the Earth's orbit are also challenging to detect. Additionally, the \textit{IBEX} observation geometry results in a correlation of the determined ISN flow parameters, the so-called parameter tube \citep{mobius_etal:12a, bzowski_etal:12a}, which is not present in the analysis of \textit{Ulysses} ISN observations \citep{witte:04, bzowski_etal:14a, wood_etal:15a}. This implies that the correlation of ISN flow parameters can be resolved with an observation scheme that includes different vantage points along the Earth orbit. Additionally, observations of the ISN flow in the upwind hemisphere, where the trajectories are not significantly affected by the gravitational bending, should allow for precise determination of the flow direction.

\begin{figure}
\includegraphics[scale=0.3]{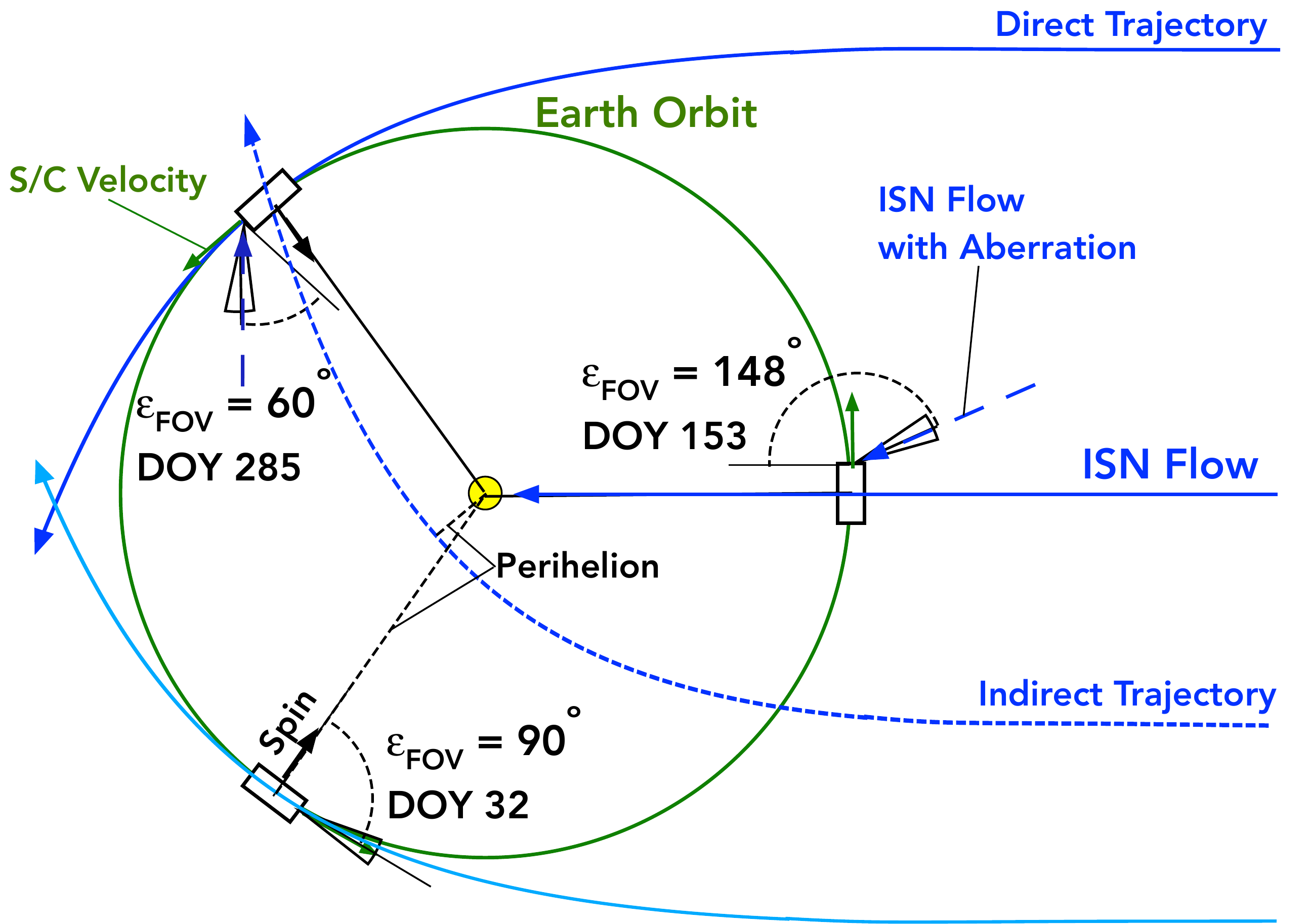}
\centering
\caption{Scheme of the observation geometry from various vantage points and elongations of the FOV at the orbit around the Sun. The Sun is in the center marked by a yellow disk; the detector orbit is illustrated by a green circle. Upwind direction is to the right from the Sun; downwind direction is to the left from the Sun. The dimensions of the objects are not in scale for illustration purposes.  See also Figure~\ref{fig:FOV} which presents the sky distribution of the flux for DOYs~153 and 285. \label{fig:obsScheme}}
\end{figure}

As mentioned in Section~\ref{sec:intro}, \textit{IMAP}-Lo is a detector similar to \textit{IBEX}-Lo that will observe the ISN gas from the vicinity of Earth's orbit in an energy range from 5~eV to 1~keV. Due to its orbit around the Sun--Earth L1 point, the measurements will not be affected by the terrestrial magnetosphere and will operate in lower backgrounds. This location provides us also with an opportunity to detect the ISN gas throughout many months of the year. \textit{IMAP}-Lo will be mounted on a moving platform with pivot angle in a range of $60\degr - 180\degr$ with respect to the spacecraft spin axis directed toward the Sun \citep{mccomas_etal:18b}. This will allow us to track the ISN flow signal in the sky (Figures~\ref{fig:obsScheme} and \ref{fig:FOV}). \textit{IMAP}-Lo will have a better collection power than \textit{IBEX}-Lo, because of increased (1) viewing time due to low backgrounds at L1, (2) geometric factor, (3) duty cycle, and (4) efficiency \citep{mccomas_etal:18b}. 

%% Figure with schematic view

\section{Methods \label{sec:methods}}
%% Figure with calculation grid
\begin{figure}
\includegraphics[scale=0.3]{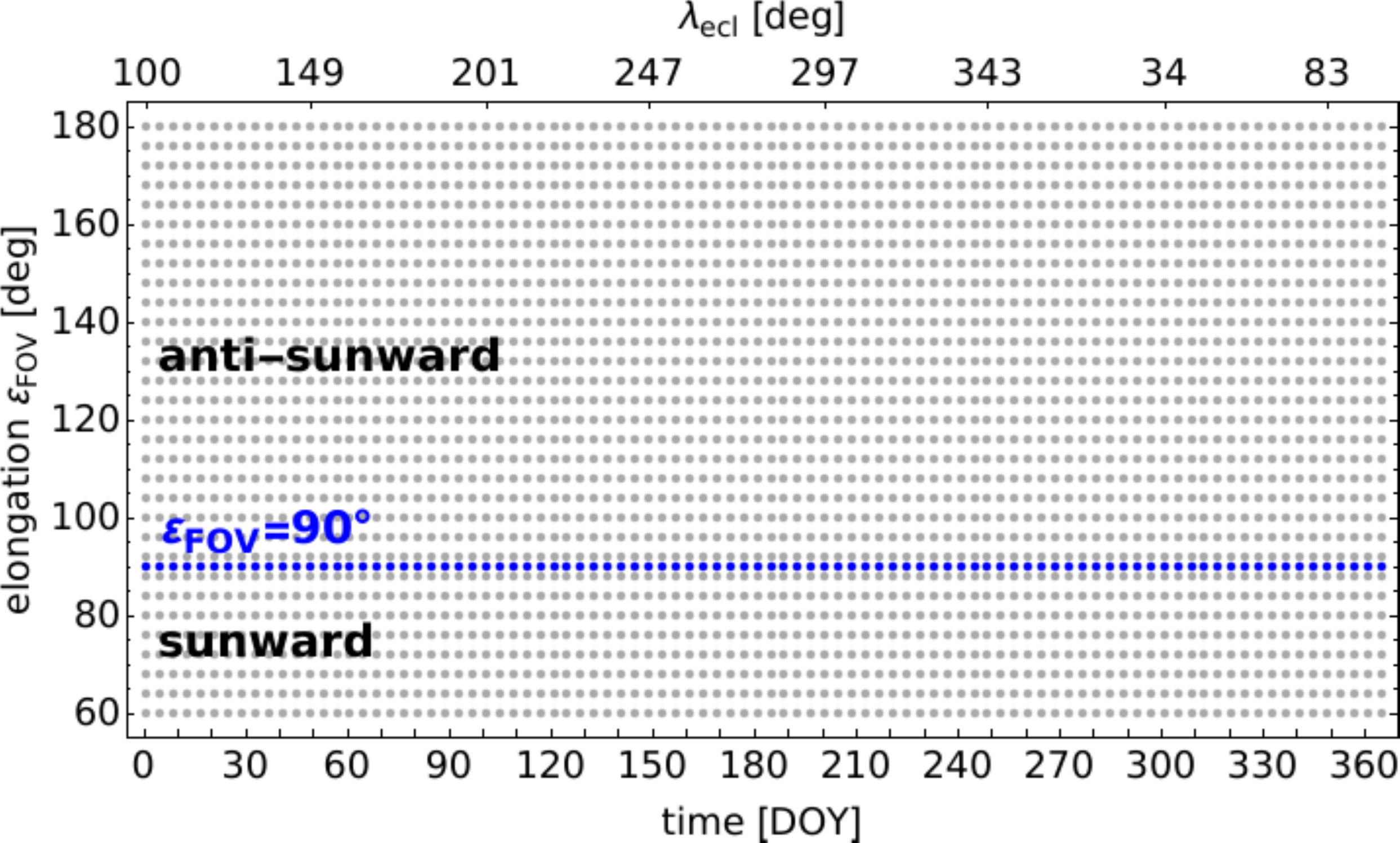}
\centering
\caption{Illustration of the simulation grid as a function of boresight elongation (\elong) and DOY. The boresight \elong$=90\degr$ (as in the \textit{IBEX} observation geometry) is marked by blue circles. For \elong$<90\degr$ the boresight is directed toward the sunward hemisphere; for \elong$>90\degr$ the boresight is directed antisunward. \label{fig:grid}}
\end{figure}

%% Figure with various orientations of FOV
\begin{figure}
\begin{tabular}{c}
\includegraphics[scale=0.25]{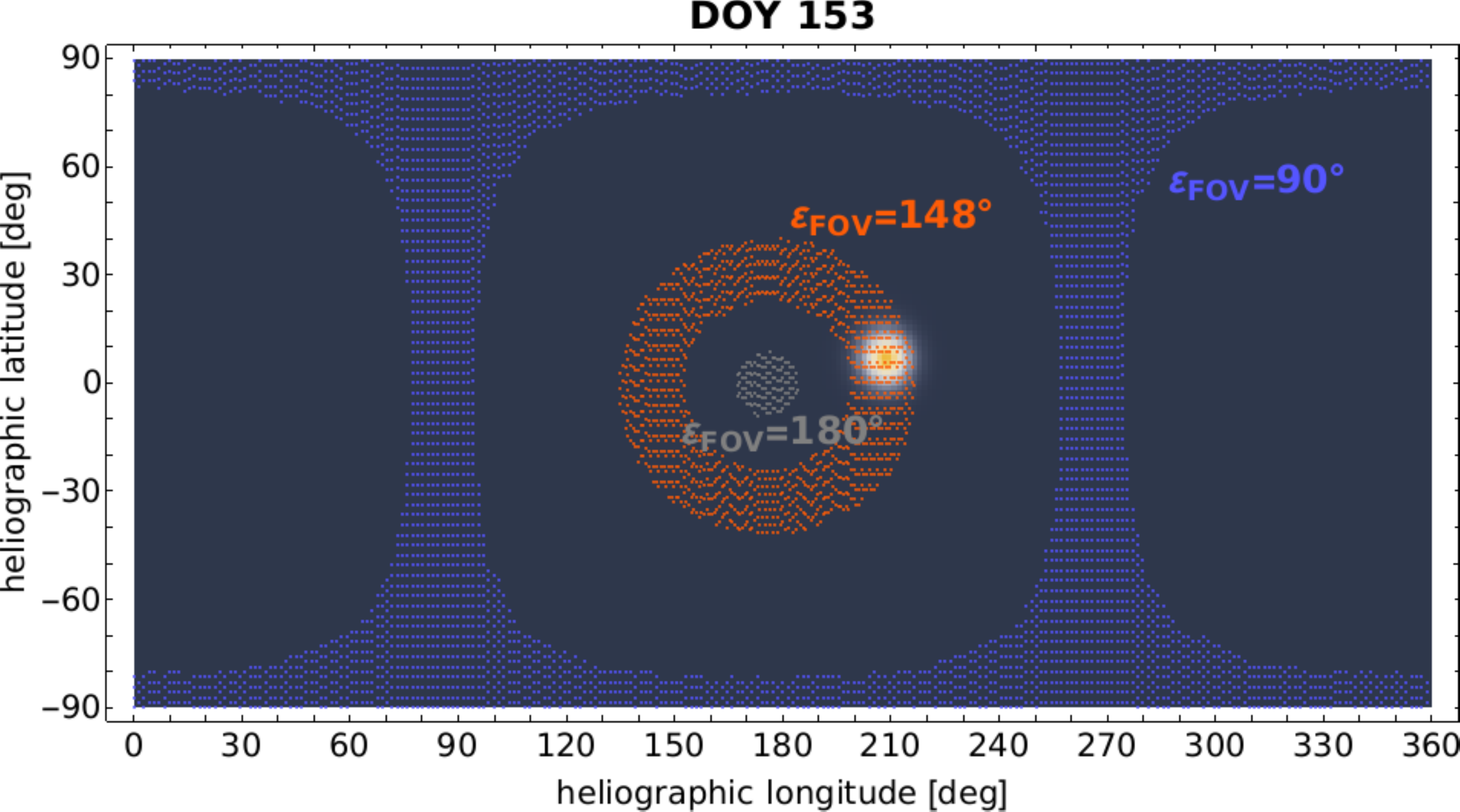} \\
\includegraphics[scale=0.25]{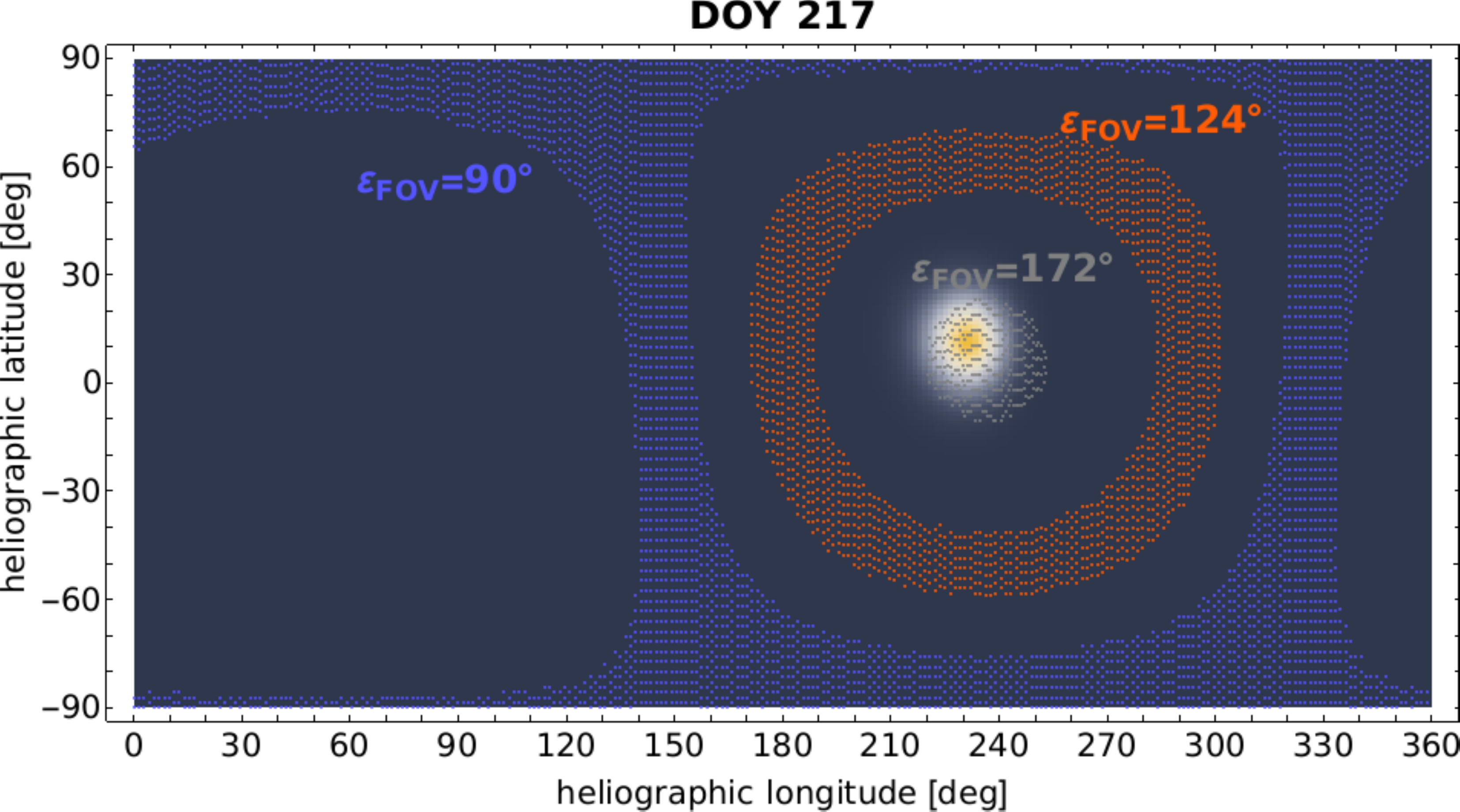} \\
\includegraphics[scale=0.25]{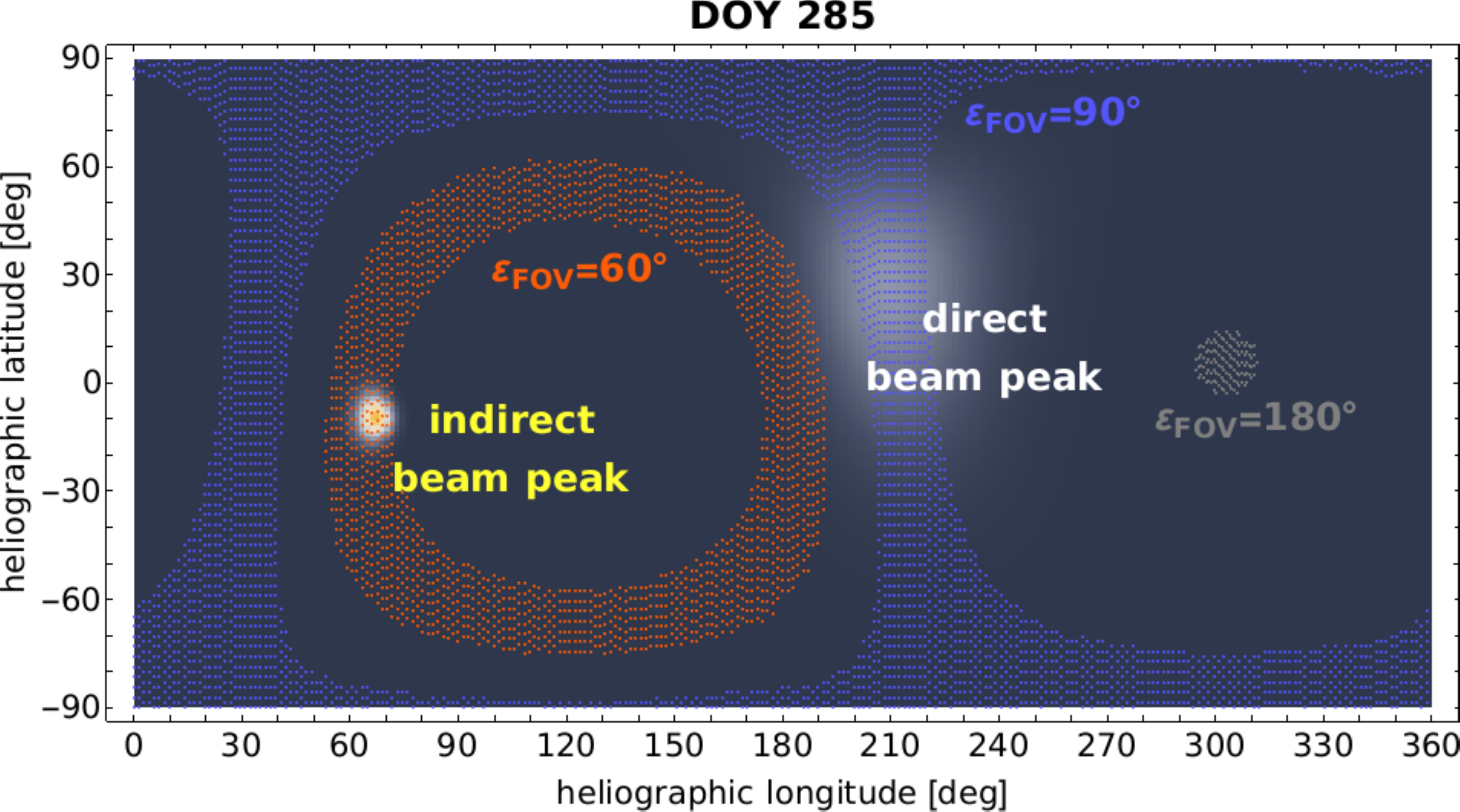} \\
\end{tabular}
\centering
\caption{Illustration of the observation geometry for selected days of the year: DOY~153 (top), DOY~217 (middle), and DOY~285 (bottom). A sky map in heliographic coordinates is presented, with examples of visibility strips encircled by various \elong\, superimposed (colored circles). The \priHe\, flux distribution in solar maximum is presented as background (the higher the flux, the brighter the color). The map for DOY~153 presents the \priHe\, distribution almost at the upwind direction, and the map for DOY~285 presents a case when both direct and indirect beams of \priHe\, are visible.}
\label{fig:FOV}
\end{figure}

We calculate the expected fluxes for ISN gas of various species using the framework of the numerical Warsaw Test Particle Model (nWTPM; \citet{sokol_etal:15b}). In this model, the so-called hot paradigm of the ISN gas distribution in the source region is applied \citep{thomas:78, wu_judge:79a, fahr:78, fahr:79a} assuming a Maxwell--Boltzmann distribution function in the VLISM with thermal speeds,\footnote{$u_T=\sqrt{\frac{2kT}{m}}$ is the thermal speed, where $m$ is the atomic mass of the species, $T$ is the temperature of a given population, and $k$ is the Boltzmann constant.} including the differences in mass of the species discussed. The trajectories of the atoms are traced back from the observer location to the source region beyond the solar wind termination shock with ionization rates and radiation pressure variable along the trajectory both in time and in space \citep{bzowski:08a, tarnopolski_bzowski:09}. The radiation pressure, which is nonnegligible for ISN~H and D atoms, is adopted from \citet{kowalska-leszczynska_etal:18a}. The ionization rates used in the study are assessed based on available solar wind and solar EUV flux measurements as described in \citet{sokol_etal:19a}. 

We consider the primary and secondary populations for ISN He, H, and D and the primary population for ISN Ne and O. The ISN flow parameters (velocity and temperature) of the primary ISN gas populations are adopted from the results of the analysis of the \textit{IBEX}-Lo direct sampling of ISN~He flow after \citet{bzowski_etal:15a}:
\begin{displaymath}
\lambda_{\mathrm{pri}}=255.745\degr, \quad \phi_{\mathrm{pri}}=5.169\degr, %\quad v_\mathrm{pri}=25.784~\mathrm{km~s}^{-1}, \quad T_{\mathrm{pri}}=7443~\mathrm{K},
\end{displaymath}
\begin{displaymath}
v_\mathrm{pri}=25.784~\mathrm{km~s}^{-1}, \quad T_{\mathrm{pri}}=7443~\mathrm{K},
\end{displaymath}
where $\lambda_{\mathrm{pri}}$ and $\phi_{\mathrm{pri}}$ are the ecliptic longitude and latitude of the flow in the source region (in ecliptic J2000 coordinates), respectively, $v_\mathrm{pri}$ is the flow speed, and  $T_{\mathrm{pri}}$ is the gas temperature. The simulation parameters used here are identical for the primary H, D, He, Ne, and O. For the secondary He population, we use the results from the study by \citet{kubiak_etal:16a}:
\begin{displaymath}
\lambda_{\mathrm{sec,He}}=251.57\degr, \quad \phi_{\mathrm{sec,He}}=11.95\degr, %\quad  v_{\mathrm{sec,He}}=11.28~\mathrm{km~s}^{-1}, \quad T_{\mathrm{sec,He}}=9480~\mathrm{K}, \quad \xi_{\mathrm{sec,He}}=0.057,
\end{displaymath}
\begin{displaymath}
 v_{\mathrm{sec,He}}=11.28~\mathrm{km~s}^{-1}, \quad T_{\mathrm{sec,He}}=9480~\mathrm{K},
\end{displaymath}
\begin{displaymath}
\xi_{\mathrm{sec,He}}=0.057,
\end{displaymath}
with $\xi_{\mathrm{sec,He}}$ being the abundance with respect to the primary population. As did \citet{kowalska-leszczynska_etal:18b}, we adopted the density, temperature, and speed of the secondary H after \citet{bzowski_etal:08a}, but with the inflow direction identical as for the secondary He:
\begin{displaymath}
\lambda_{\mathrm{sec,H}}=\lambda_{\mathrm{sec,He}}, \quad \phi_{\mathrm{sec,H}}=\phi_{\mathrm{sec,He}},% \quad v_{\mathrm{sec,H}}=18.744~\mathrm{km~s}^{-1}, \quad T_{\mathrm{sec,H}}=16300~\mathrm{K}, \quad \xi_{\mathrm{sec,H}}=1.75.
\end{displaymath}
\begin{displaymath}
\quad v_{\mathrm{sec,H}}=18.744~\mathrm{km~s}^{-1}, \quad T_{\mathrm{sec,H}}=16300~\mathrm{K},
\end{displaymath}
\begin{displaymath}
\xi_{\mathrm{sec,H}}=1.75.
\end{displaymath}
The angles between the inflow directions of the primary and secondary populations of He, according to the parameter determination by \citet{kubiak_etal:16a}, and of H, based on the kinetic-MHD model of the heliosphere by \citet{izmodenov_alexashov:15a}, are both about $8\degr$. Additionally, the deflection angle for primary H from the unperturbed inflow direction of ISN~He is about $1\degr$ (according to the model by \citet{izmodenov_alexashov:15a}). Therefore, the adopted assumptions about the inflow directions of primary and secondary H should not bias the results, especially when these two flow directions have not been fully resolved by observations yet. Following \citet{kubiak_etal:13a}, the parameters of the primary and secondary~D are adopted as for H, with the densities multiplied by the interstellar abundance $\xi_{\mathrm{D}}=15.6$~ppm \citep{linsky_etal:06a}.

The adopted densities of the ISN species in the source region (assumed upwind at the solar wind termination shock) are the following: $ n_{\mathrm{ISN,H}}=8.5\times10^{-2}$~\cmD\,\citep{bzowski_etal:08a}, $ n_{\mathrm{ISN,D}}=\xi_{\mathrm{D}} n_{\mathrm{ISN,H}},$ \citep{kubiak_etal:13a}, $ n_{\mathrm{ISN,He}}=1.5\times10^{-2}$~\cmD\, \citep{witte:04}, $ n_{\mathrm{ISN,Ne}}=5.8\times10^{-6}$~\cmD\, \citep{slavin_frisch:07a, slavin_frisch:08a}, and $ n_{\mathrm{ISN,O}}=5.0\times10^{-5}$~\cmD\, \citep{slavin_frisch:07a, slavin_frisch:08a}. From here on, we use the following naming convention: \priHe, \priH, and \priD\, for the primary population of He, H, and D; \secHe, \secH, and \secD\, for the secondary populations; \isnHe, \isnH, and \isnD\, to name the sum of the respective primary and secondary populations. In the case of Ne and O we consider only one: the primary, population. The secondary population of ISN~O has been detected based on \textit{IBEX}-Lo measurements by \citet{park_etal:16a, baliukin_etal:17a, park_etal:19a}. However, the abundance ratios of secondary to primary O obtained from data analysis and modeling differ significantly. \citet{park_etal:19a} derived the abundance ratio of 0.5, which is twice the ratio estimated by \citet{baliukin_etal:17a} and higher than the ratio predicted by \citet{izmodenov_etal:97a}. As discussed by \citet{park_etal:19a}, this discrepancy indicates a necessity for reevaluation of the secondary O production model in the OHS. Therefore, we refrain from calculation of the secondary O population in this study.

The nWTPM software, which was used in the analysis of \textit{IBEX}-Lo measurements \citep[e.g.,][]{bzowski_etal:15a, kubiak_etal:16a}, has been adjusted to the present study. The modifications include variation of the viewing direction with respect to the spacecraft spin axis (the varying FOV elongation angle, \elong) and enlarging of the collimator FOV to $\sim9\degr$ at FWHM while keeping its hexagonal shape from \textit{IBEX}-Lo. The collimator transmission function, as described in \citet{sokol_etal:15b}, but adjusted to larger FOV, is included. For each observation direction the velocity distribution function of the gas is calculated and integrated in the reference frame of a detector moving in the Earth's orbit (the Earth velocity is included). Next, the moments of the velocity distribution function are calculated. We focus on flux and average kinetic energy computed from the mean-square speed of the atoms relative to the spacecraft calculated from integration of the velocity distribution function over the full relative speed range determined as described in \citet{sokol_etal:15b}. No instrument energy bands or energy sensitivity functions of the detector are considered.

The simulations cover a year of observations with a 4-day time resolution for solar minimum and solar maximum conditions as in 2009 and 2015, respectively. We selected these two years because they follow the solar activity forecast for the coming solar cycle \citep{pesnell_schatten:18a}. As the time unit we use day of year (DOY). We performed the calculations for various elongations of the FOV (\elong), defined as the angular distance between the boresight of the FOV and the direction toward the Sun. The calculation grid is presented in Figure~\ref{fig:grid}. In the simulations, the elongation angle varies from $60\degr$ to $180\degr$ with a spacing of $4\degr$. We also simulate the \textit{IBEX}-Lo observation geometry, \elong$=90\degr$. For \elong$<90\degr$, the boresight is directed into the sunward hemisphere, and  for \elong$>90\degr$ it is directed into the antisunward hemisphere. However, elongations \elong$\gtrsim160\degr$ should be considered with cautious because of possible contamination of the signal by the Earth's exosphere \citep{baliukin_etal:19a}. For each DOY-\elong\, point in the calculation grid (Figure~\ref{fig:grid}), we calculate the flux as it would be observed by a detector on a spinning spacecraft; thus the signal is collected from a circle in the sky (great circle for \elong$=90\degr$ and small circles for \elong$\neq90\degr$). A full range of spin angles ($360\degr$) from the northernmost direction with a resolution of $6\degr$ is considered.

Figures~\ref{fig:obsScheme} and \ref{fig:FOV} illustrate the advantage of the adjustable boresight for ISN studies.  Figure~\ref{fig:obsScheme} presents a bird's-eye view of the observation geometry. The ISN gas flows from the right-hand side (it is the upwind direction). In each point of the spacecraft orbit, two beams of ISN gas intersect, one at $\theta$ and the other at the $360\degr - \theta$ angle measured between the space vector and the initial velocity vector \citep{fahr:68a, fahr:68b, axford:72}. The trajectories with two different impact parameters belonging to the same space point are referred to as direct and indirect trajectories. The elongation of the arrival direction of the ISN beam in the spacecraft frame depends on the inflow direction and speed of the ISN gas and the location of the spacecraft in its orbit around the Sun. A comparison of the portion of the sky sampled by the detector's FOV for various \elong\, including \elong$=90\degr$ as on \textit{IBEX}-Lo, is presented in Figure~\ref{fig:FOV}. A detector with an adjustable boresight (like \textit{IMAP}-Lo) will be able to probe the flux not accessible to \textit{IBEX}, like on DOYs~153 (almost at the upwind direction) and 217. The polar regions are accessible for observations only with \elong\, close to $90\degr$, and the portion of the sky covered by the FOV during one spacecraft spin decreases with an increase of angular distance to \elong$=90\degr$, with almost point-like measurements for \elong$>170\degr$. Additionally, as illustrated in the bottom panel of Figure~\ref{fig:FOV}, on DOY~285 the indirect beam peak is accessible with the elongation angle set to $60\degr$ from the sunward direction (see also Figure~\ref{fig:obsScheme}), while with \elong$=90\degr$ the faint direct beam is covered. As will be further discussed in Section~\ref{sec:twoBeams}, the direct and indirect beams are not accessible for detection at the same time owing to the energy threshold for detection.

Due to the detection method (Section~\ref{sec:ibexImap}), the following limitations need to be considered when planning the ISN observations. The first limitation is given by the energy of atoms hitting the detector in the spacecraft frame. As demonstrated with \textit{IBEX}-Lo, only atoms with the incoming energies greater than $\sim10$~eV are observable for \isnH, and greater than $\sim20$~eV for \isnHe\, \citep{galli_etal:15a, galli_etal:19a, sokol_etal:15a}. The other limitation is defined by a minimum flux to provide a sufficient count statistics. We assumed that fluxes greater than 100~\fluxU\, are measurable.

\section{Observations with Adjustable Boresight Directions \label{sec:peaks}}
An adjustable boresight direction provides the opportunity to observe various ISN gas species and populations from a range of vantage points as the spacecraft orbits the Sun. This includes tracking of the maximum flux in a first step, which arrives from different directions in the sky at different longitudes.
\subsection{Maximum Fluxes \label{sec:peaksImap}}
%% Figure with peak locations for IMAP
\begin{figure*}
\begin{tabular}{cc}
\includegraphics[scale=0.3]{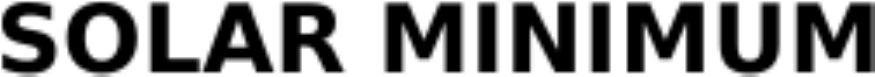} & \includegraphics[scale=0.3]{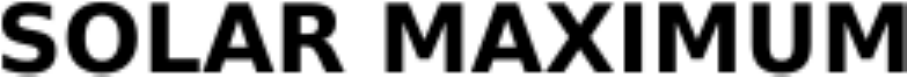} \\
\includegraphics[scale=0.3]{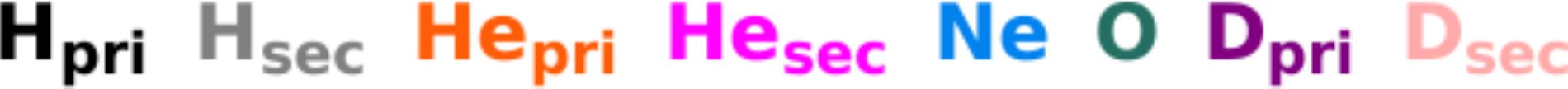} & \includegraphics[scale=0.3]{fig_legendSpecies.eps} \\
\includegraphics[scale=0.3]{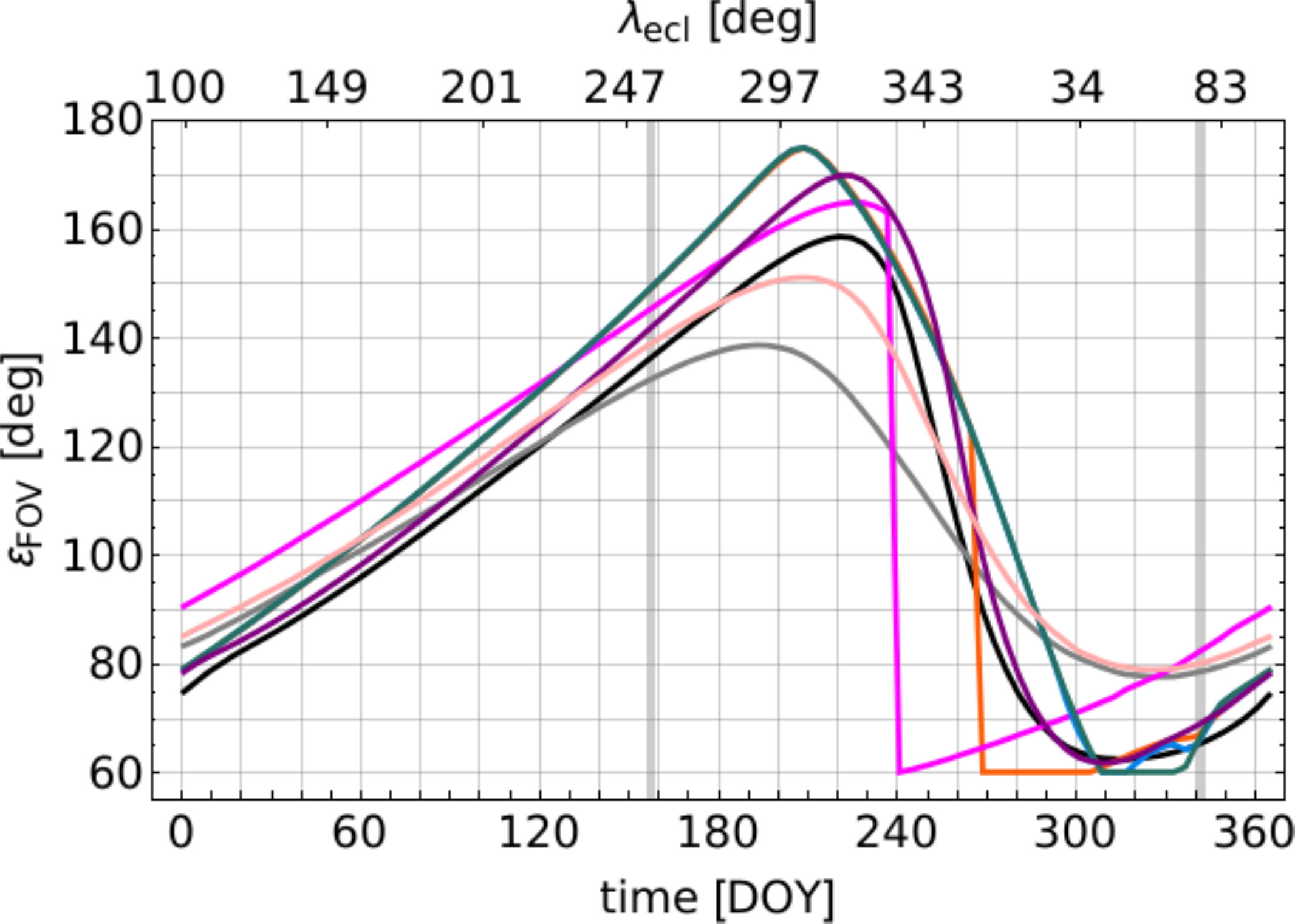} & \includegraphics[scale=0.3]{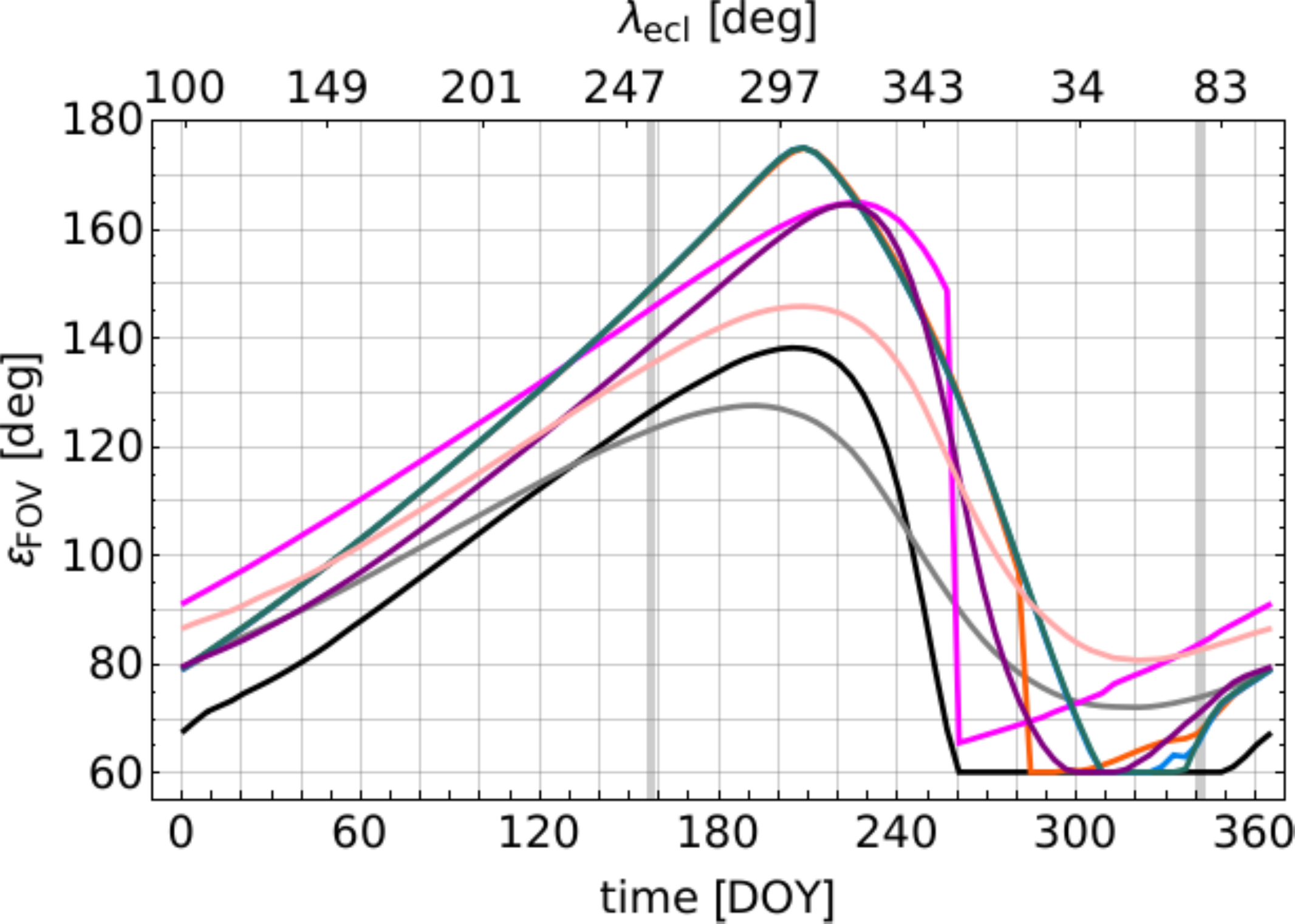} \\
\includegraphics[scale=0.3]{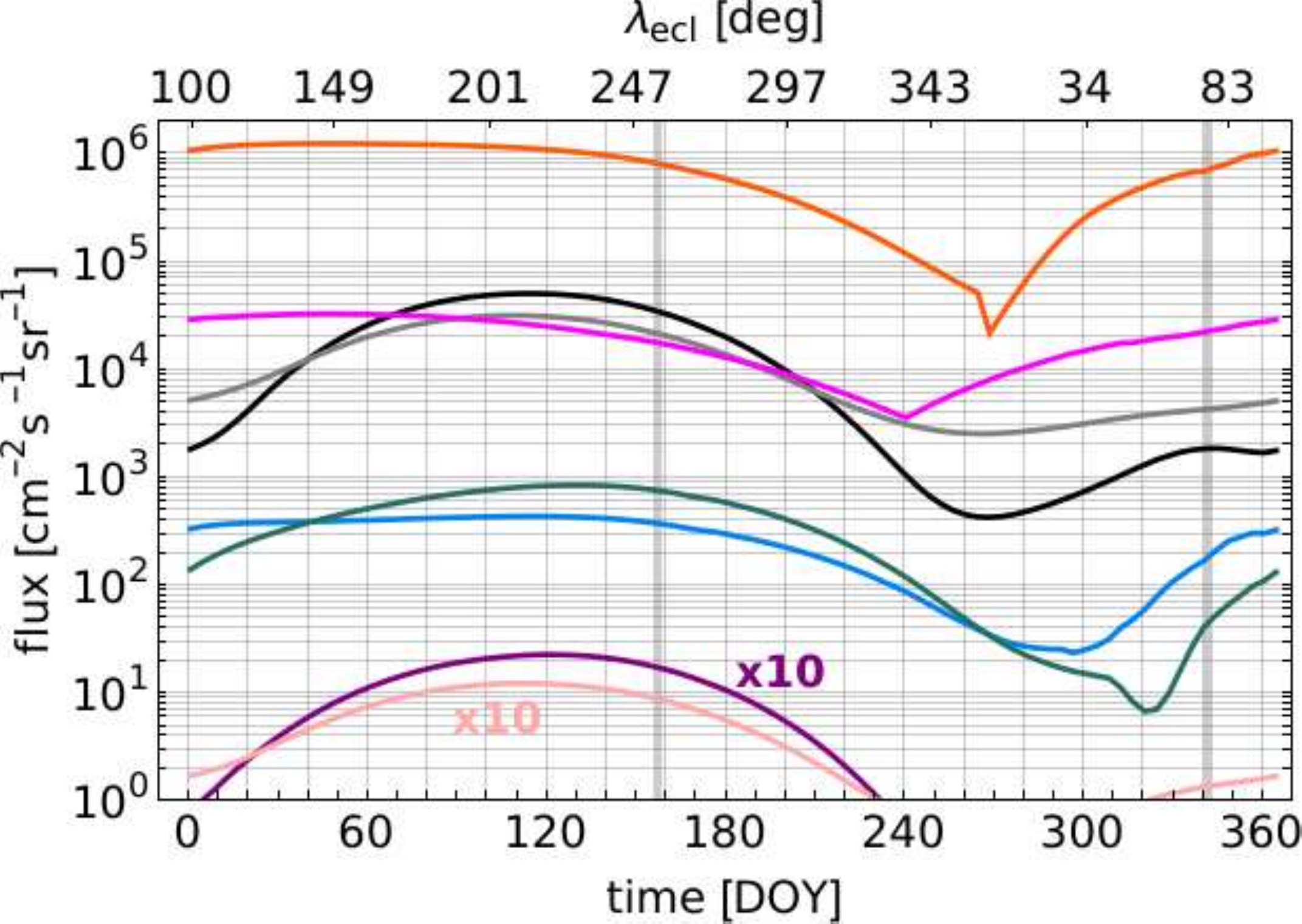} & \includegraphics[scale=0.3]{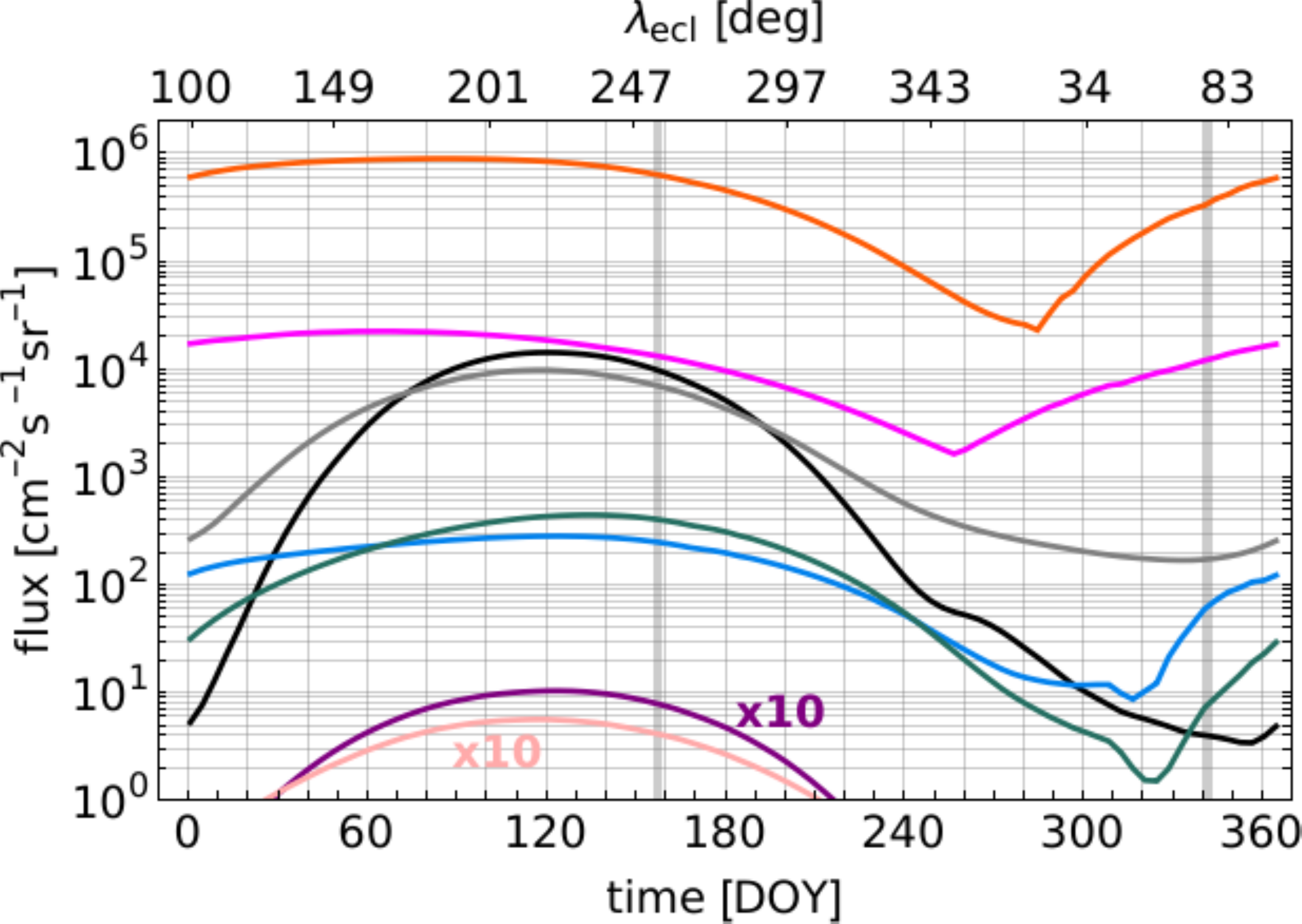} \\
\includegraphics[scale=0.3]{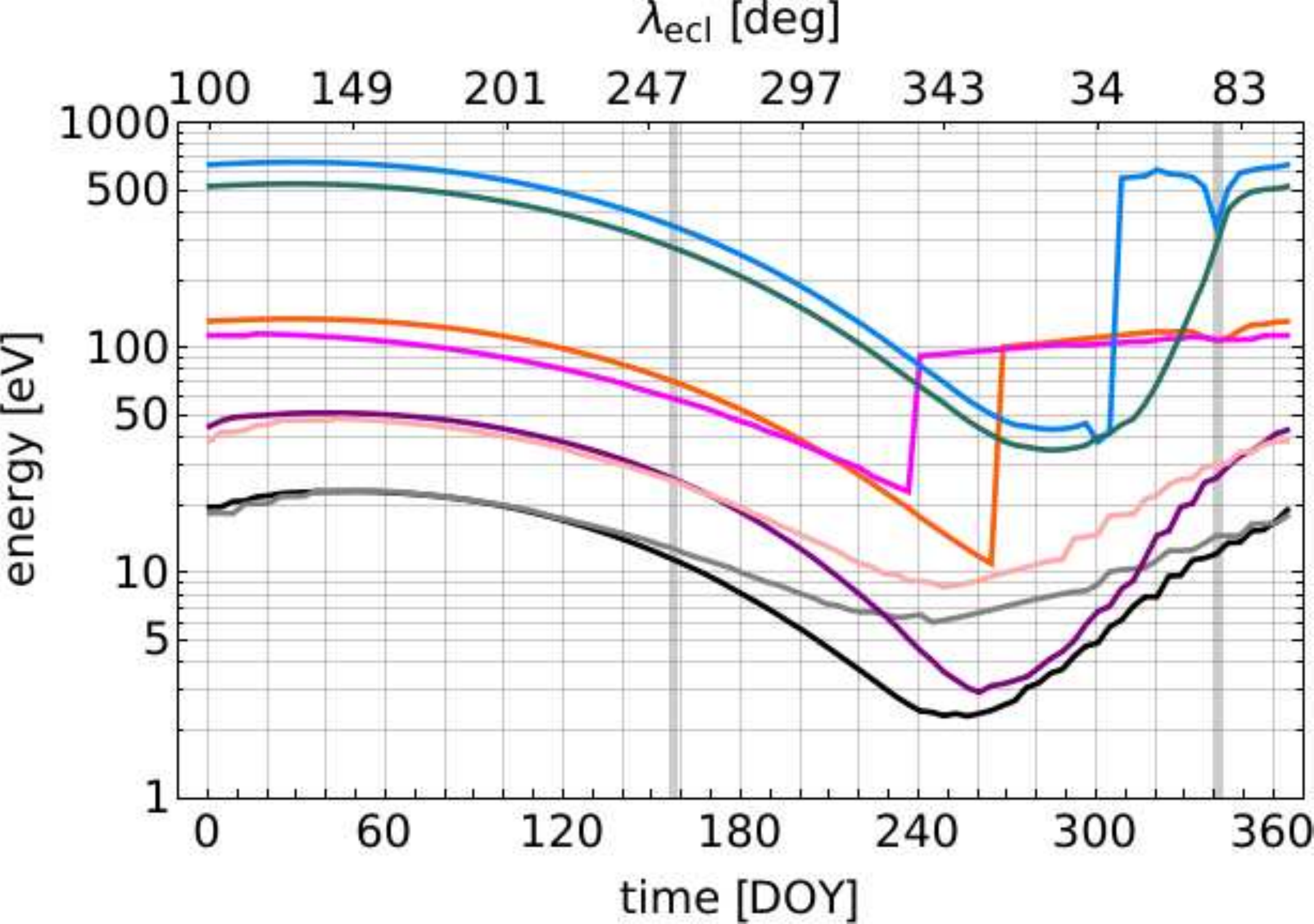} & \includegraphics[scale=0.3]{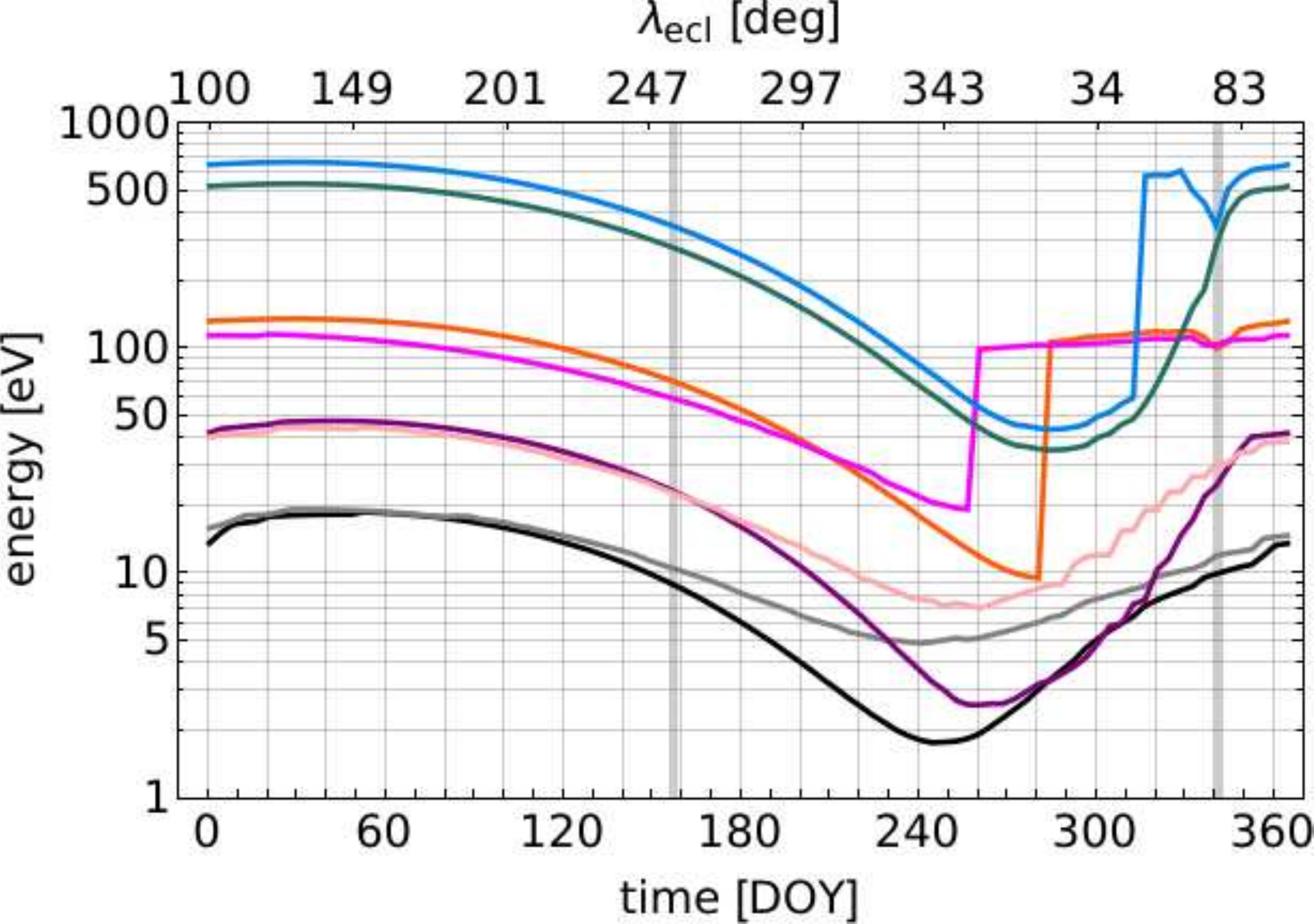} \\
\end{tabular}
\centering
\caption{Top row: maximum flux location as a function of an elongation angle (\elong) and a time (DOY) for solar minimum (left columns) and solar maximum (right column) for He, H, D (the primary and secondary populations separately), Ne and O. Middle row: expected maximum flux of the ISN gas as a function of time during the year for the \elong\, values from the top panel. Bottom row: average kinetic energy of the beam contributing to the maximum flux presented in the middle panels in the reference frame of detector (please note that these are not the maximum energies). Please note that the \priD\, and \secD\, fluxes are multiplied by 10.}
\label{fig:peakImap}
\end{figure*}

The distribution of the ISN gas inside the heliosphere is governed by the gravitational force (and modified by the radiation pressure for H and D) and the solar ionizing environment; thus, different observation geometries are needed to observe the flux maximum for different species throughout the year and during different phases of the solar activity cycle. The solar cycle modulates the flux by ionization and, for H and D, by radiation pressure, which also results in a shift of the peak location, as demonstrated for H \citep{kowalska-leszczynska_etal:18b, galli_etal:19a,  rahmanifard_etal:19a}.

Figure~\ref{fig:peakImap} presents the resulting maximum fluxes of selected species and populations. The elongation angles \elong\, (top panels), the fluxes (middle panels), and corresponding average kinetic energies of the maximum flux beam (bottom panels) as a function of time during solar minimum (left) and solar maximum (right) are presented. In general, the elongation angles of the maximum flux are close to about $80\degr$ at the beginning of the year, and they increase toward the antisunward direction around DOY~200. Then, the maximum flux elongations decrease steeply to \elong$<90\degr$ around DOY~240 or later, depending on the species and population, when the indirect beam flux dominates over the direct beam flux. 

Different species and populations have different peak locations, which is a consequence of different flow parameters and of different modulation of the bulk flow trajectory and the flux magnitude by the solar-cycle effects. The most pronounced differentiation in the peak location between the species and populations occurs in the second half of a year (Figure~\ref{fig:peakImap}). Typically, the elongations for maximum flux of different species and populations gather in a corridor with a width of $\sim15\degr$ during the first few months and the last two months of a year during solar minimum conditions. For the remaining months, the corridor width extends because of larger separation of the peak locations. For most of the time, the corridor width is determined by the peaks of \priH\, (lower limit) and \secHe\, (upper limit). During solar maximum, the corridor width is wider, mostly due to the lower elongation of the \priH\, peak caused by the solar-cycle variation of the radiation pressure \citep[see, e.g.,][]{tarnopolski_bzowski:09, kowalska-leszczynska_etal:18b} and differential ionization losses \citep{bzowski_etal:97}. The trajectories of the remaining species (He, Ne, O) are not affected by the solar-cycle changes. Ionization losses do not change the atom kinematics, but only the magnitude of fluxes inside the heliosphere. Thus, the flux peaks for those species are expected at similar locations during different phases of solar activity, with small variations as a result of differential ionization. 

The separation of the ISN flux peak locations between species and populations could mean that the boresight direction of the detector needs to be adjusted every day depending on the target. However, continuous pointing toward the peak location will not necessarily result in satisfactory observations because two other factors need to be considered: the energy of the incoming atoms in the reference frame of detector, and the magnitude of the flux. The middle row panels of Figure~\ref{fig:peakImap} present the expected maximum flux in absolute units as seen by the detector. The \priHe\, flux is the largest, and it dominates by more than one order of magnitude over the maximum fluxes of the remaining species regardless of the phase of the solar activity cycle. The \priHe\, flux at the detector exceeds $3\times10^{5}$~atoms  \fluxU\, for about eight to nine months during a year, with a decrease to less than $5\times10^{4}$~\fluxU\, between September and October. Thus, it should be possible to observe \priHe\, throughout almost the entire year with appropriately adjusted \elong\, direction.

The next-highest flux is that of \secHe, about $4\times10^{-2}$ smaller than that of \priHe. A flux with a similar magnitude to \secHe\, is obtained for \isnH, which, however, varies during a year and with the solar cycle. During solar minimum, the \isnH\, signal dominates over \secHe\, by as much as a factor of $\sim3$ from February to August, while during solar maximum the \isnH\, flux is higher than \secHe\, only from April to June, and only up to a factor of $\sim 1.3$. The disappearance of the \isnH\, signal during the solar maximum in the \textit{IBEX}-Lo observations was discussed by \citet{saul_etal:13a} and \citet{galli_etal:19a}. It is related to the increase of the ionization rate and the radiation pressure with solar activity. The preferable boresight orientations for probing the secondary populations are discussed further in Section~\ref{sec:secondaries}.

The ISN~Ne and O gas co-flow with \priHe, and the highest flux ratios of Ne and O with respect to \isnHe\, are expected in September, when the Ne maximum flux is $(6-7.5)\times10^{-4}$ of the \isnHe\, flux, and in July/August, when the O maximum flux is $(7-10)\times10^{-4}$ of the \isnHe. However, the overall maximum fluxes of Ne and O are expected at the beginning of May at elongations ranging from $128\degr$ to $136\degr$ (see Figure~\ref{fig:peakImap}). The difference in time between the flux peak and the maximum ratio with respect to \isnHe\, is due to the decrease of \isnHe\, flux in time, while the Ne and O fluxes are still high. Moreover, from June to September, the flux peak elongations are greater than 150$\degr$, which means that a small portion of the sky is covered by the FOV (see Figure~\ref{fig:FOV}). The signal is confined to a small solid angle, which provides an opportunity to increase the statistics of measurements. Such observation geometry is especially beneficial for Ne and O, whose fluxes are generally low.

The bottom panels of Figure~\ref{fig:peakImap} present the energy of impacting atoms of the maximum flux presented in the middle panels in the detector reference frame. In general, the highest energies are expected for Ne and O atoms (greater than 500~eV), while the lowest energies are expected for H atoms (about 20~eV). The greatest energies with respect to the detector are from December to March, when the ISN gas flows against the detector and the velocities of the detector and the atoms sum up. The lowest energies are around August/September, when the detector moves with the gas. The energy of atoms entering the detector system might significantly influence the determination of the observation season for specific objectives. For example, around DOY~269, at \elong$=116\degr$, the flux of \priHe\, is above $10^{4}$~\fluxU, but the energy is about 10~eV, which is definitely too low to satisfy detection with the technology available nowadays. However, at the same DOY, the maximum flux of \secHe\, is observed at \elong$=64\degr$ with $\sim 8\times10^3$~\fluxU\, and an energy of about 100~eV, which satisfies detection (see also Figure~\ref{fig:secondariesHe}). Therefore, only \secHe\, will be observable that day, and thus the boresight of the detector needs to be adjusted accordingly. 

The determination of the best seasons to observe various ISN gas species depends on several observational conditions. The first two, already mentioned, are the sufficiently high flux and the energy higher than the threshold for detection. Another important aspect is the presence of atoms of other species in the FOV and an ability to differentiate between species in the detector system. Figure~\ref{fig:peakImap} illustrates that when tracking the ISN species along Earth's orbit, \priHe\, has the highest fluxes among the species in question. However, its energies decrease from May to October, making the detection more challenging. In the case of \isnH, both fluxes and energies are much lower than those of \isnHe, with the energies close to the detection threshold and with the flux magnitude variable by almost two orders of magnitude over the year during solar maximum. 

Ne and O have the highest energies because of their higher masses, but their fluxes are three orders of magnitude lower than that for \isnHe, because of lower abundance in the VLISM and higher ionization losses inside the heliosphere (see, e.g., Figure~8 of \citet{sokol_etal:19a}). The highest energies for Ne and O are expected at the end of January. However, the highest fluxes are expected at the beginning of May each year, when the energies are reduced by about $70\%$. As further discussed in Section~\ref{sec:NeO}, the Ne and O beams are narrow also in spin angle, which causes an additional challenge for detection. 

The most challenging species to detect is \isnD, which, despite relatively high energies, has very low flux. Please note that the peak fluxes for \isnD\, presented in Figure~\ref{fig:peakImap} are multiplied by 10. Again, the highest flux of \isnD\, is expected when the energy decreases and is predicted to be of the order of a few atoms per \fluxU\, between April and May. Similarly to \isnH, the expected fluxes are reduced during solar maximum, up to about $50\%$ in the case of \isnD. The discussion of optimization for the  \isnD\, observations is presented in Section~\ref{sec:D2H}. 

\subsection{Comparison with a Fixed Boresight Direction\label{sec:compIBEX}}
%% Figure with peak locations for IBEX
\begin{figure*}
\begin{tabular}{cc}
\includegraphics[scale=0.3]{fig_legendSolarMin.eps} & \includegraphics[scale=0.3]{fig_legendSolarMax.eps} \\
\includegraphics[scale=0.3]{fig_legendSpecies.eps} & \includegraphics[scale=0.3]{fig_legendSpecies.eps} \\
\includegraphics[scale=0.3]{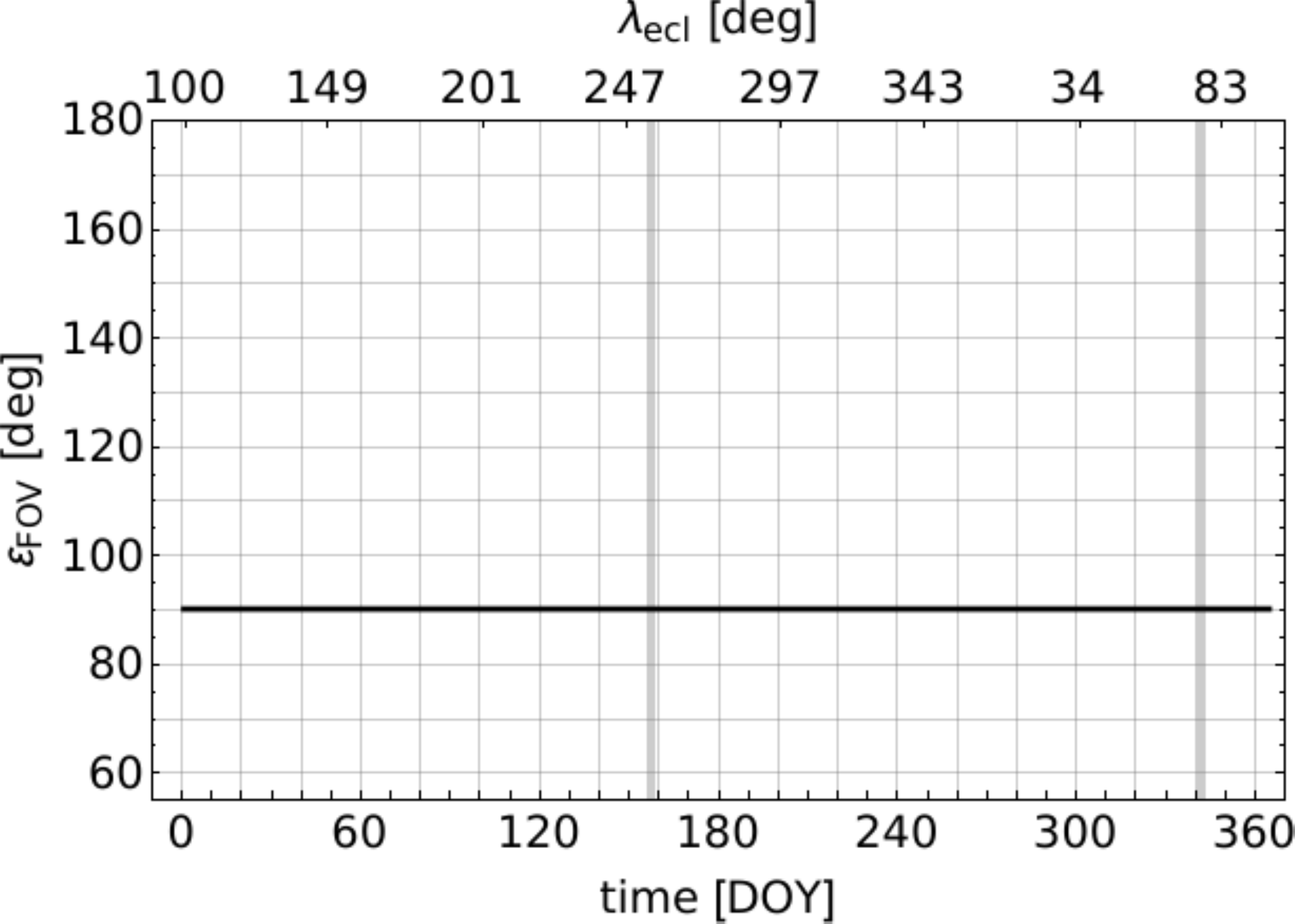} & \includegraphics[scale=0.3]{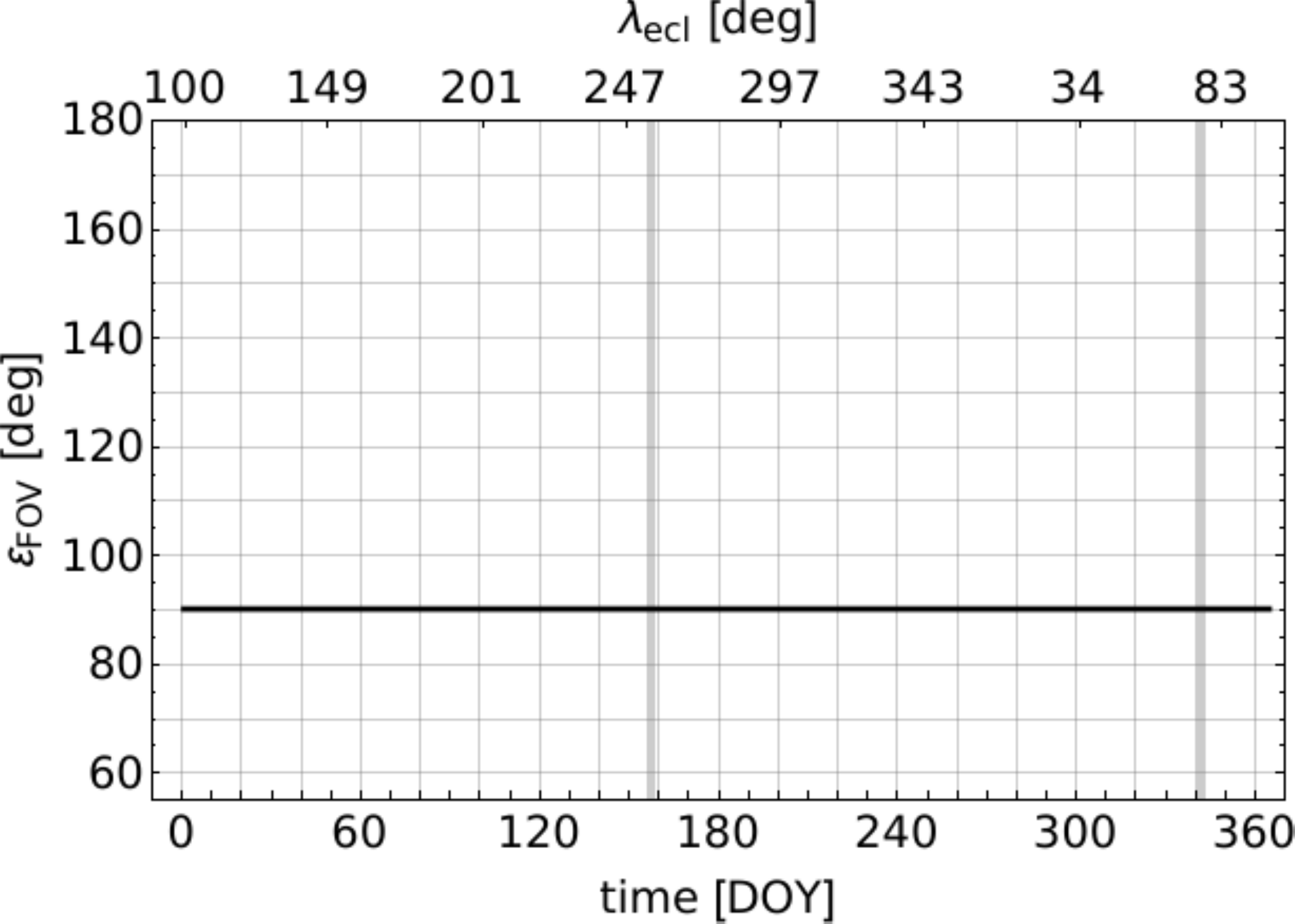} \\
\includegraphics[scale=0.3]{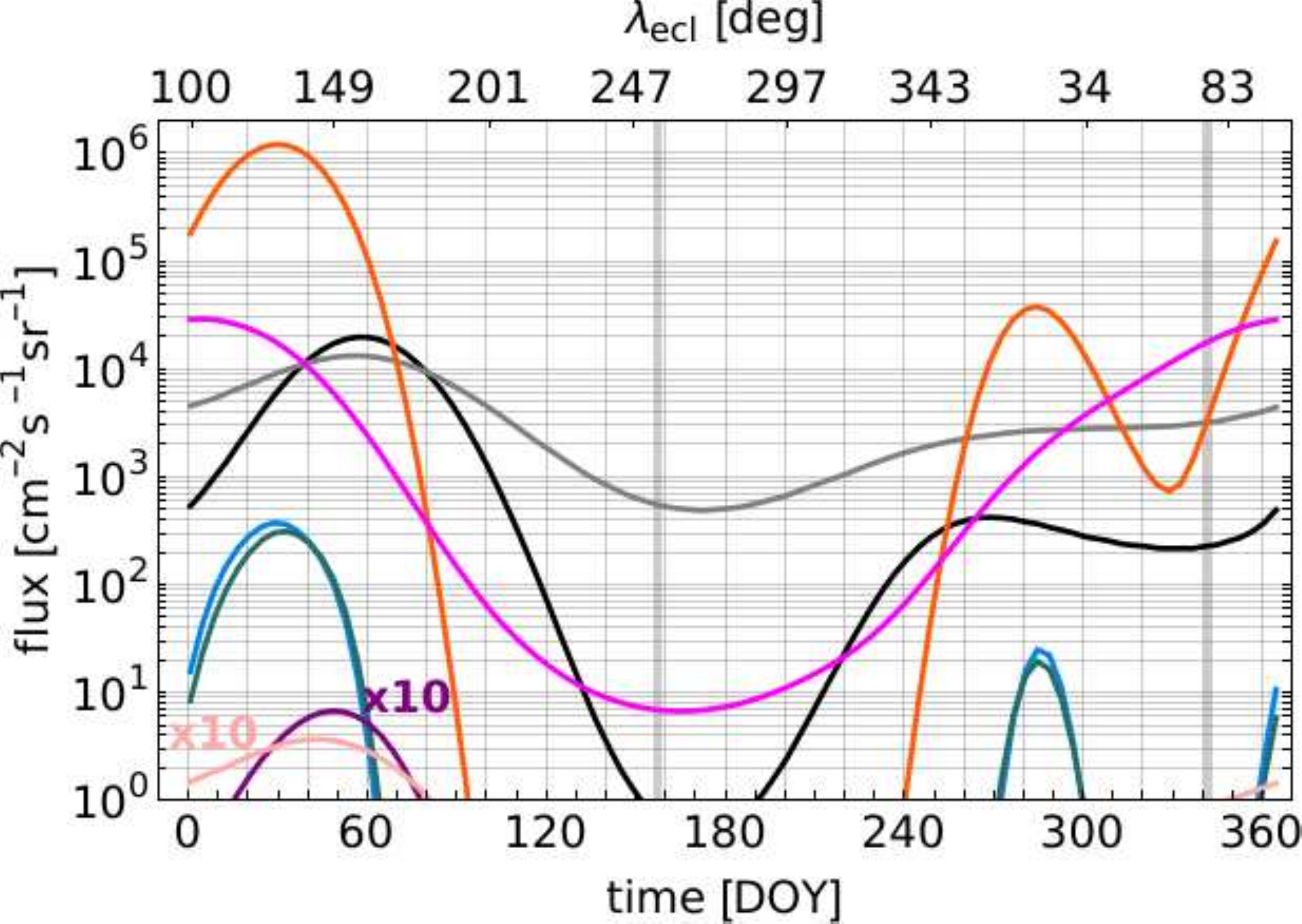} & \includegraphics[scale=0.3]{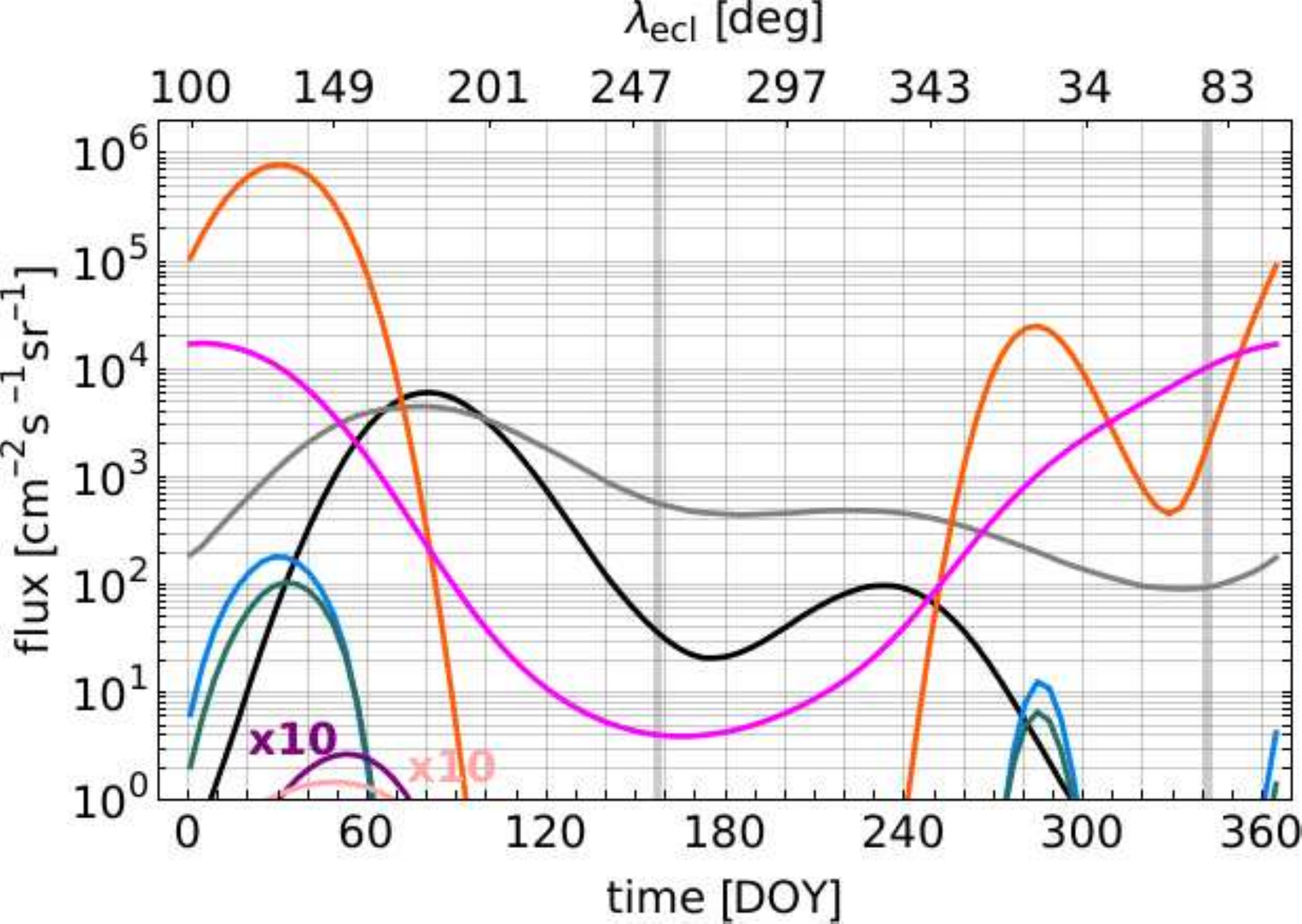} \\
\includegraphics[scale=0.3]{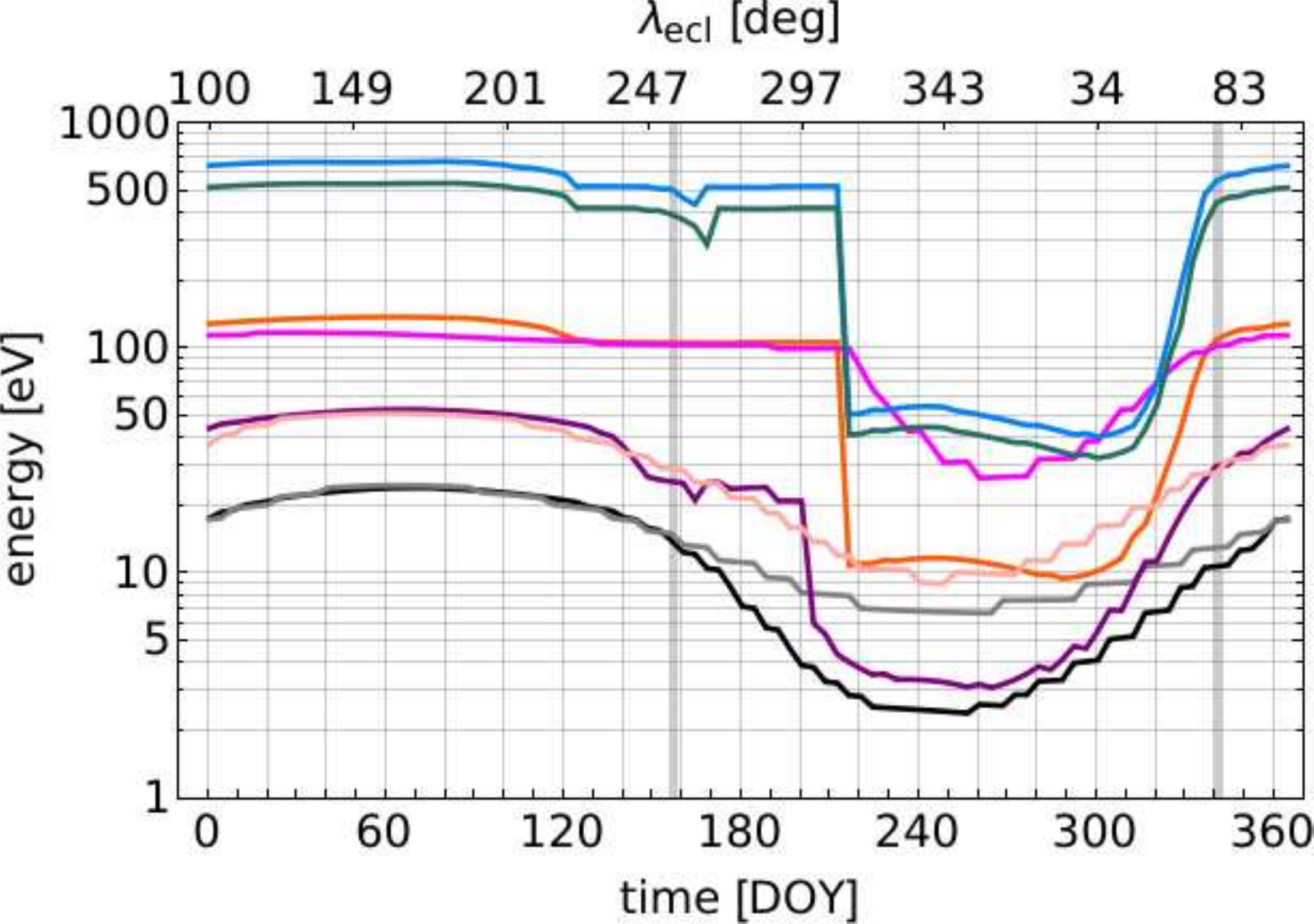} & \includegraphics[scale=0.3]{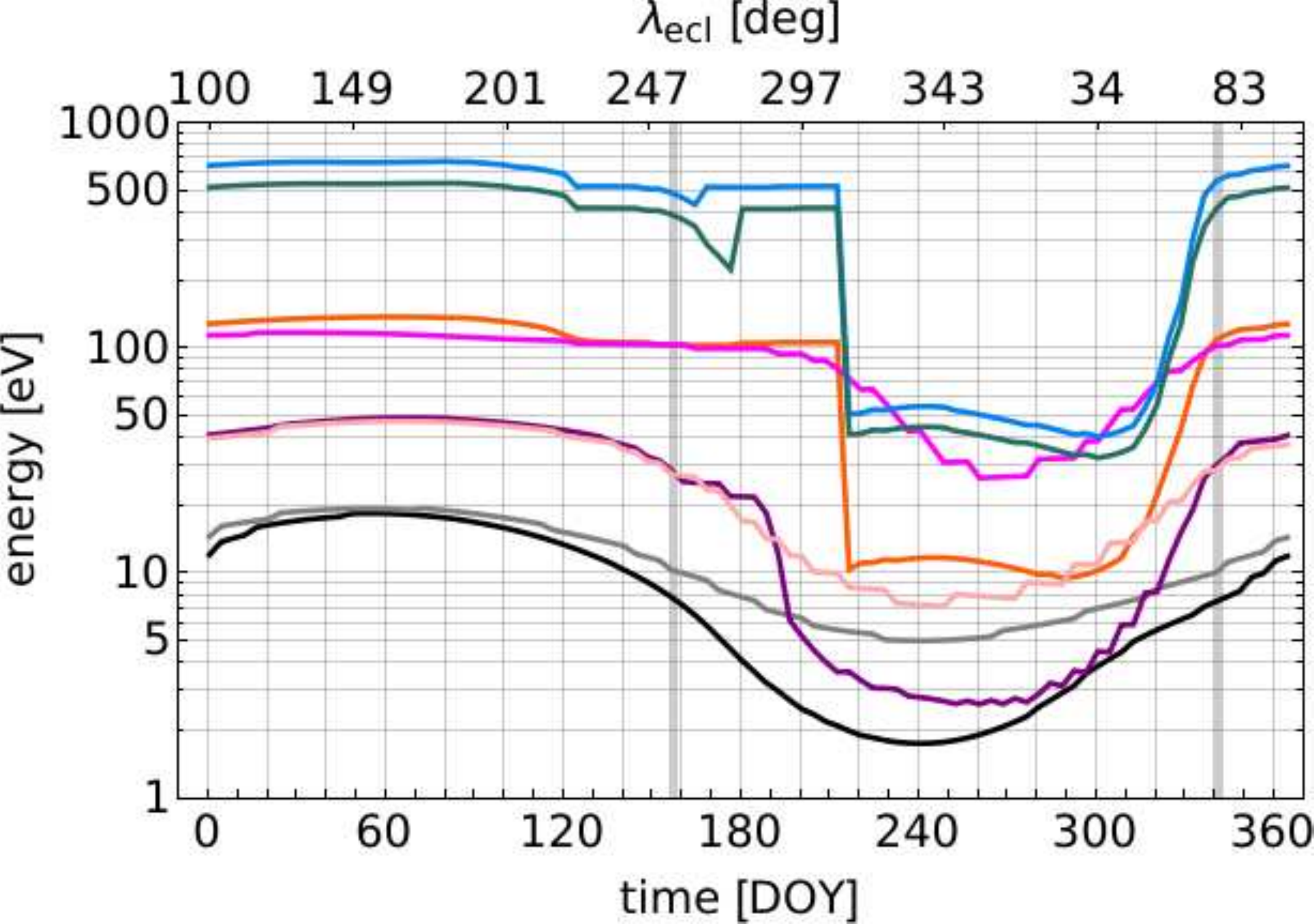} \\
\end{tabular}
\centering
\caption{Same as in Figure~\ref{fig:peakImap} but with the boresight direction kept constant at \elong$=90\degr$ during the entire year (the \textit{IBEX}-Lo observation geometry).}
\label{fig:peakIbex}
\end{figure*}

The maximum fluxes of the ISN species calculated with the boresight direction fixed at \elong$=90\degr$ are presented in Figure~\ref{fig:peakIbex}. The first and most striking difference, when compared with the adjustable boresight pointing, is the reduction of observation time for all species and populations. Also, the time periods  when the maximum flux is observed are limited to the first three months of each year. Next, the maximum fluxes are reduced, differently for different species. Note that the simulations for \elong$=90\degr$ and \elong$\neq90\degr$ were carried out assuming an identical FOV size. 

Tables~\ref{tab:peaks} and \ref{tab:peaks2} summarize the peak locations for the species and populations discussed for two variants of the orientation of the boresight, with \elong\, adjusted to track the maximum flux of the ISN flow and with \elong$=90\degr$ during solar minimum and solar maximum, respectively. Together with the time and ecliptic longitude of the peak, the ratio of the maximum fluxes measured with these two orientations is presented. The largest gain in the ISN flux observed with the boresight directed toward the flux maximum is expected for D, O, and H, which is up to $70\%$. The gain is about $10\%$ for \secHe\, and Ne during solar minimum and about $20-30\%$ during solar maximum. In the case of \priHe, the gain in the maximum flux sampling is $\sim1\%$ during solar minimum and $\sim10\%$ during solar maximum. However, the available time for probing the high flux of both \priHe\, and \secHe\, is greatly extended, from one month with \elong $=90\degr$ to almost a whole year with \elong$\neq90\degr$. Hence, one of the main advantages of the adjustable boresight direction for observation of the ISN gas is an improvement in the statistics due to much longer observation times during the year.

\begin{deluxetable}{c|c|c|c|c|c|c|c}
\centering
\tablecaption{The DOY--\elong\, Combinations for the Maximum Flux of ISN Gas during Solar Minimum.  \label{tab:peaks}}
%\tablewidth{20pt}
\tablecolumns{8}
\tablehead{
\multicolumn{8}{c}{Solar Minimum}\\ \hline
\colhead{} & \colhead{\elongF} & \colhead{DOY$_1$} & \colhead{$\lambda_{\mathrm{ecl,1}}$} & \colhead{\elongT} & \colhead{DOY$_2$} & \colhead{$\lambda_{\mathrm{ecl,2}}$} & \colhead{$F_1/F_2$}
}
\startdata
 \priH & 116$\degr$ & 113 & 212$\degr$ & 90$\degr$ & 57 & 157$\degr$ & 2.52  \\
 \secH & 116$\degr$ & 109 & 208$\degr$ & 90$\degr$ & 57 & 157$\degr$ & 2.35  \\
 \isnH & 116$\degr$ & 109 & 208$\degr$ & 90$\degr$ & 57 & 157$\degr$ & 2.53  \\ \hline
 \priHe & 96$\degr$ & 45 & 145$\degr$ & 90$\degr$ & 29 & 129$\degr$ & 1.01  \\
 \secHe & 104$\degr$ & 45 & 145$\degr$ & 90$\degr$ & 5 & 104$\degr$ & 1.11  \\
 \isnHe & 96$\degr$ & 45 & 145$\degr$ & 90$\degr$ & 29 & 129$\degr$ & 1.01  \\ \hline
 Ne & 128$\degr$ & 117 & 216$\degr$ & 90$\degr$ & 29 & 129$\degr$ & 1.13  \\ \hline
 O & 136$\degr$ & 133 & 232$\degr$ & 90$\degr$ & 33 & 133$\degr$ & 2.64  \\ \hline
 \priD & 124$\degr$ & 121 & 220$\degr$ & 90$\degr$ & 49 & 149$\degr$ & 3.31 \\
 \secD & 120$\degr$ & 109 & 208$\degr$ & 90$\degr$ & 45 & 145$\degr$ & 3.30  \\
 \isnD & 124$\degr$ & 121 & 220$\degr$ & 90$\degr$ & 49 & 149$\degr$ & 3.38  \\
\enddata
\tablecomments{$F$ stands for flux, $\lambda_{\mathrm{ecl}}$ is the ecliptic longitude of the observer. Subscript "1" ("2") refers to observation geometry with \elong$\neq90\degr$ (\elong$=90\degr$). The DOY-\elong combinations are determined within the grid presented in Figure~\ref{fig:grid}, thus $\Delta$\elong$=4\degr$ and $\Delta$DOY$=4$ days.}
\end{deluxetable}
\begin{deluxetable}{c|c|c|c|c|c|c|c}[h!]
\centering
\tablecaption{Same as Table~\ref{tab:peaks}, but during Solar Maximum. \label{tab:peaks2}}
\tablecolumns{8}
\tablehead{
\multicolumn{8}{c}{Solar Maximum}\\ \hline
\colhead{} & \colhead{\elongF} & \colhead{DOY$_1$} & \colhead{$\lambda_{\mathrm{ecl,1}}$} & \colhead{\elongT} & \colhead{DOY$_2$} & \colhead{$\lambda_{\mathrm{ecl,2}}$} & \colhead{$F_1/F_2$}}
\startdata
 \priH & 112$\degr$ & 121 & 220$\degr$ & 90$\degr$ & 81 & 181$\degr$ & 2.34 \\
 \secH & 112$\degr$ & 117 & 216$\degr$ & 90$\degr$ & 77 & 177$\degr$ & 2.18 \\
 \isnH & 112$\degr$ & 121 & 220$\degr$ & 90$\degr$ & 81 & 181$\degr$ & 2.35 \\ \hline
 \priHe & 112$\degr$ & 81 & 181$\degr$ & 90$\degr$ & 29 & 129$\degr$ & 1.13 \\
 \secHe & 112$\degr$ & 65 & 165$\degr$ & 90$\degr$ & 5 & 104$\degr$ & 1.29 \\
 \isnHe & 112$\degr$ & 81 & 181$\degr$ & 90$\degr$ & 29 & 129$\degr$ & 1.14 \\ \hline
 Ne & 132$\degr$ & 125 & 224$\degr$ & 90$\degr$ & 29 & 129$\degr$ & 1.54 \\ \hline
 O & 136$\degr$ & 133 & 232$\degr$ & 90$\degr$ & 33 & 133$\degr$ & 4.17 \\ \hline
 \priD & 124$\degr$ & 125 & 224$\degr$ & 90$\degr$ & 53 & 153$\degr$ & 3.85 \\
 \secD & 120$\degr$ & 117 & 216$\degr$ & 90$\degr$ & 49 & 149$\degr$ & 3.79 \\
 \isnD & 120$\degr$ & 117 & 216$\degr$ & 90$\degr$ & 53 & 153$\degr$ & 3.98 \\
\enddata
\end{deluxetable}

\subsection{Accessible Observation Geometries \label{sec:pxlMaps}}
%% Figures with pixel maps
\begin{figure*}
\begin{tabular}{cc}
\includegraphics[scale=0.25]{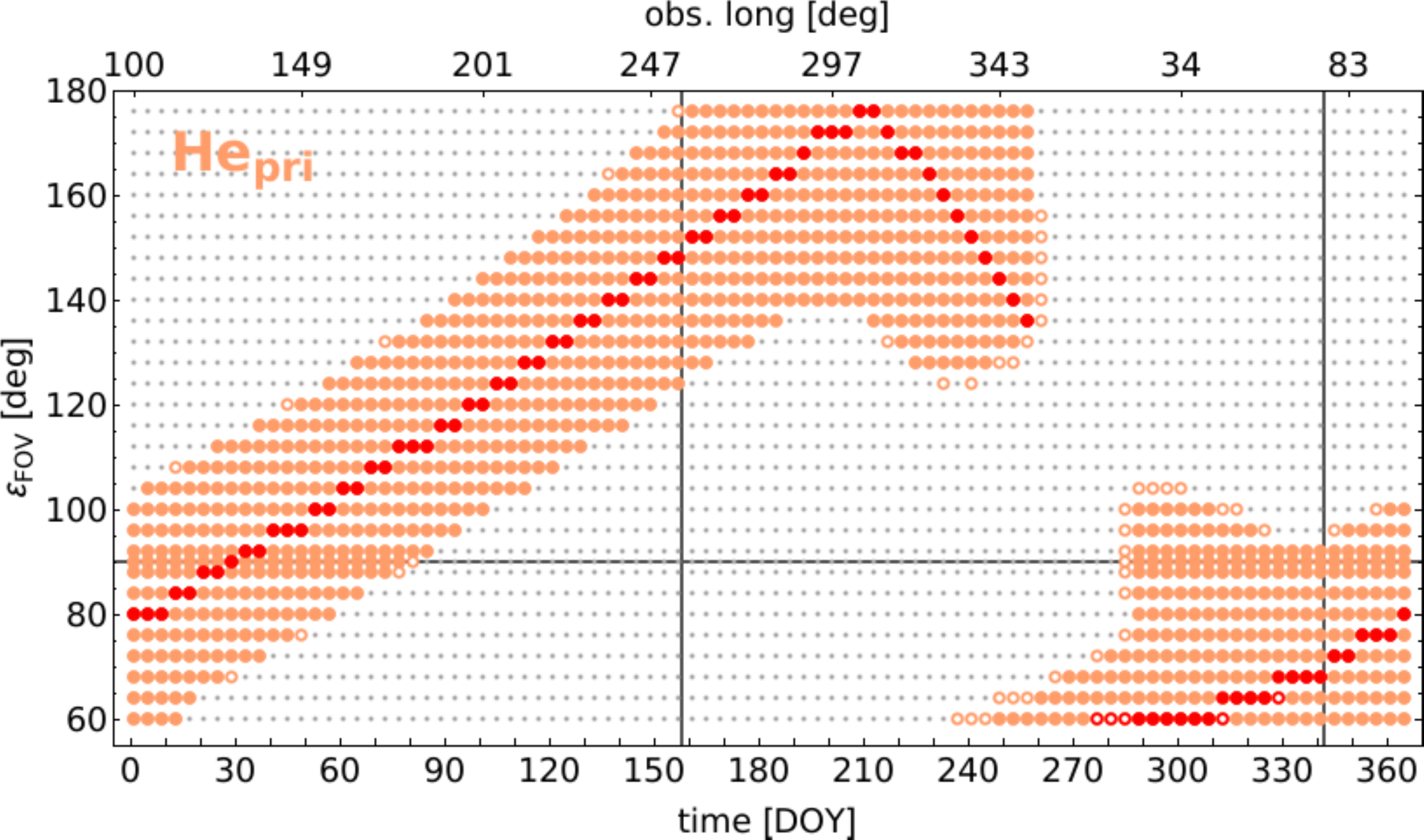} & \includegraphics[scale=0.25]{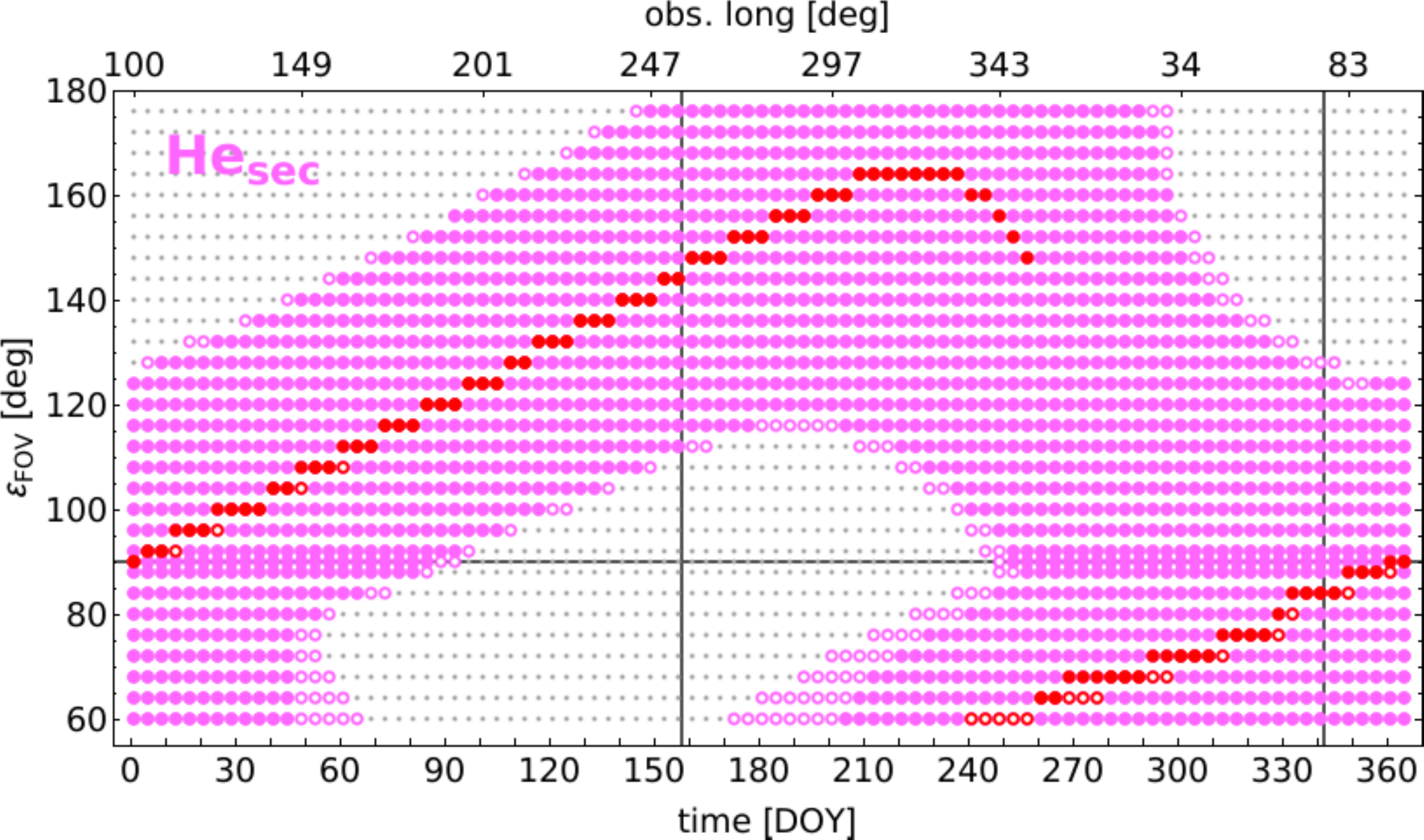} \\
\includegraphics[scale=0.25]{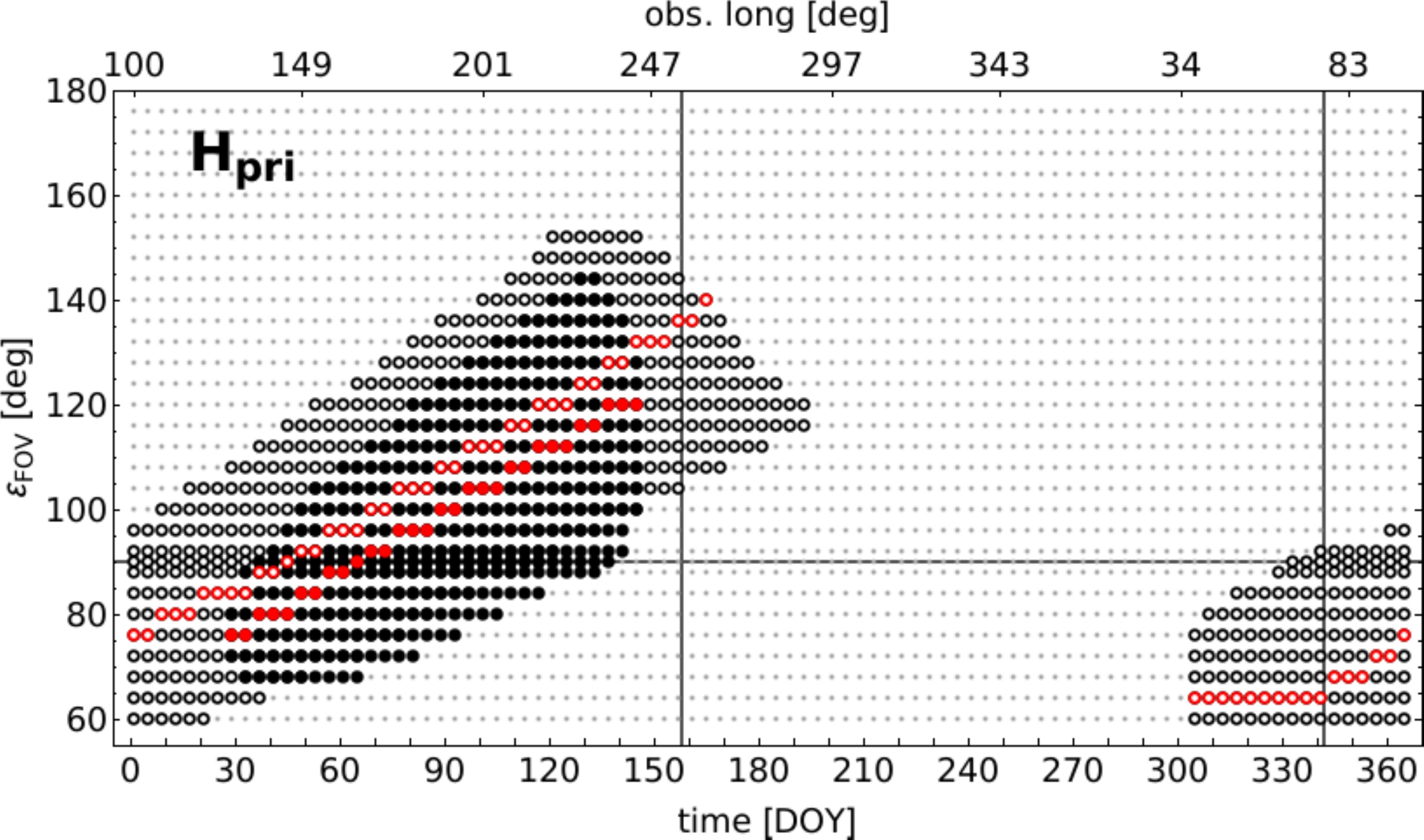} & \includegraphics[scale=0.25]{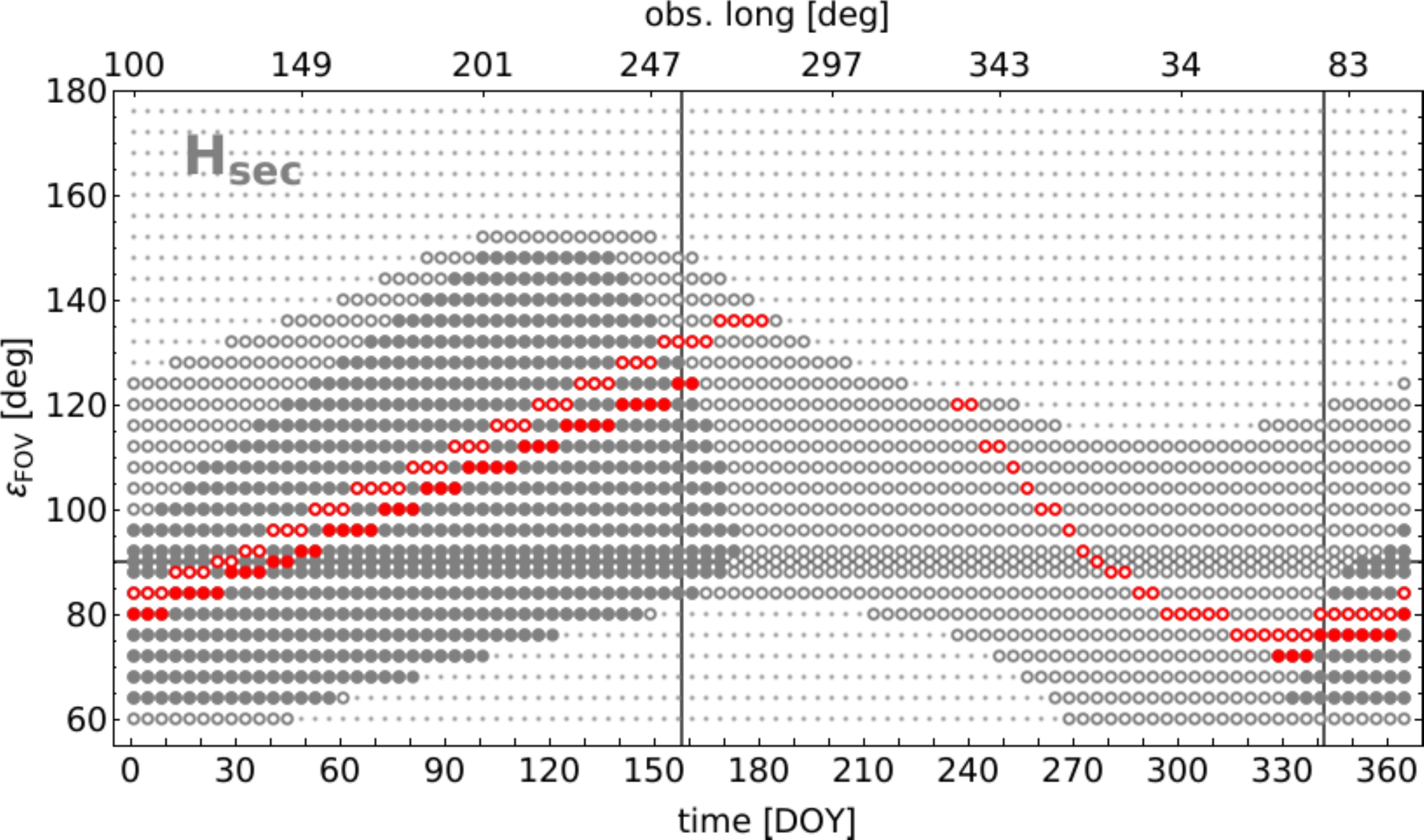} \\
\includegraphics[scale=0.25]{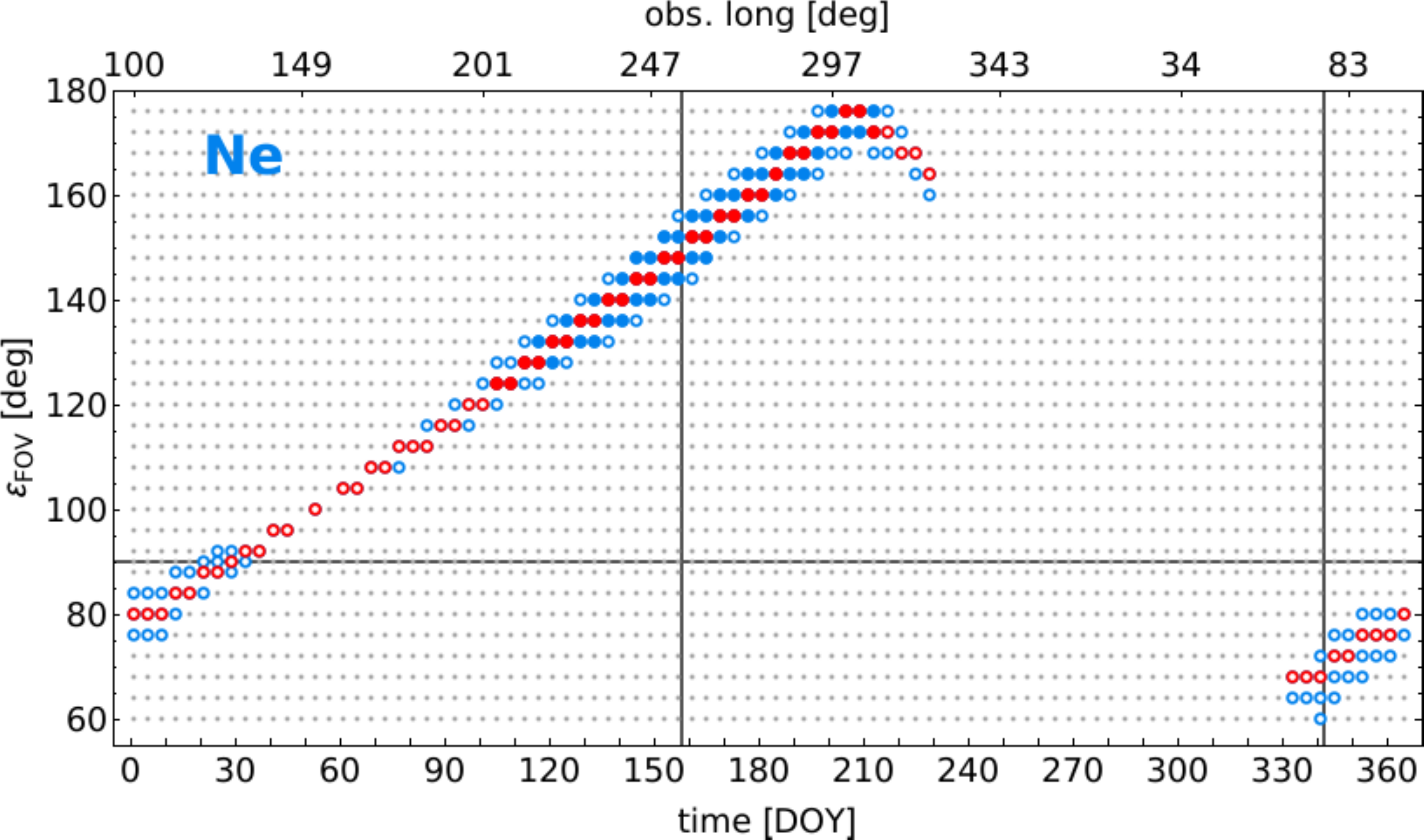} & \includegraphics[scale=0.25]{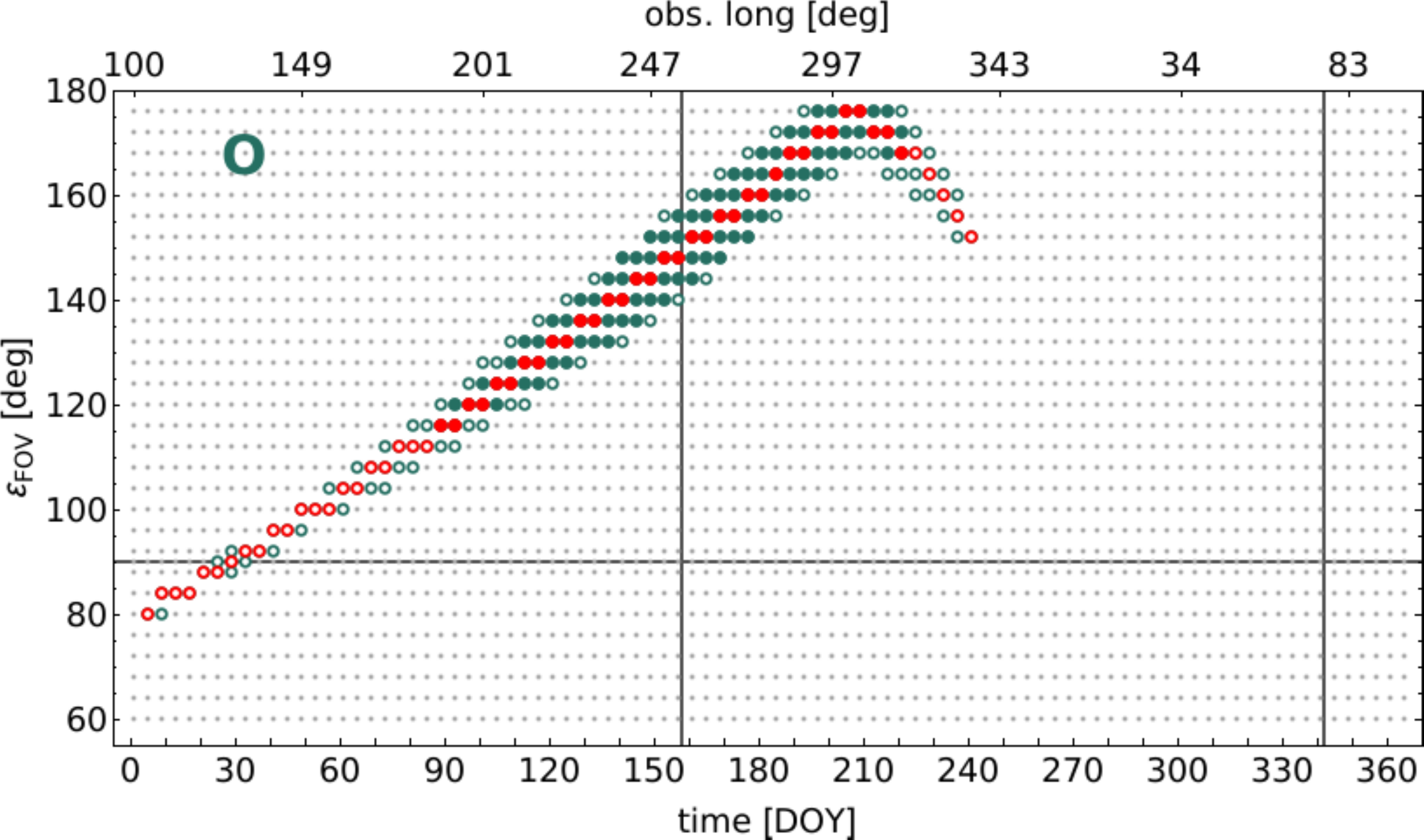} \\
\includegraphics[scale=0.25]{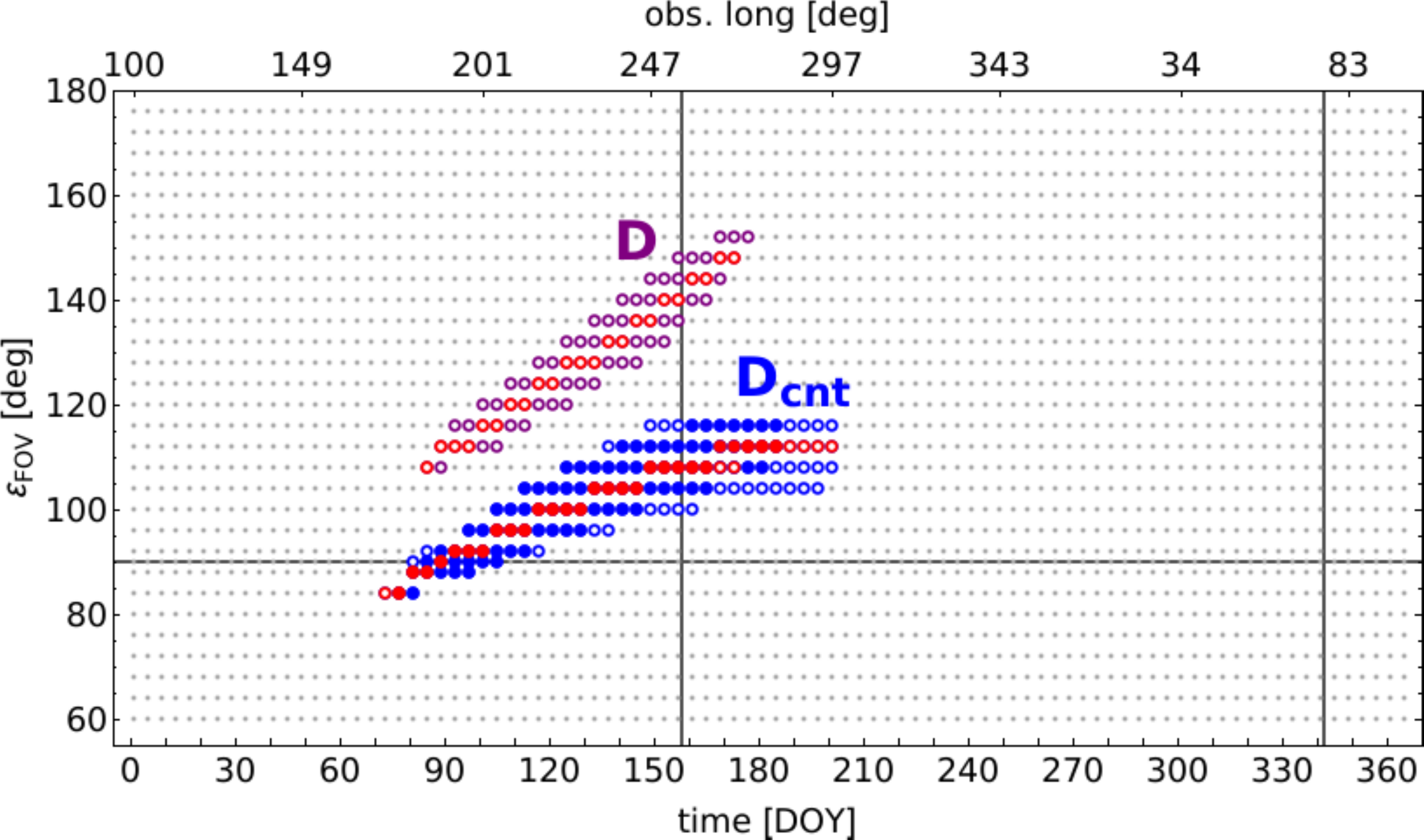} & \\
\end{tabular}
\centering
\caption{Maps of DOY--\elong\, combinations when the ISN gas of a given species is detectable in at least three spin angle bins. Filled circles correspond to solar maximum, and open dots correspond to solar minimum. The energy and flux requirements are adopted as stated in Table~\ref{tab:summary}. The blue circles in the \isnD\, panel present the observation region determined by the analysis of the \isnD\, count rates as described in Section~\ref{sec:D2H}. The red circles in each panel present the DOY--\elong\, combination for the maximum flux within the region showed. 
\label{fig:pxlMaps}}
\end{figure*}

Tracking the ISN flux maximum is not the only possible observation scenario to fulfill the SOs listed in Section~\ref{sec:sciOpp}. With an instrument with a capability to adjust the boresight direction, the \elong\, can be set off the peak and provide successful observations for most of the species discussed. Figure~\ref{fig:pxlMaps} shows DOY--\elong\, combinations accessible with an adjustable boresight direction of the FOV for given species and populations through the entire year. We applied similar criteria to select these combinations according to the limitations in energy and flux as described in Section~\ref{sec:methods} (and summarized in Table~\ref{tab:summary}). A given DOY--\elong\, combination was included, if the signal fulfills the flux and energy criteria for more than three spin angle bins. Open circles illustrate results for solar minimum, and filled circles illustrate results for solar maximum. In each panel the location of maximum flux is shown in red. A comparison of the open and filled circles indicates how much the potential observation seasons can be extended during the solar minimum, which is the most pronounced for \isnH. Also, the observation season for Ne is longer during solar minimum, with detection possible in the cone region starting from about DOY~330. 

With a carefully defined observation geometry, both \priHe\, and \secHe\, might be accessible for detection throughout the entire year during both solar minimum and solar maximum. However, as further discussed in Section~\ref{sec:secondaryHe}, \secHe\, is observed in the wings of the spin angle distribution for many DOY--\elong\, combinations, and to fully resolve it from \priHe\, a dedicated observation geometry needs to be applied. The same applies to \secH\, as further discussed in Section~\ref{sec:secondaryH}. The preferable observation season for \priH\, and \secH\, is the solar minimum, when \secH\, can be tracked during the entire year (see Figure~\ref{fig:pxlMaps}). During solar maximum, the season for \isnH\, is limited to about 100 days for \priH\, and about 180 days for \secH, but the contribution from \secHe\, in the detection process also needs to be taken into account (see Section~\ref{sec:secondaryHe}). 

In contrast to \isnHe\, and \isnH, for which the range of \elong\, accessible for observations is wide, for Ne and O the \elong\, needs to be set precisely to obtain the peak flux. As illustrated in Figure~\ref{fig:pxlMaps}, the Ne and O observation season is limited to the first eight months of a year, with an additional month of observations for Ne at the end of a year during solar minimum. The observation of \isnD\, is possible during the second quarter of a year with \elong\, directed either to the maximum of the flux or to the direction where the contribution from the \isnHe\, is negligibly small (more in Section~\ref{sec:D2H}). In contrast to the first scenario, the latter one allows for observations also during solar maximum. However, in that case, the observations are hampered by very low fluxes and thus very low counting statistics; a discussion of possible methods to observe \isnD\, is further provided in Section~\ref{sec:D2H}.

%% Summary table
\begin{deluxetable*}{c c c c c}[t]
\centering
\tablecaption{Available Observation Geometries for Various ISN Species and Populations throughout a Year during Solar Minimum and Solar Maximum. \label{tab:summary}}
\tablecolumns{5}
\tablehead{
\colhead{Target} &
\colhead{DOY} &
\colhead{\elong\,} & \colhead{Limitations} & \colhead{Comment} \\
\colhead{} & \colhead{} &
\colhead{[$\degr$]} & \colhead{E [eV], F \fluxU} & \colhead{} 
}
\startdata
 \priHe, sol. min & 1--365 & 60--176 & E$\geq$20, F$\geq$100 & Section~\ref{sec:peaks} \\
 \priHe, sol. max & 1--365 & 60--176 & E$\geq$20, F$\geq$100 & Section~\ref{sec:peaks} \\
\hline
 \secHe, sol. min & 1--365 & 60--176 & E$\geq$20, F$\geq$100 & Section~\ref{sec:secondaryHe} \\
 \secHe, sol. max & 1--365 & 60--176 & E$\geq$20, F$\geq$100 & Section~\ref{sec:secondaryHe} \\
\hline
 \priH, sol. min & 1--193, 305--365 & 60--152 & E$\geq$10, F$\geq$100 & Section~\ref{sec:peaks} \\
 \priH, sol. max & 29--145 & 68--144 & E$\geq$10, F$\geq$100 & Section~\ref{sec:peaks} \\
\hline
 \secH, sol. min & 1--365 & 60--152 & E$\geq$10, F$\geq$100 & Section~\ref{sec:secondaryH} \\
 \secH, sol. max & 1--173, 329--365 & 64--148 & E$\geq$10, F$\geq$100 & Section~\ref{sec:secondaryH} \\
\hline
 Ne, sol. min & 1--229, 333--365 & 60--176 & E$\geq$10, F$\geq$100 & Section~\ref{sec:NeO} \\
 Ne, sol. max & 105--213 & 124--176 & E$\geq$10, F$\geq$100 & Section~\ref{sec:NeO} \\
\hline
 O, sol. min & 5--241 & 80--176 & E$\geq$10, F$\geq$100 & Section~\ref{sec:NeO} \\
 O, sol. max & 89--221 & 116--176 & E$\geq$10, F$\geq$100 & Section~\ref{sec:NeO} \\
\hline
 \isnD, sol. min (max. flux) & 85--177 & 108--152 & E$\geq$10, F$\geq$1 & Section~\ref{sec:D2H}\\
 \isnD, sol. max (max. flux) & -- & -- & E$\geq$10, F$\geq$1 & Section~\ref{sec:D2H} \\
\hline
 \isnD, sol. min (\ionD\, ratio) & 73--201 & 84--116 & E$\geq$10, F$\geq$1 & Section~\ref{sec:D2H}\\
 \isnD, sol. max (\ionD\, ratio) & 77--185 & 84--116 & E$\geq$10, F$\geq$1 & Section~\ref{sec:D2H} \\
\hline
 indirect beam \priHe, sol. min  & 237--321 & 60--72 & E$\geq$20, F$\geq$100 &  Section~\ref{sec:twoBeams} \\
 indirect beam \priHe, sol. max & 249--321 & 60--72 & E$\geq$20, F$\geq$100 &  Section~\ref{sec:twoBeams} \\
 \hline
 indirect beam \secHe, sol. min & 173--289 & 60--72 & E$\geq$20, F$\geq$100 &  Section~\ref{sec:twoBeams} \\
 indirect beam \secHe, sol. max & 205--289 & 60--72 & E$\geq$20, F$\geq$100 &  Section~\ref{sec:twoBeams} \\
\enddata
\tablecomments{The observational limitations for flux and energy (in detector reference frame) and visibility in minimum three spin angle bins are applied. The content of the table corresponds to Figure~\ref{fig:pxlMaps}. }
\tablecomments{Indirect beams of \priHe\, (\secHe\,) flux are observed up to about DOY~321 (DOY~289) because later direct and indirect beams start to mix up. }
\end{deluxetable*}

\section{Discussion \label{sec:discussion}}
Figure~\ref{fig:pxlMaps} shows the accessible DOY--\elong\, combinations for various species and populations separately. However, to successfully address the SOs described in Section~\ref{sec:sciOpp} (Figure~\ref{fig:graph}), a dedicated observation geometries need to be determined. In the following sections, we discuss observation geometries favorable for observations of secondary populations of H and He (Section~\ref{sec:secondaries}), ISN Ne and O gas (Section~\ref{sec:NeO}), the observation possibilities for \isnD\, (Section~\ref{sec:D2H}), and for the indirect beams of \isnHe\, (Section~\ref{sec:twoBeams}).

\subsection{Secondary Populations \label{sec:secondaries}}
The secondary populations are products of charge exchange between the ISN atoms and ions in OHS \citep{baranov_malama:93}. The secondary populations of H \citep{lallement_etal:05a}, He \citep{bzowski_etal:12a, kubiak_etal:14a}, and O \citep{park_etal:16a} have been discovered in the heliosphere. Thus far, the secondary populations could only be analyzed based on the \textit{IBEX}-Lo \citep{kubiak_etal:16a, bzowski_etal:17a, baliukin_etal:17a, kubiak_etal:19a, park_etal:19a} and \textit{Ulysses} \citep{wood_etal:17a} measurements. In Sections~\ref{sec:secondaryHe} and \ref{sec:secondaryH} we focus on how to observe the ISN gas to see secondary populations of He and H separately from the primary flow populations with measurements from the Earth's orbit. 

\subsubsection{Secondary He \label{sec:secondaryHe}}
%% Figure with secondaries options for He
\begin{figure*}
\begin{tabular}{ccc}
\includegraphics[scale=0.3]{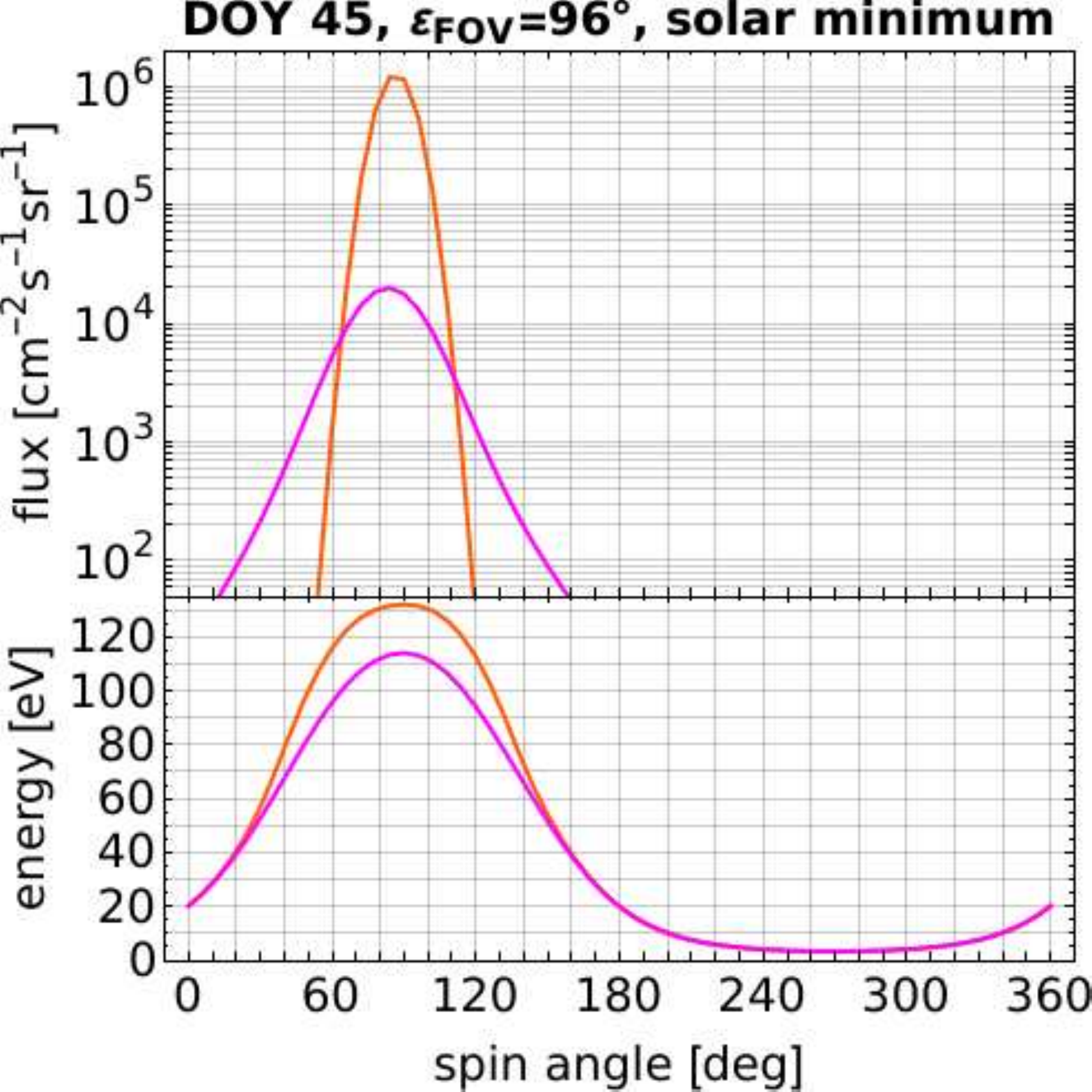} & \includegraphics[scale=0.3]{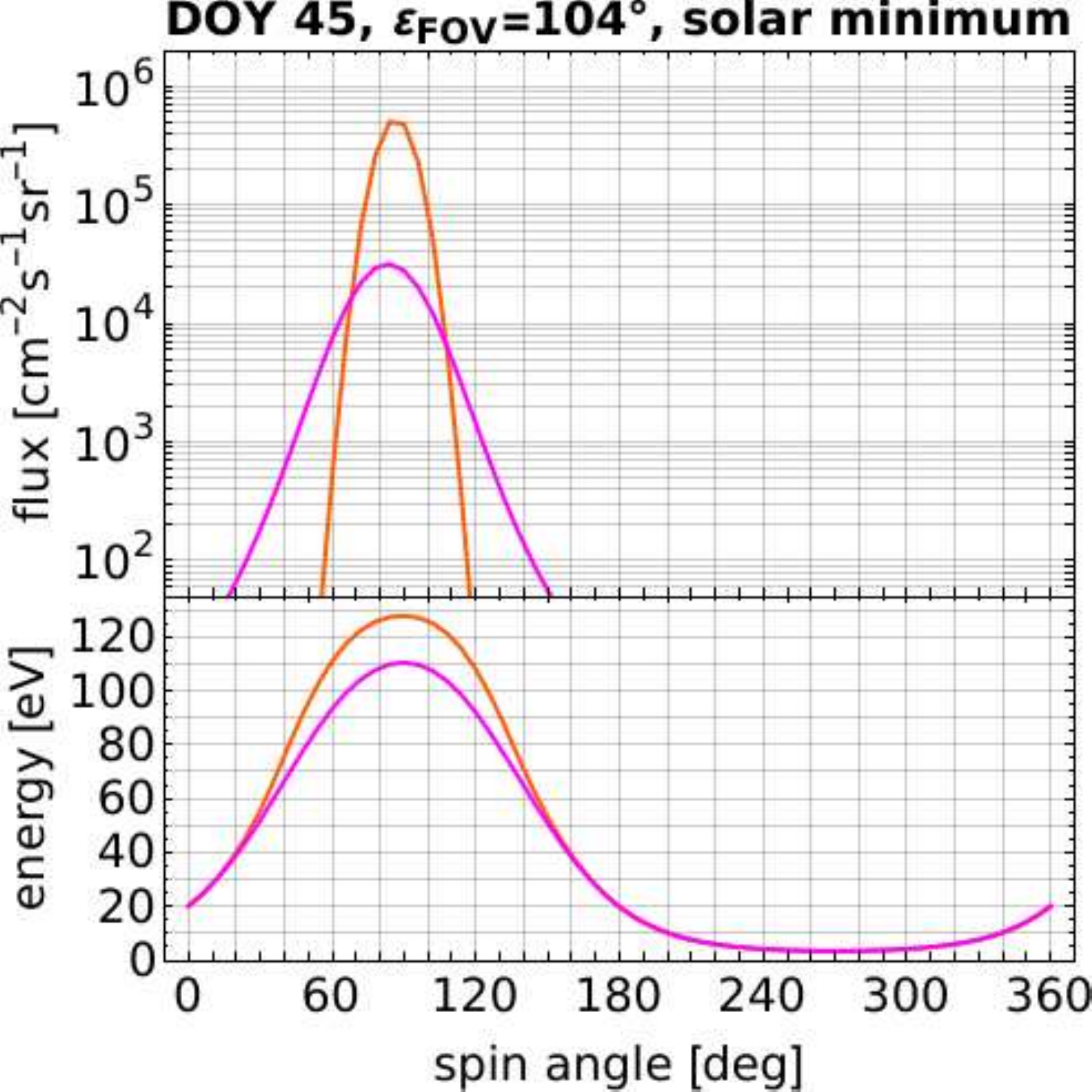} & \includegraphics[scale=0.3]{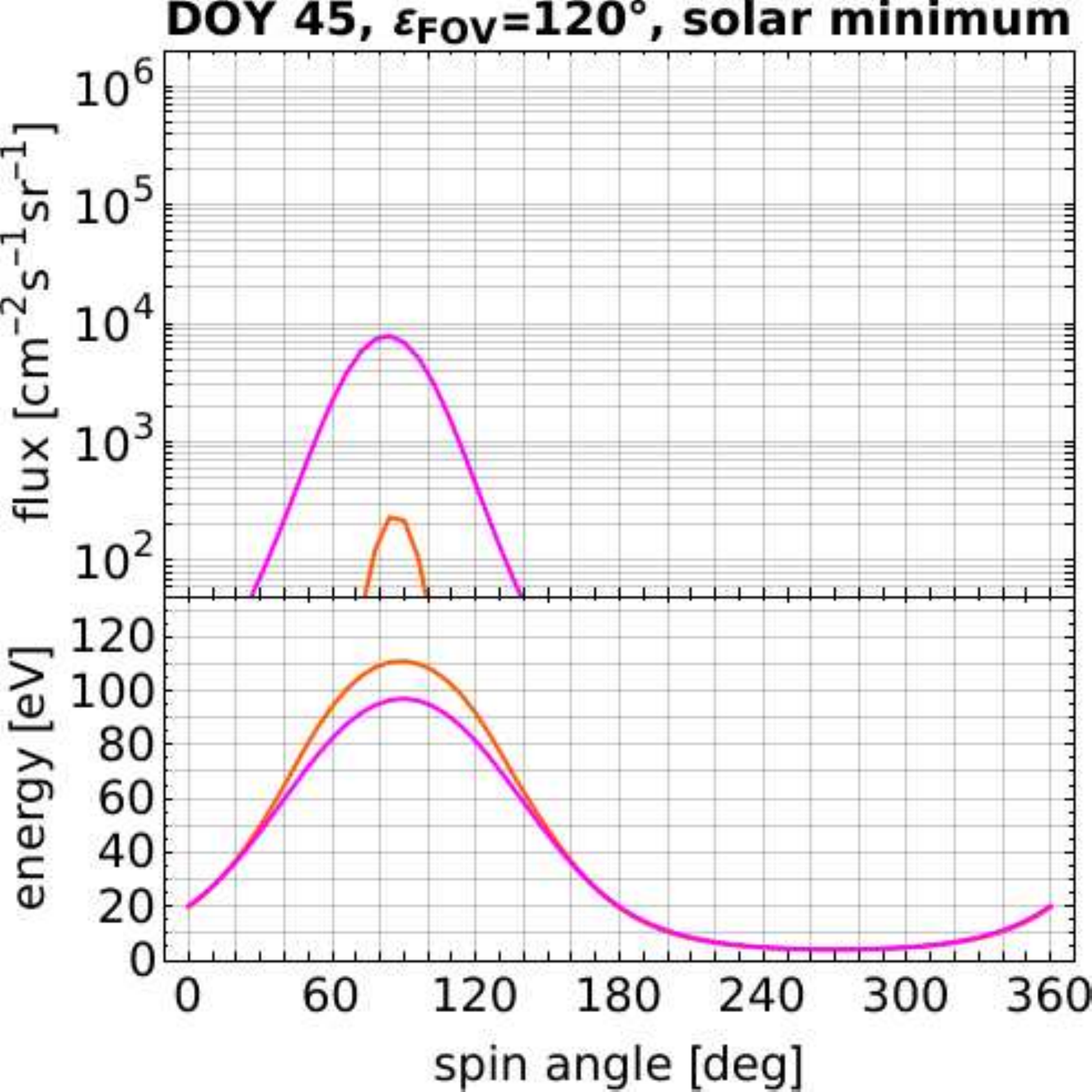} \\
\includegraphics[scale=0.3]{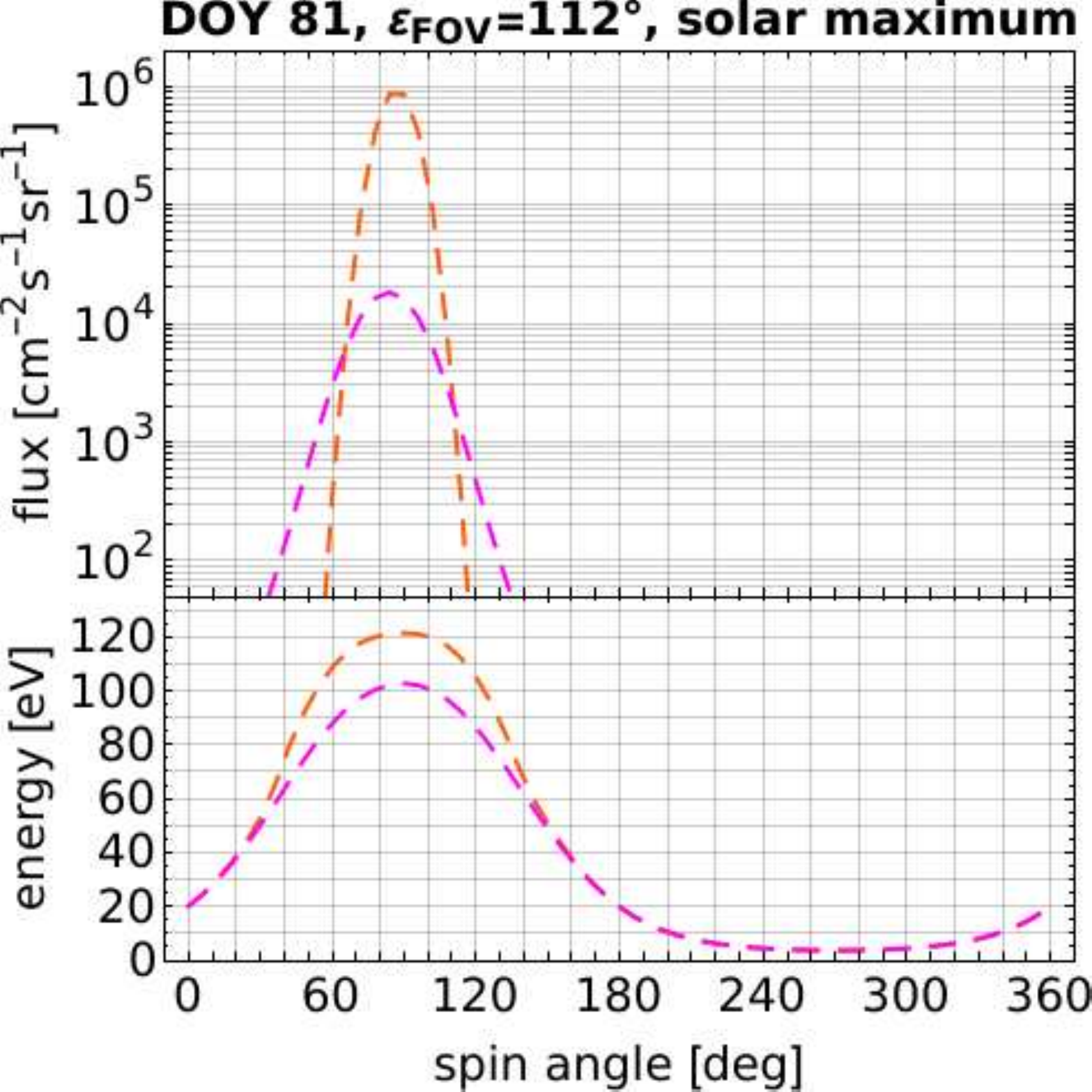} & \includegraphics[scale=0.3]{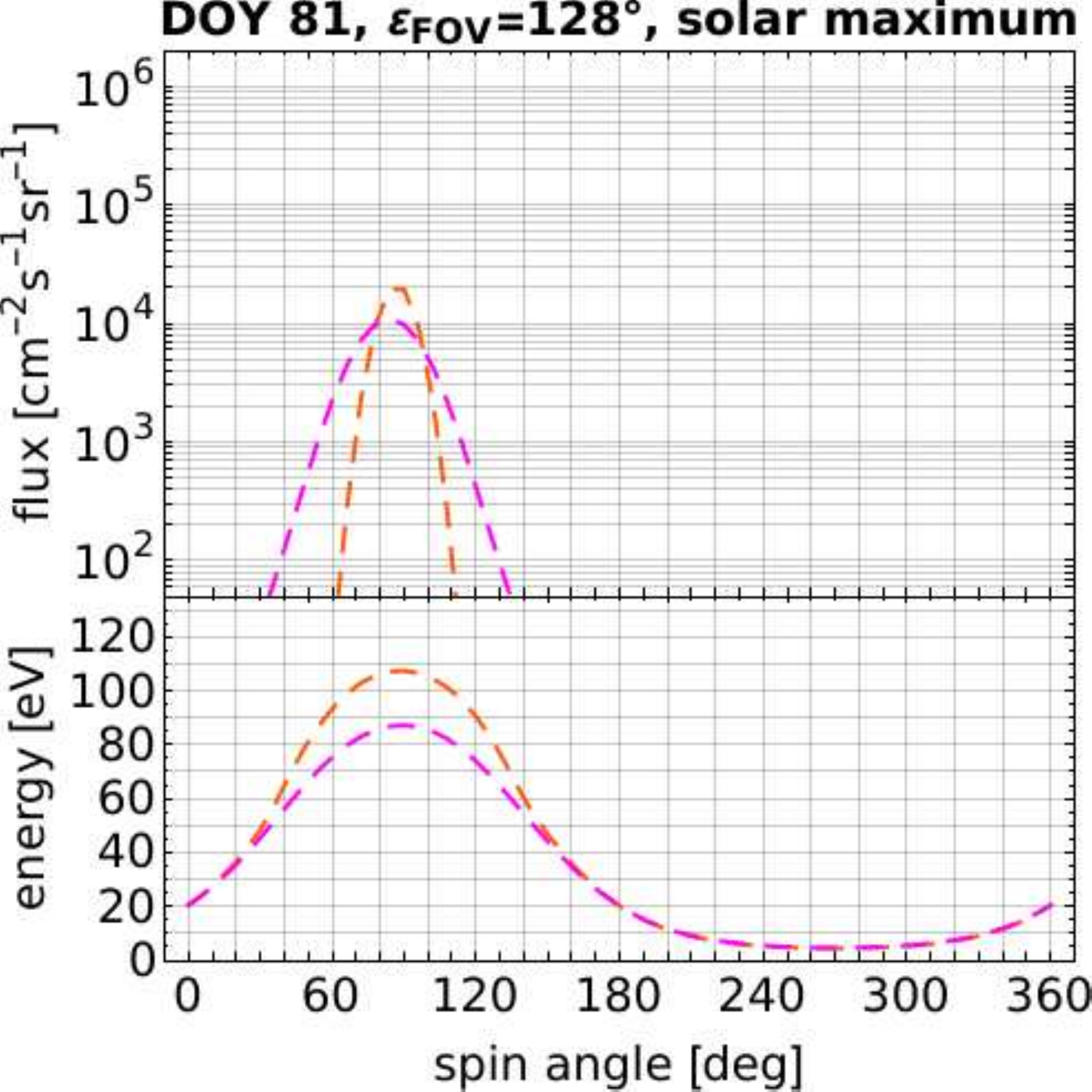} & \includegraphics[scale=0.3]{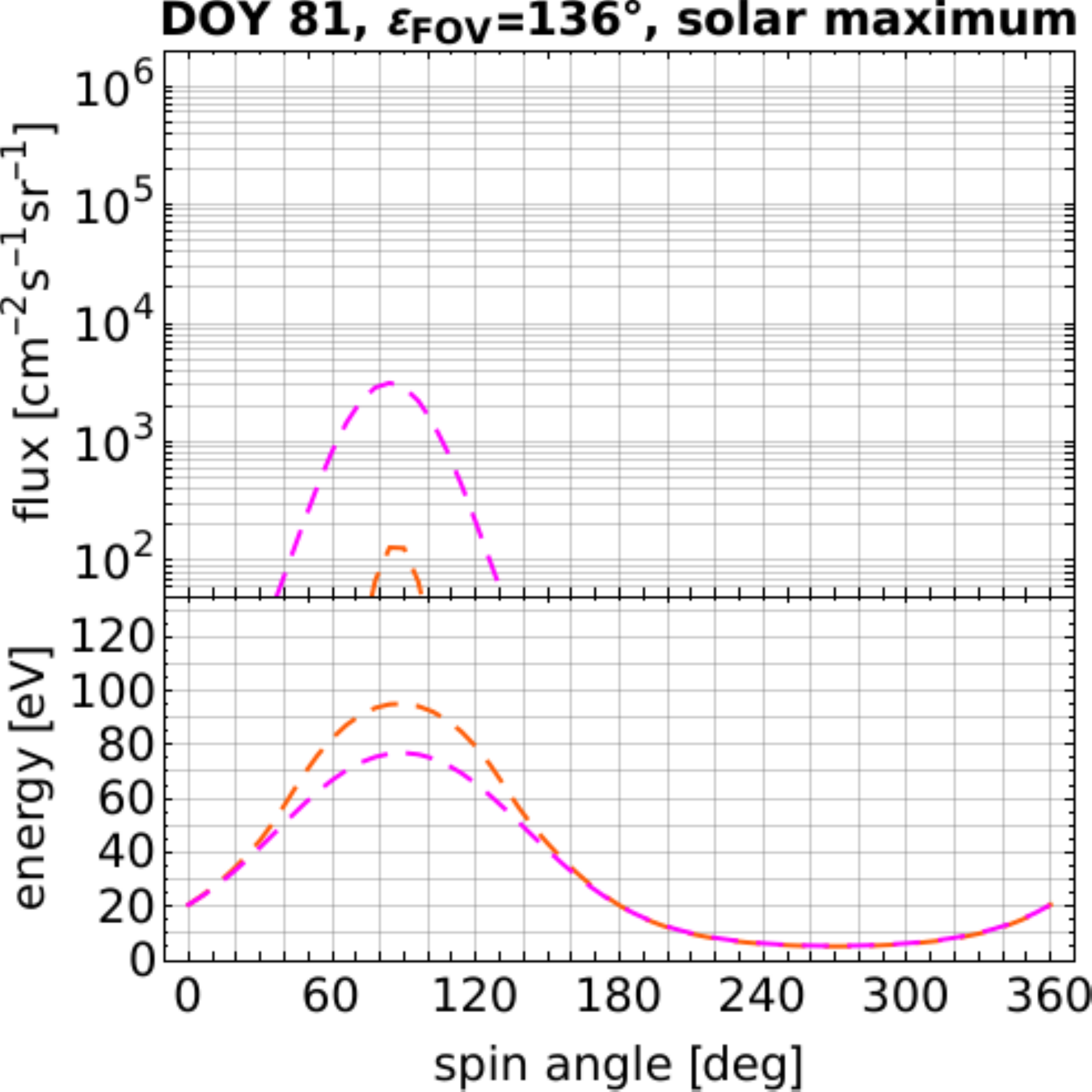} \\
\includegraphics[scale=0.3]{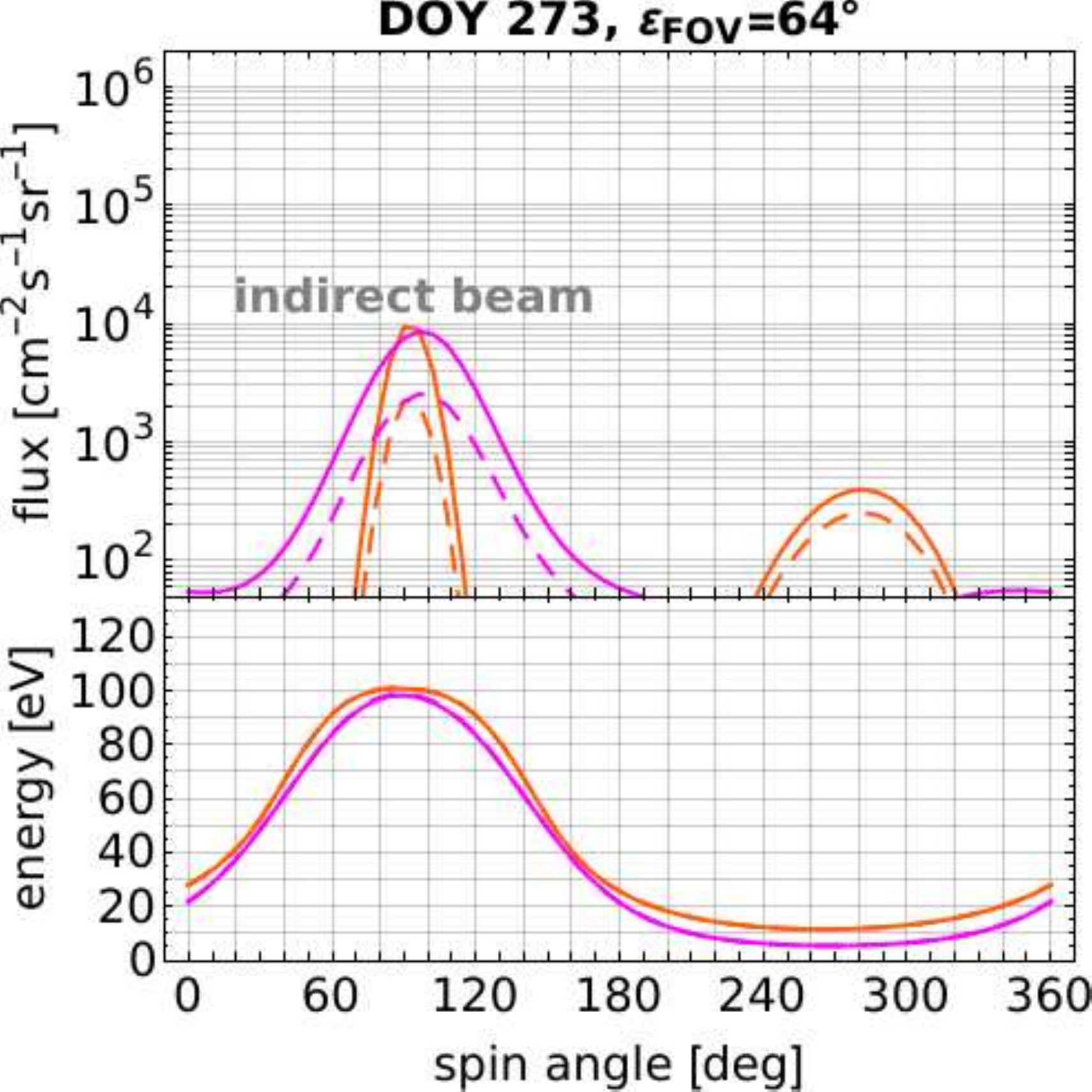} & \includegraphics[scale=0.3]{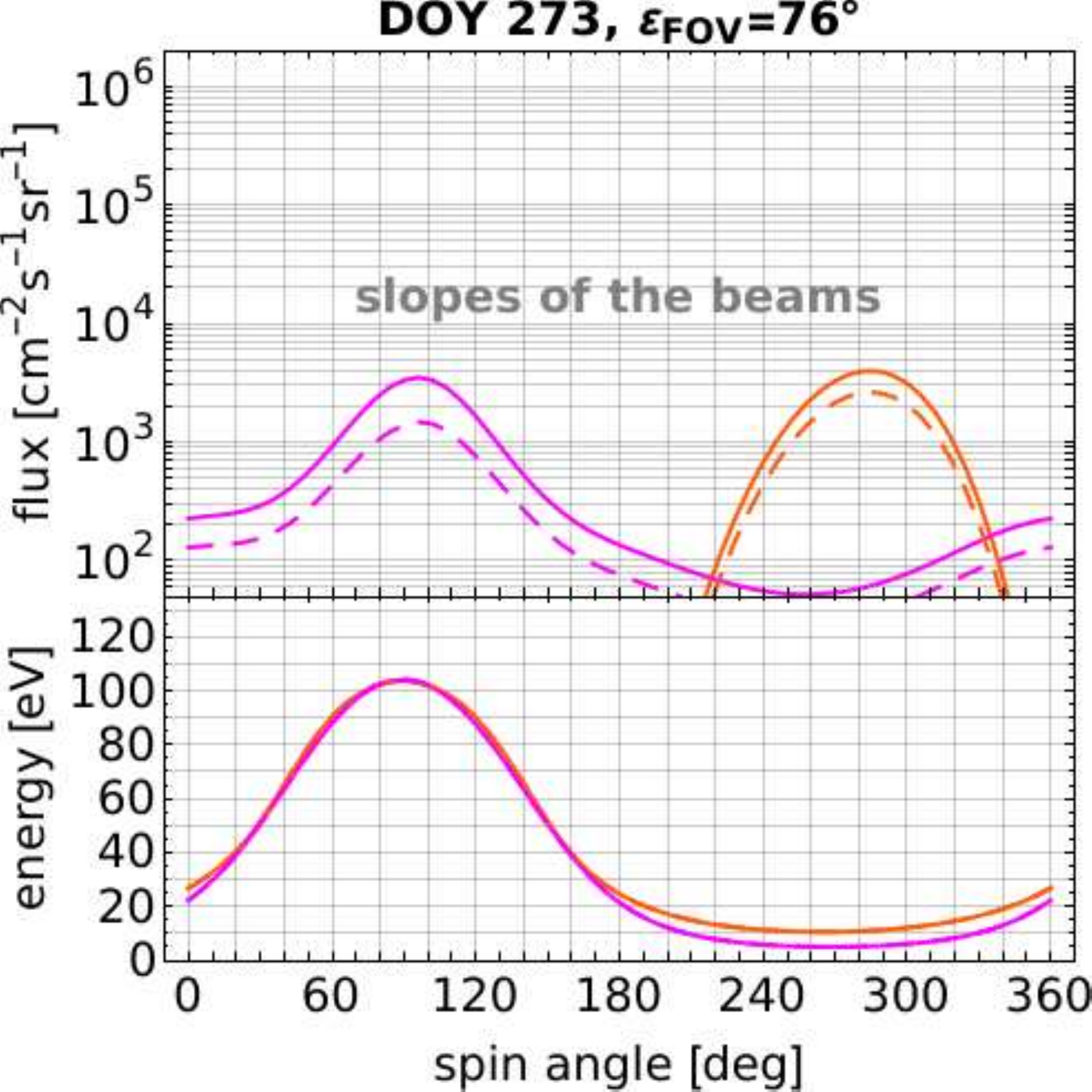} & \includegraphics[scale=0.3]{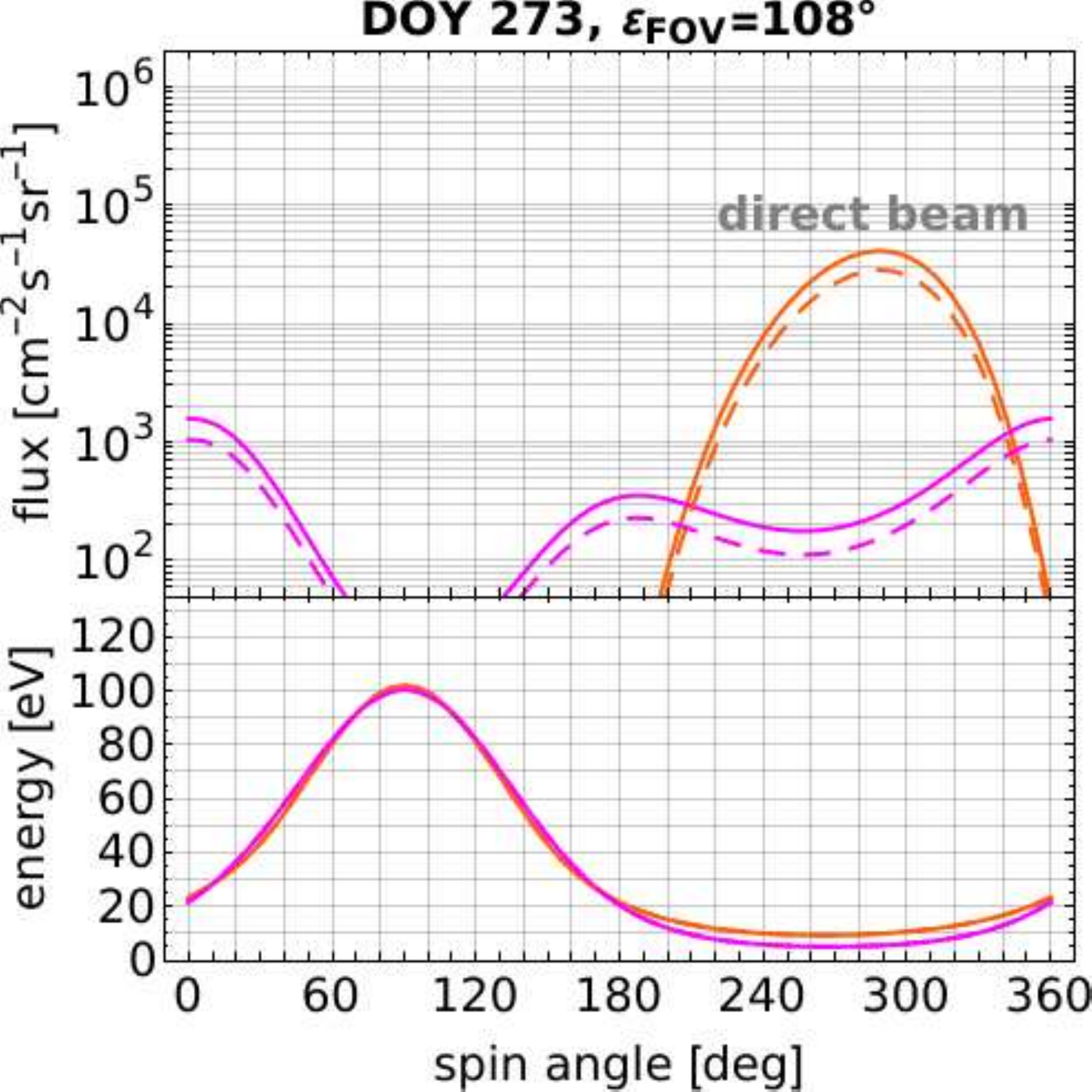} \\
\end{tabular}
\centering
\caption{Fluxes and corresponding energies of \priHe\, (orange) and \secHe\, (magenta) as a function of spin angle for various DOY--\elong\, combinations. Top row: the maximum flux of \isnHe\, (sum of \priHe\, and \secHe) is observed on DOY~45 at \elong$=96\degr$ (due to the maximum flux of \priHe) during solar minimum, while the maximum of \secHe\, is observed at \elong$=104\degr$ on the same day.  Middle row: the maximum flux of \isnHe\, (sum of \priHe\, and \secHe) is observed on DOY~81 at \elong$=112\degr$ during solar maximum, while the maximum of \secHe\, is observed  at \elong$=116\degr$ on this day (not shown). Bottom row: various \elong\, orientations on DOY~273 during solar minimum (solid lines) and solar maximum (dashed lines) illustrating the capability to observe the direct and indirect beams. \label{fig:secondariesHe}}
\end{figure*}

The flow of \secHe\, detected from the vicinity of Earth's orbit appears deflected in angle by $7.9\degr$ relative to \priHe\, (see Section~\ref{sec:methods}). That apparent spatial separation of fluxes allows \textit{IBEX}-Lo to observe \secHe\, every November/December, when the \secHe\, flux exceeds the \priHe\, flux. However, when running a campaign to track the ISN flow with the detector's boresight of an instrument with \textit{IMAP}-Lo capabilities, the \secHe\, will remain below the high \priHe\, flow (see Figure~\ref{fig:peakImap}). Thus, a dedicated observation campaign for \secHe\, needs to be defined as well. 

Favorable observation geometries for the \secHe\, can be implemented in multiple ways. As a first possibility, the boresight could be pointed toward the secondary population peak. However, then the peak of \secHe\, is accompanied by a high flux of \priHe\, and thus both populations are observed simultaneously, with \priHe\, in the center of the distribution and \secHe\, in its wings as a function of spin angle. That case is presented in the top and middle rows of Figure~\ref{fig:secondariesHe}. The maximum flux of \isnHe\, (sum of \priHe\, and \secHe) is observed on DOY~45 at \elong$=96\degr$ during solar minimum (top row) and on DOY~81 at \elong$=112\degr$ during solar maximum (middle row). During these DOYs, the maximum flux of \secHe\, is observed at \elong$=104\degr$ and \elong$=116\degr$ (not shown), respectively. However, the peak of \secHe\, flux is covered by the \priHe\, flux and dominates over \priHe\, only in the wings of the distribution in spin angle. Thus, as a second possibility, the boresight could be pointed toward directions where the \secHe\, flux is separated substantially from the \priHe\, flux in spin angle, as observed on DOY~273 at \elong$=76\degr$ in the middle panel of the bottom row of Figure~\ref{fig:secondariesHe}. In that case the peaks of \secHe\, and \priHe\, are separated in spin angle, and additionally the energies are high enough for detection only for \secHe. A third possibility is to select DOY--\elong combinations for which the \secHe\, flux dominates significantly over the \priHe\, flux, as on DOY~45 at \elong$=120\degr$ or on DOY~81 at \elong$=136\degr$, where it is a few degrees off the maximum flux. There are DOY--\elong\, combinations for which both \priHe\, and \secHe\, fluxes are comparable, such as on DOY~81 at \elong$=128\degr$ (middle panel of the middle row in Figure~\ref{fig:secondariesHe}). The difference in the distribution width in spin angle of the populations (\secHe\, is wider than \priHe) allows us to distinguish the contributions to the measured flux from these two populations by means of model-assisted data analysis.

The preferred viewing direction depends on the adopted observation goal. The study shows that to separate the \secHe\, from the \priHe\, the FOV should be directed off the peak of the populations. The fluxes are overall lower than in the peak, but within the observability limit, and the separation is possible because the flux of \secHe\, is much higher than that of \priHe\, (e.g., $\sim 50$ times on DOY~45 at \elong$=120\degr$ or $\sim 30$ times on DOY~81 at \elong$=136\degr$ in Figure~\ref{fig:secondariesHe}).  The best observation strategy to study the \secHe\, flux can be to detect it either in the wings of the spin angle distribution of the \isnHe\, flux with \priHe\, flux in the core (as done on \textit{IBEX}-Lo) or with \elong\, at few degrees off the maximum flux of both \priHe\, and \secHe\, where differences in flux magnitude and energy of the atoms facilitate detection. A comparison of DOY--\elong\, combination maps for \priHe\, and \secHe\, in Figure~\ref{fig:pxlMaps} shows that \secHe\, is accessible to observation for a larger number of DOY--\elong\, combinations than \priHe. Thus, it is possible to select such offset angles of \secHe\, from \priHe, which allow for their separation throughout the entire year.

\subsubsection{Secondary H \label{sec:secondaryH}}
%% Figure with secondaries options for H
\begin{figure*}
\begin{tabular}{ccc}
\includegraphics[scale=0.3]{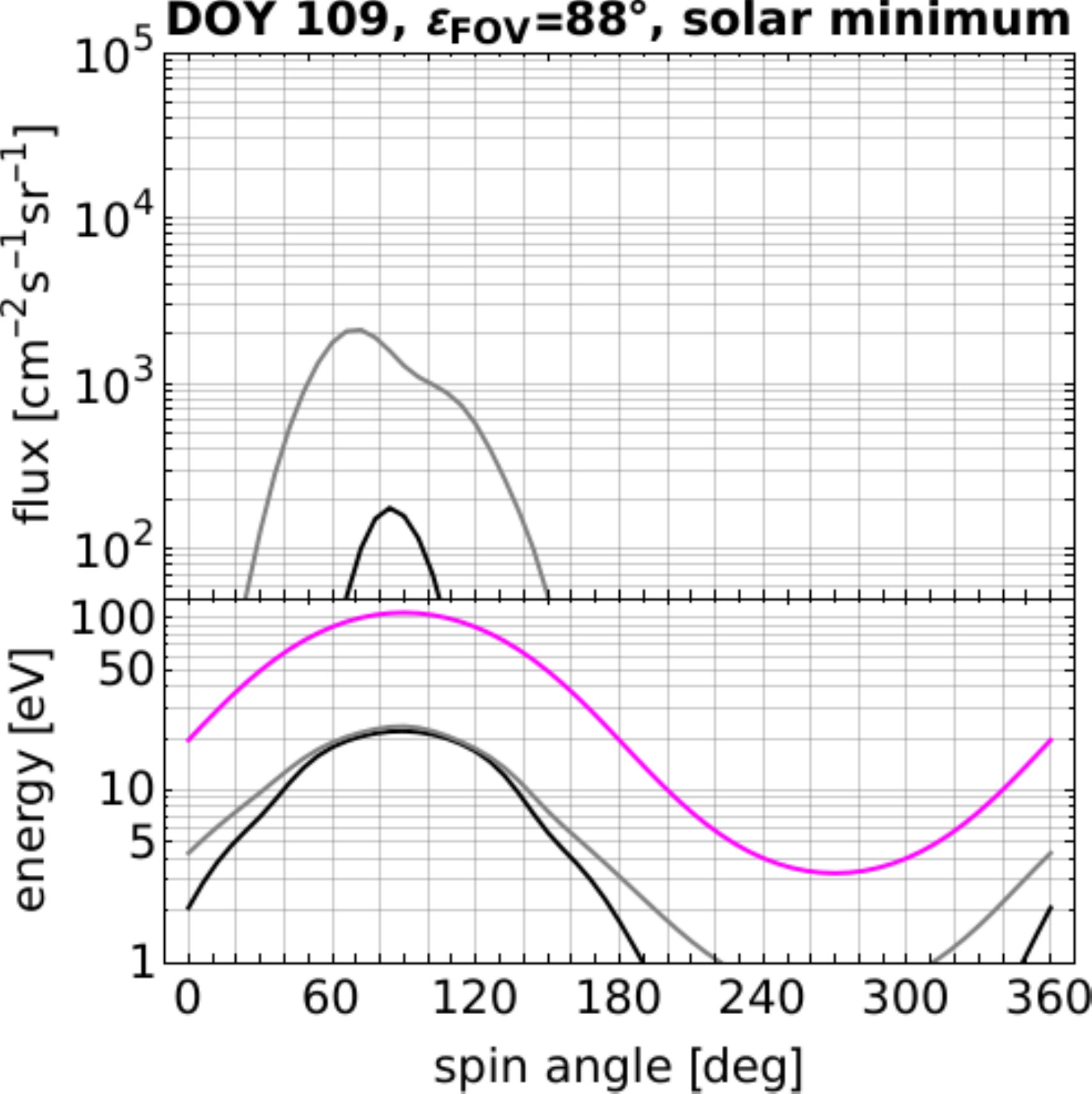} & \includegraphics[scale=0.3]{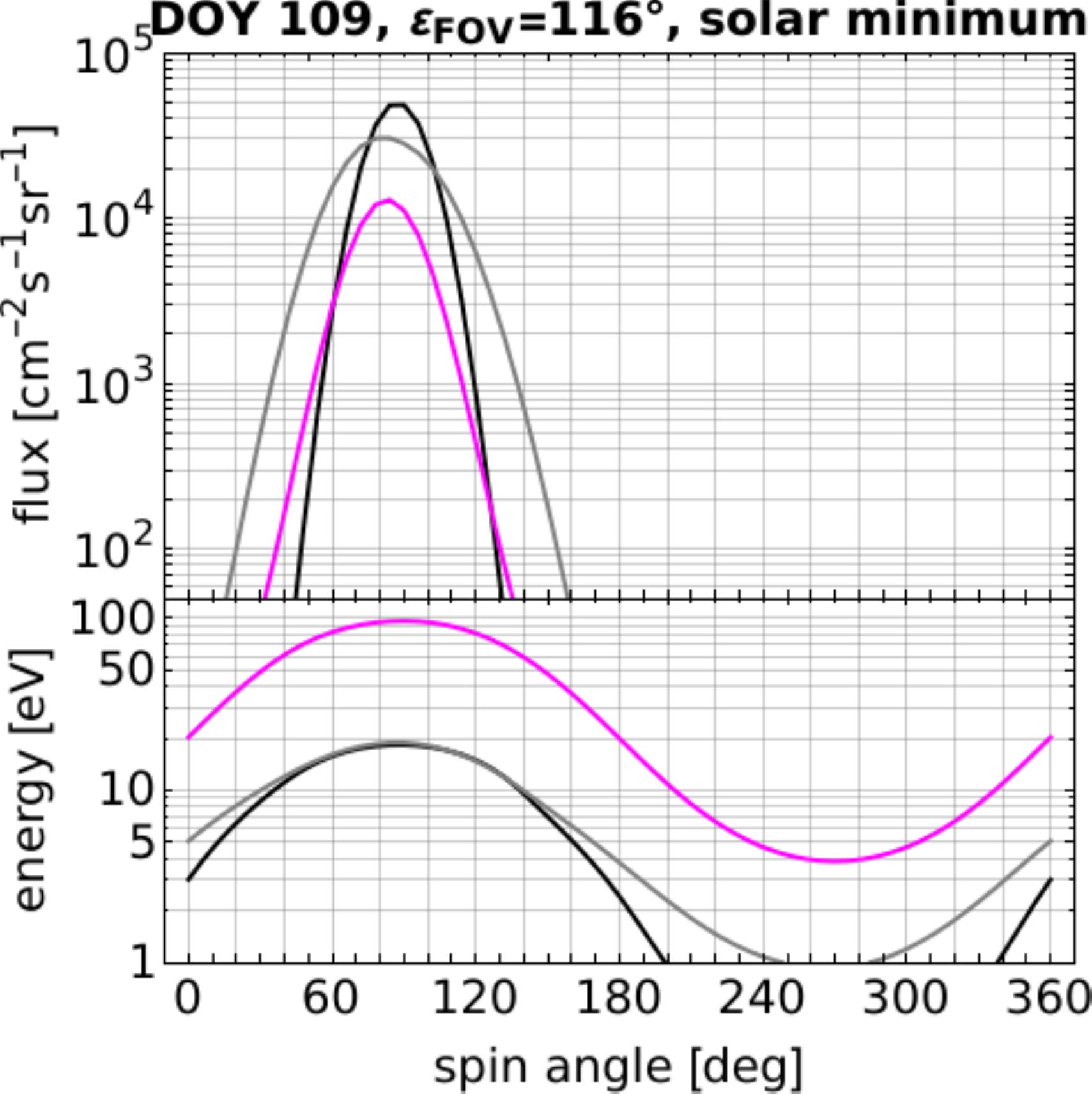} & \includegraphics[scale=0.3]{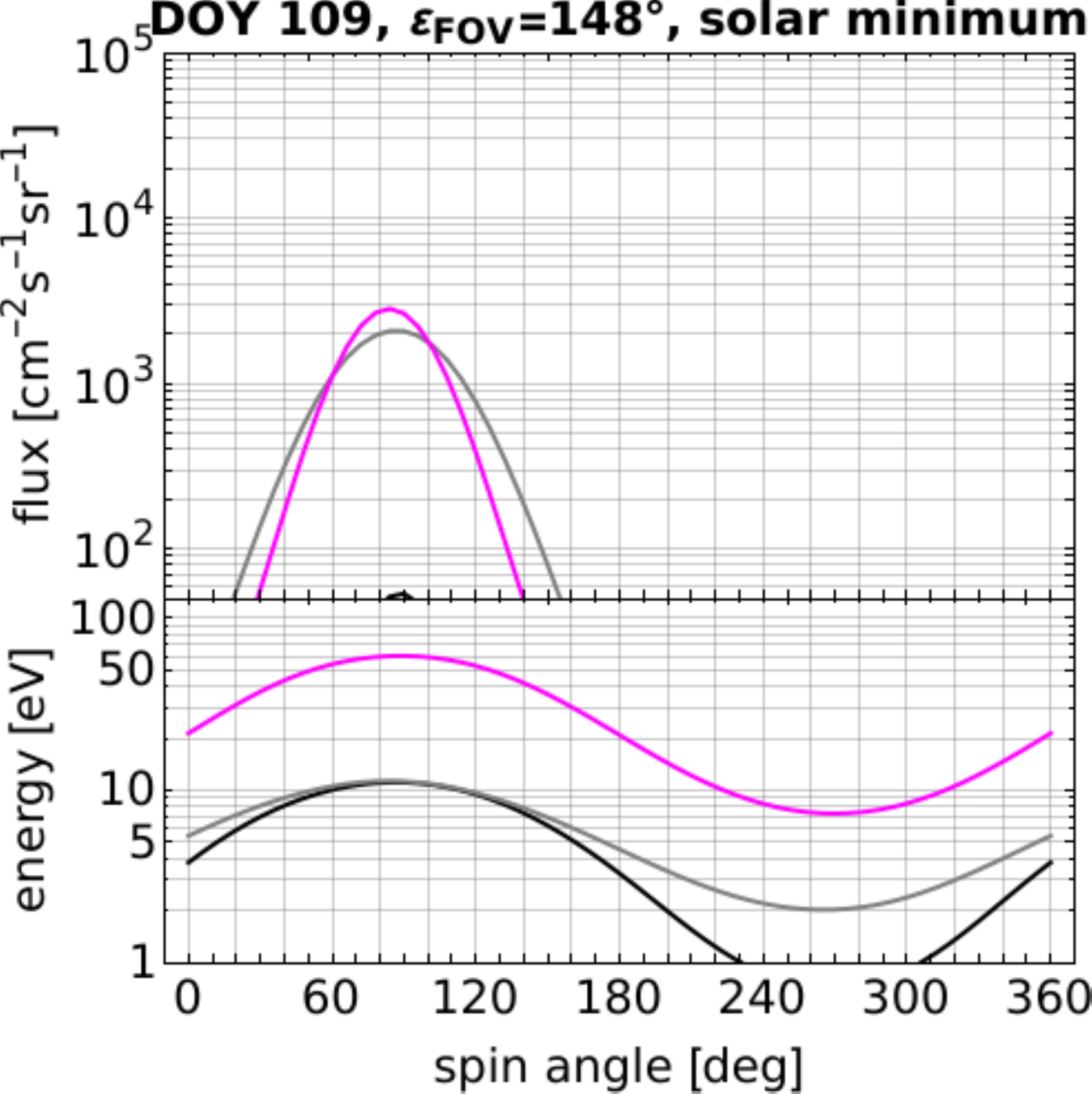} \\
\includegraphics[scale=0.3]{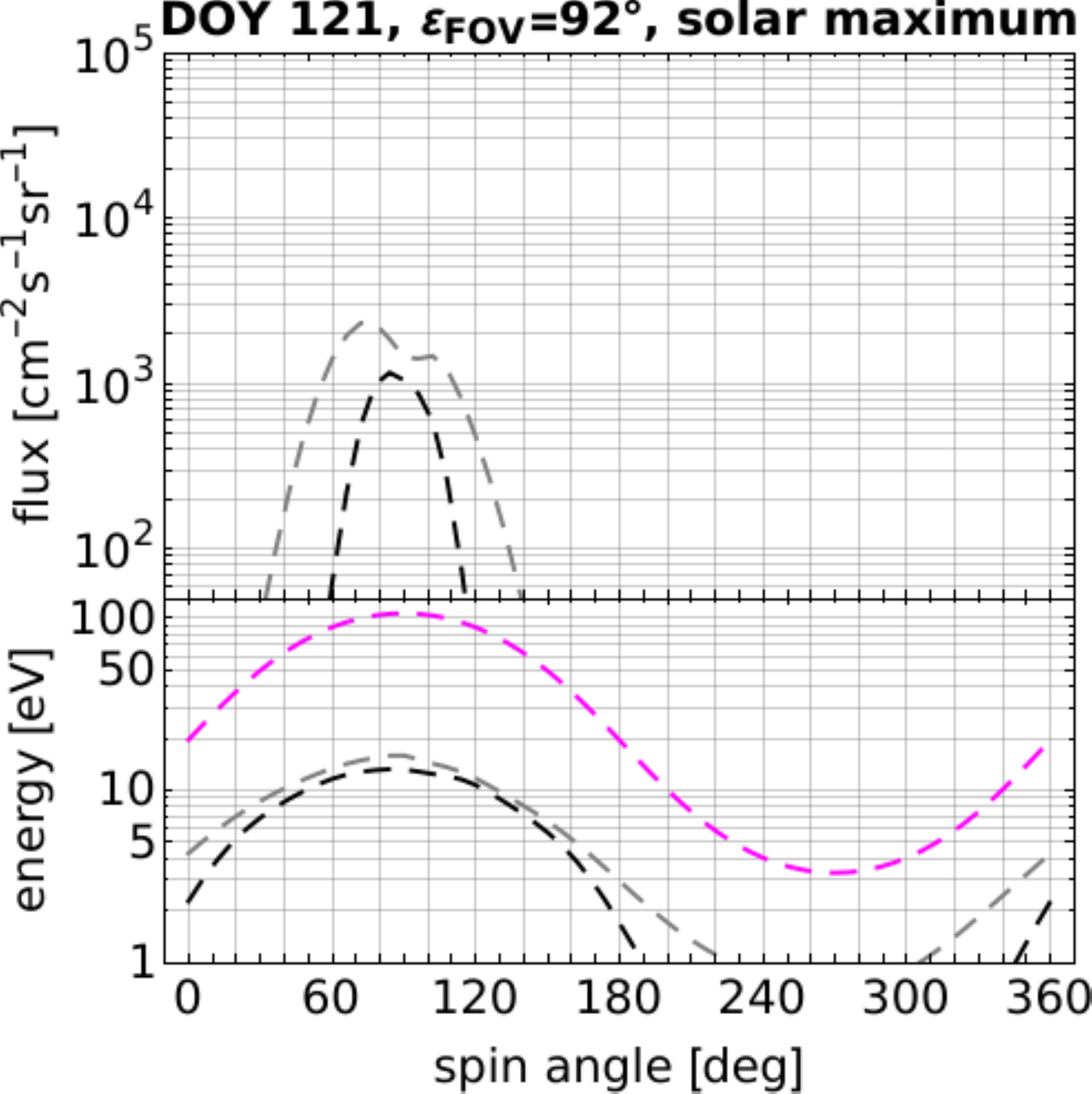} & \includegraphics[scale=0.3]{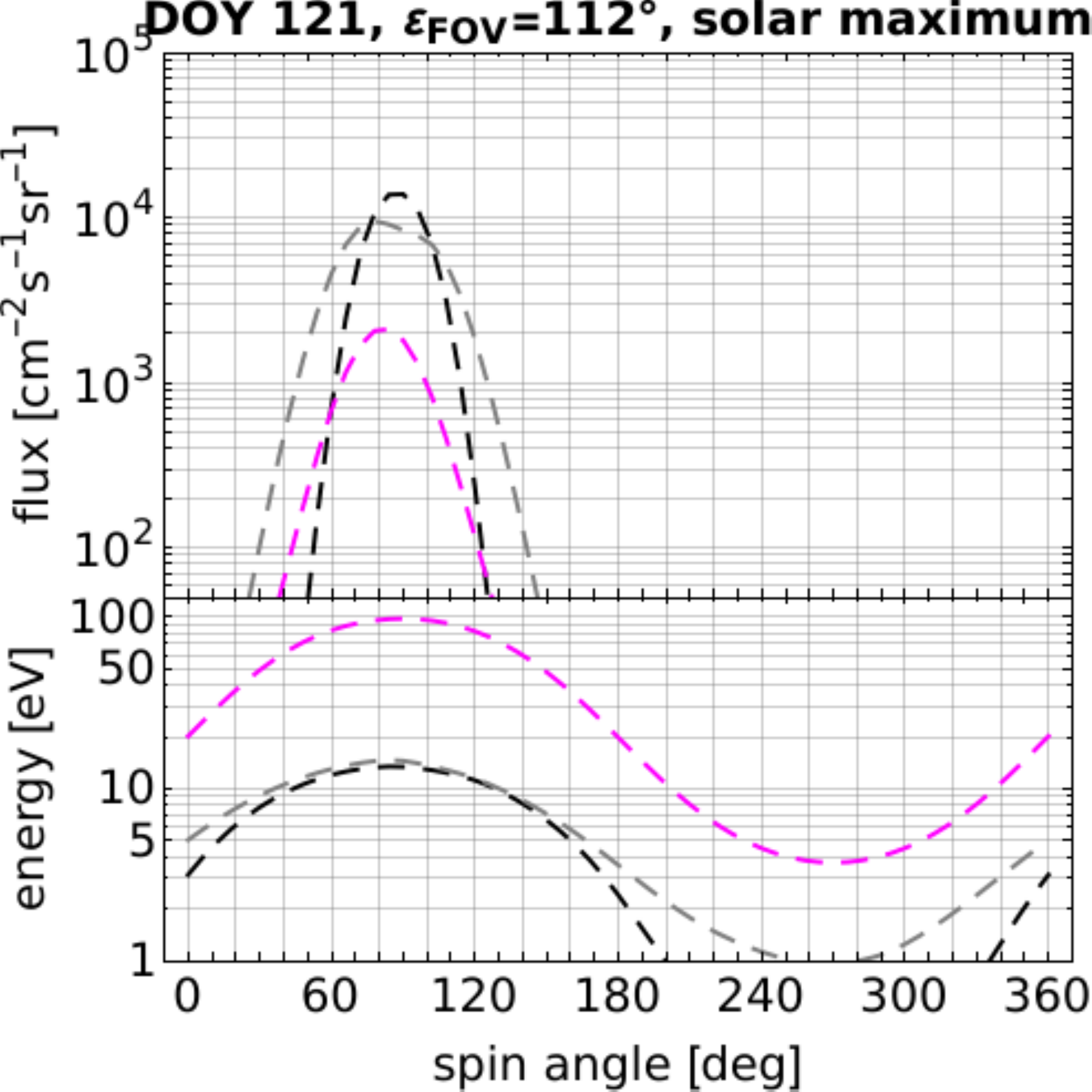} & \includegraphics[scale=0.3]{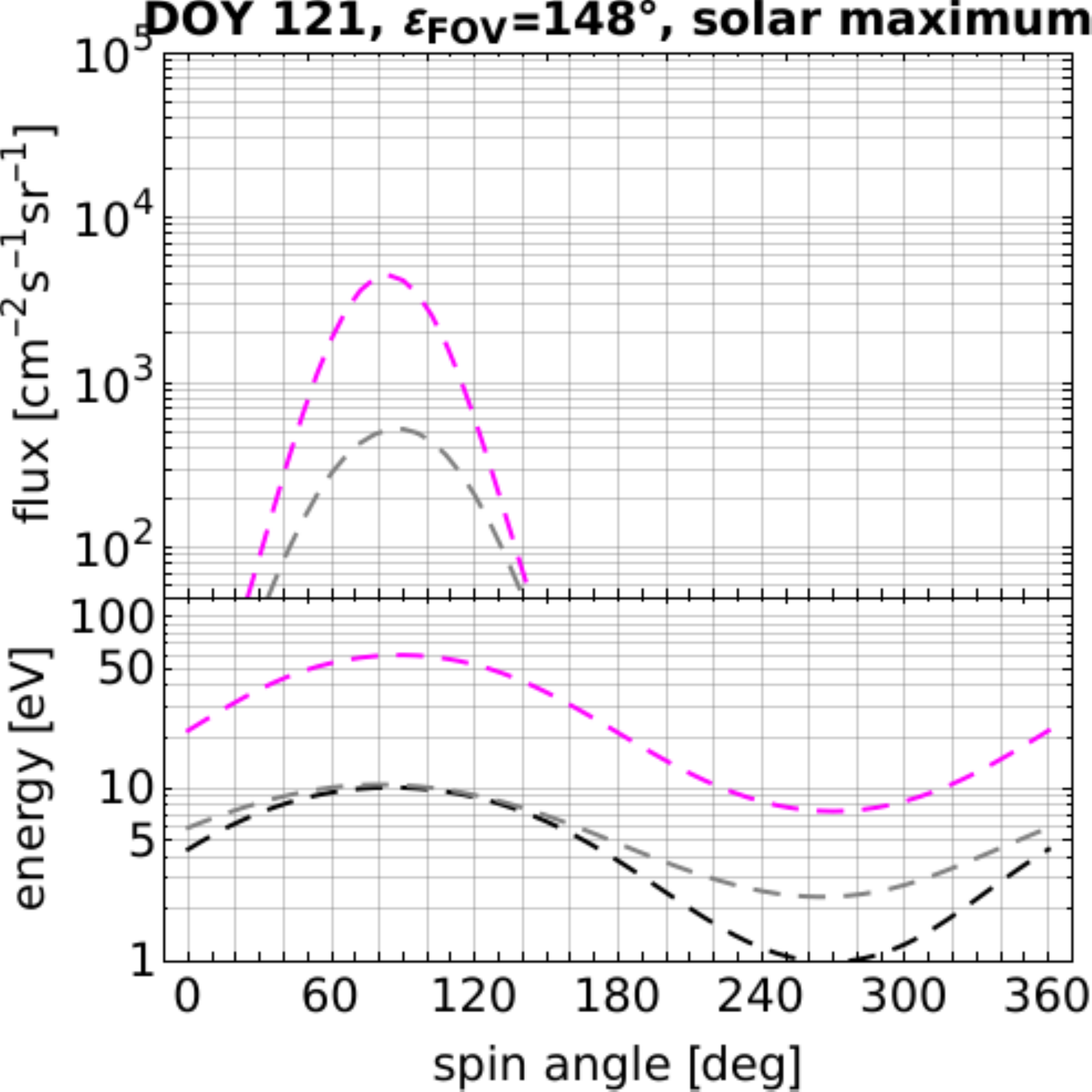} \\
\includegraphics[scale=0.3]{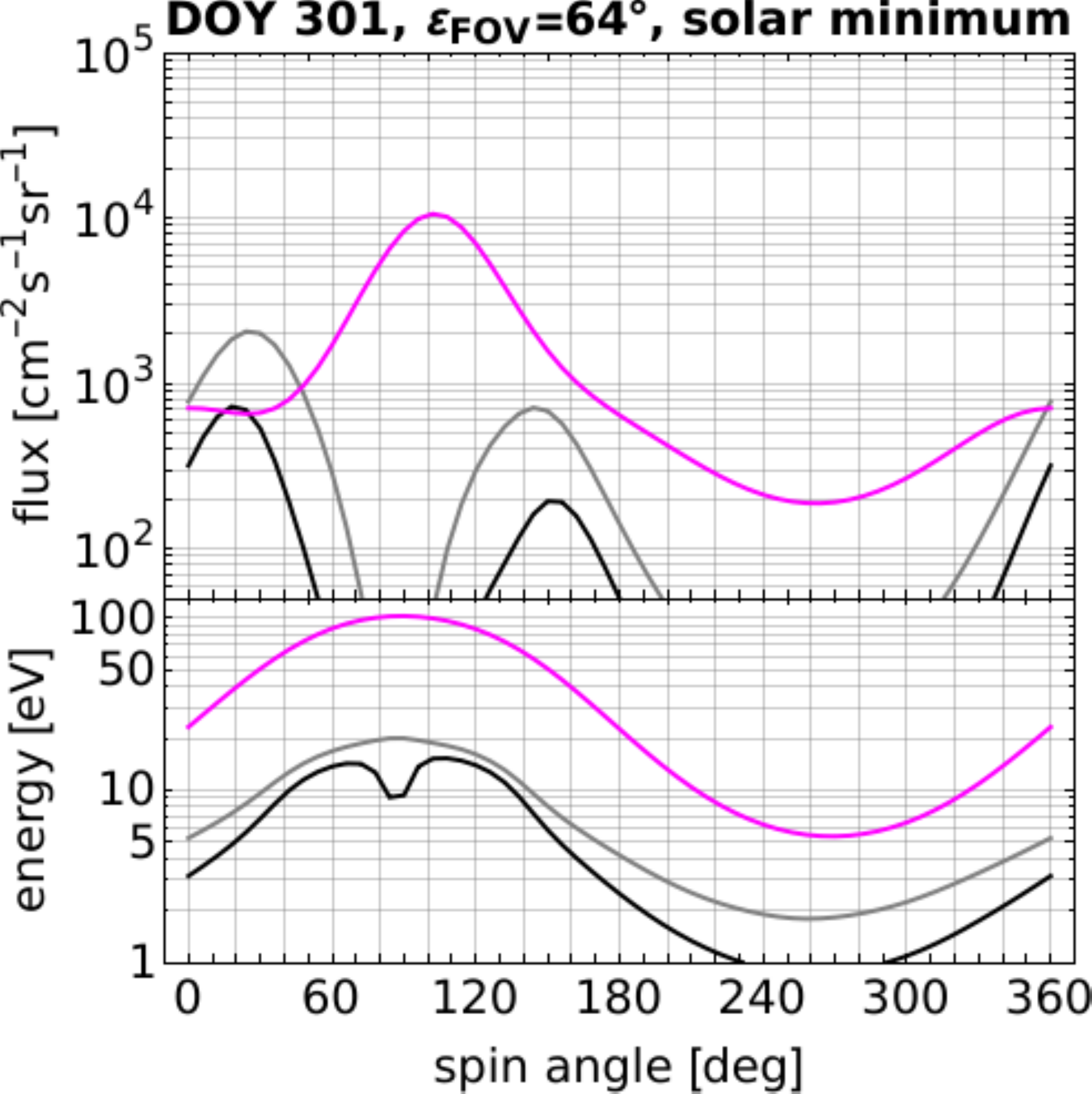} & \includegraphics[scale=0.3]{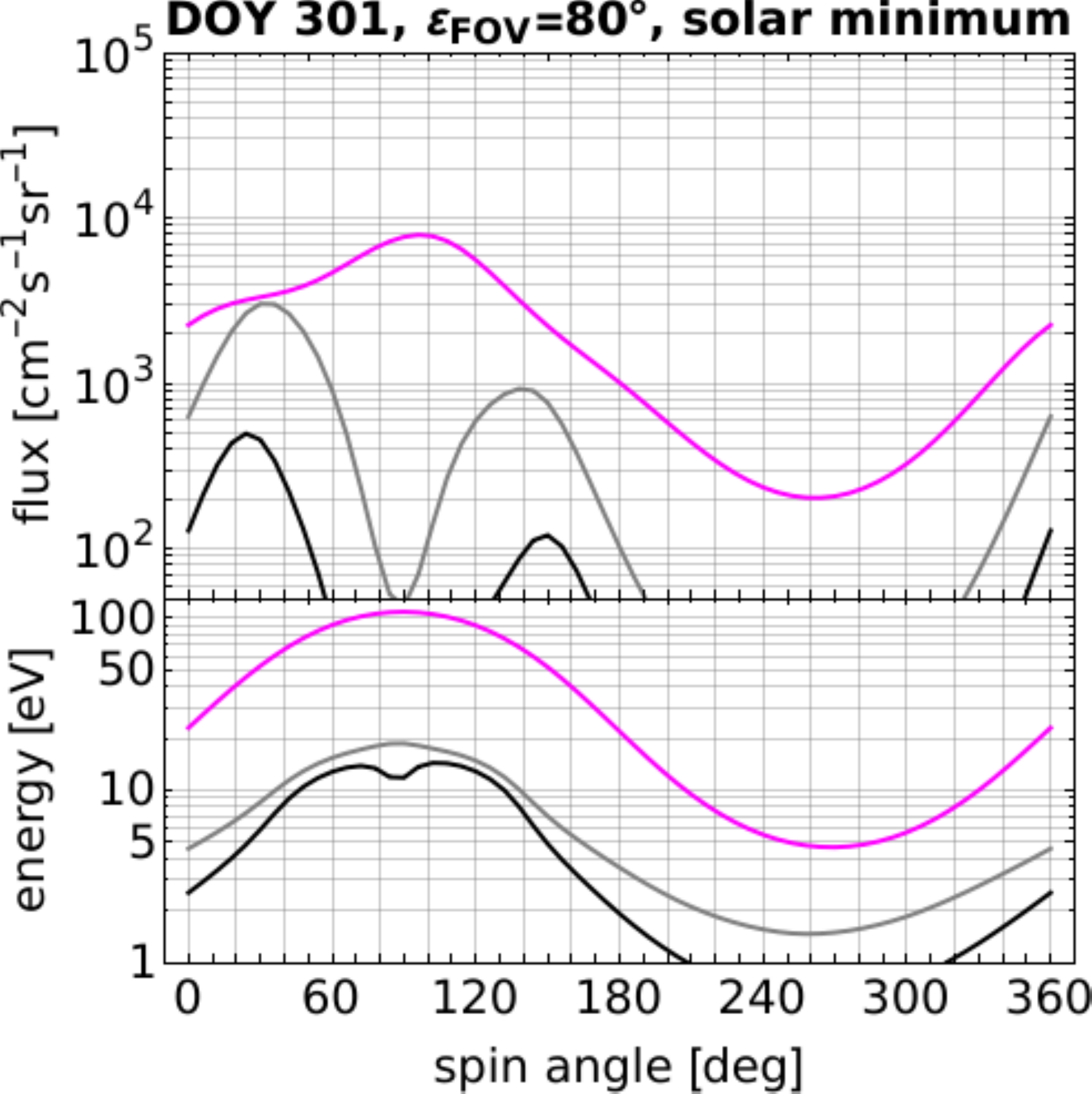} & \includegraphics[scale=0.3]{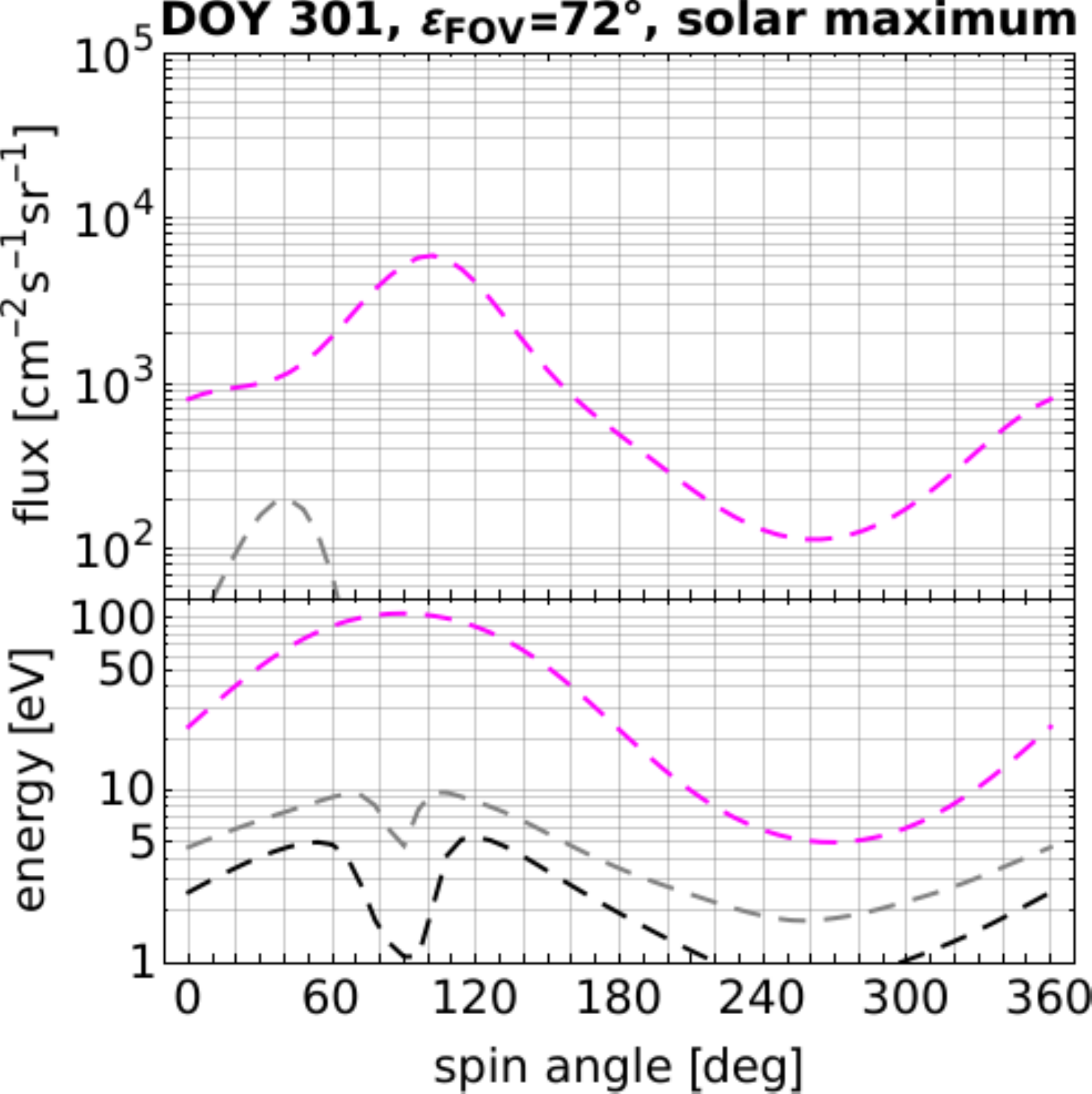} \\
\end{tabular}
\centering
\caption{Fluxes and corresponding energies of \priH\, (black) and \secH\, (gray) and \secHe\, (magenta) as a function of spin angle for various DOY--\elong\, combinations. Top row: the maximum flux of \isnH\, (sum of \priH\, and \secH) is observed on DOY~109 at \elong$=116\degr$ during solar minimum.  Middle row: the maximum flux of \isnH\, is observed on DOY~121 at \elong$=121\degr$ during solar maximum. Bottom row: various \elong\, orientations on DOY~301 during solar minimum (solid lines) and solar maximum (dashed lines) illustrating cases when the \isnH\, is not accessible for detection owing to either flux or energy limitations.  \label{fig:secondariesH}}
\end{figure*}

Figure~\ref{fig:peakImap} shows that \secH\, flux dominates the \priH\, flux during the first few weeks of the year and in the second half of the year, when, unfortunately, for most of the time the energies of both populations are below 10~eV,  which is the energy threshold of the ISN detector. Inspection of the distribution of the \priH\, and \secH\, fluxes as a function of spin angle shows that the \priH\, beam is narrower than that of \secH. The \priH\, flux exceeds the \secH\, flux only for a few spin angle bins with the \secH\, at the wings of the spin angle distribution, as clearly visible in the middle column of Figure~\ref{fig:secondariesH}. This means that when the boresight is pointed toward the maximum flux of \isnH, both populations are detectable, similar to the case of \isnHe\, populations. Thus, to observe \secH\, without contribution from \priH\, a special observation campaign should be applied. In general, the possibilities considered for \secHe\, in Section~\ref{sec:secondaryHe} can be applied. 

The maximum flux of the \isnH\, is observed on DOY~109 at \elong$=109\degr$ during solar minimum and on DOY~121 at \elong$=112\degr$ during solar maximum, as presented in Figure~\ref{fig:secondariesH}. These are cases when the narrower \priH\, dominates the flux in the core and the wider \secH\, occupies the wings of the distribution in the spin angle. The boresight should be shifted about $20\degr$ off the peak direction to observe only the \secH\, without a significant contribution from the \priH\, flux (see left and right panels in the top and middle rows of Figure~\ref{fig:secondariesH}). However, shifting the boresight toward greater \elong\, angles leads to the \secHe\, also being in the FOV. \secH\, and \secHe\, are separated in energy (bottom panels of Figure~\ref{fig:secondariesH}), but the detection mechanism for He atoms by products sputtered from the conversion surface (see Section~\ref{sec:ibexImap}) leads to a contribution of ions sputtered by \secHe\, in the energy channel occupied by the converted \secH\, ions \citep{galli_etal:19a}. As a consequence, the \secHe\, present during the \secH\, observations may affect the interpretation of the measurements, especially when the H atom energies are close to the detection limit. Thus, from the two possibilities to shift the \elong\, off the peak direction for separation of \secH\, from \priH, only the one that satisfies the criterion to detect solely \secH\, is favorable. 

The preferable observation time for \isnH\, is the solar minimum when the maximum fluxes and energies are high. During low solar activity, the total ionization rates are low and the radiation pressure is less effective, and thus more \isnH\, atoms survive to closer distances to the Sun. The bottom panels of Figure~\ref{fig:secondariesH} present \elong\, settings on DOY~301, when the maximum flux for both \priH\, and \secH\, is observed for \elong$<90\degr$. These are conditions when the indirect beams of \isnHe\, are expected. However, as the results show, this part of the orbit is favorable for \isnH\, observations neither during solar minimum nor during solar maximum. During solar minimum, the fluxes would be sufficient for detection only for those spin angles where the corresponding energies are too low for detection. For spin angles where the energies are high enough, the fluxes are too small. The situation is the same for both the flux peak of the \priH\, (\elong$=64\degr$) and the flux peak of \secH\, (\elong$=80\degr$). During solar maximum, only \secH\, flux is high enough, with the maximum flux observed at \elong$=72\degr$, but the energies are too low to fulfill the detection conditions.

Additionally, comparison of the filled (solar maximum) and empty (solar minimum) pixels of the DOY--\elong\, combination maps in Figure~\ref{fig:pxlMaps} for \priH\, and \secH\, illustrates the difference in observation possibilities for \isnH\, during the solar cycle. During solar maximum, the observation opportunities for \isnH\, are limited to the first six months of the year, whereas during solar minimum, observations of \secH\, are accessible throughout the entire year with properly adjusted boresight directions. These should provide successful observation of \isnH\, together with better counting statistics  with observations in less foreground, as on \textit{IMAP}-Lo. 

\subsection{ISN Ne and O \label{sec:NeO}}
%% Figure with secondaries options for H
\begin{figure*}
\begin{tabular}{ccc}
\includegraphics[scale=0.3]{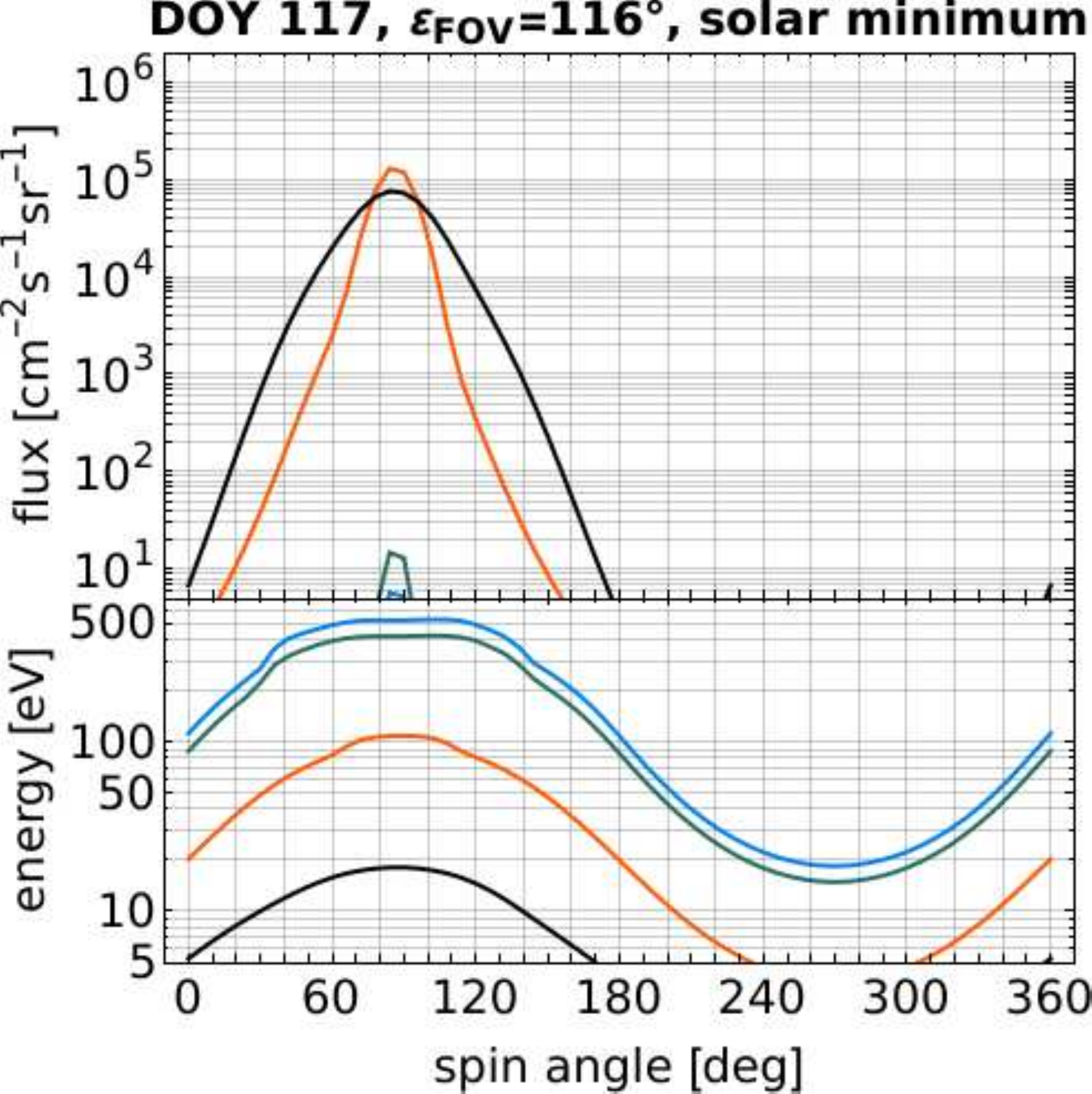} & \includegraphics[scale=0.3]{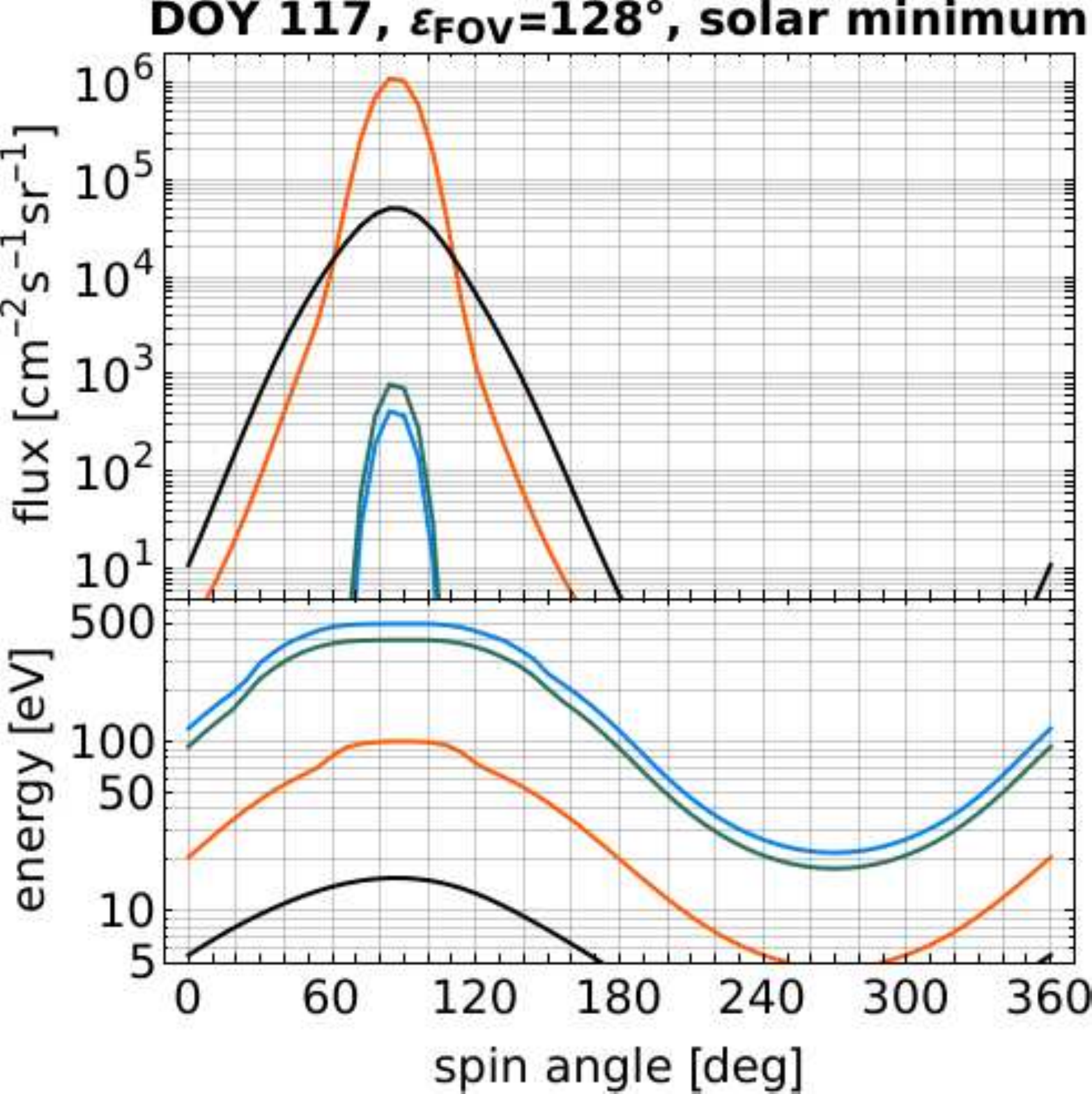} & \includegraphics[scale=0.3]{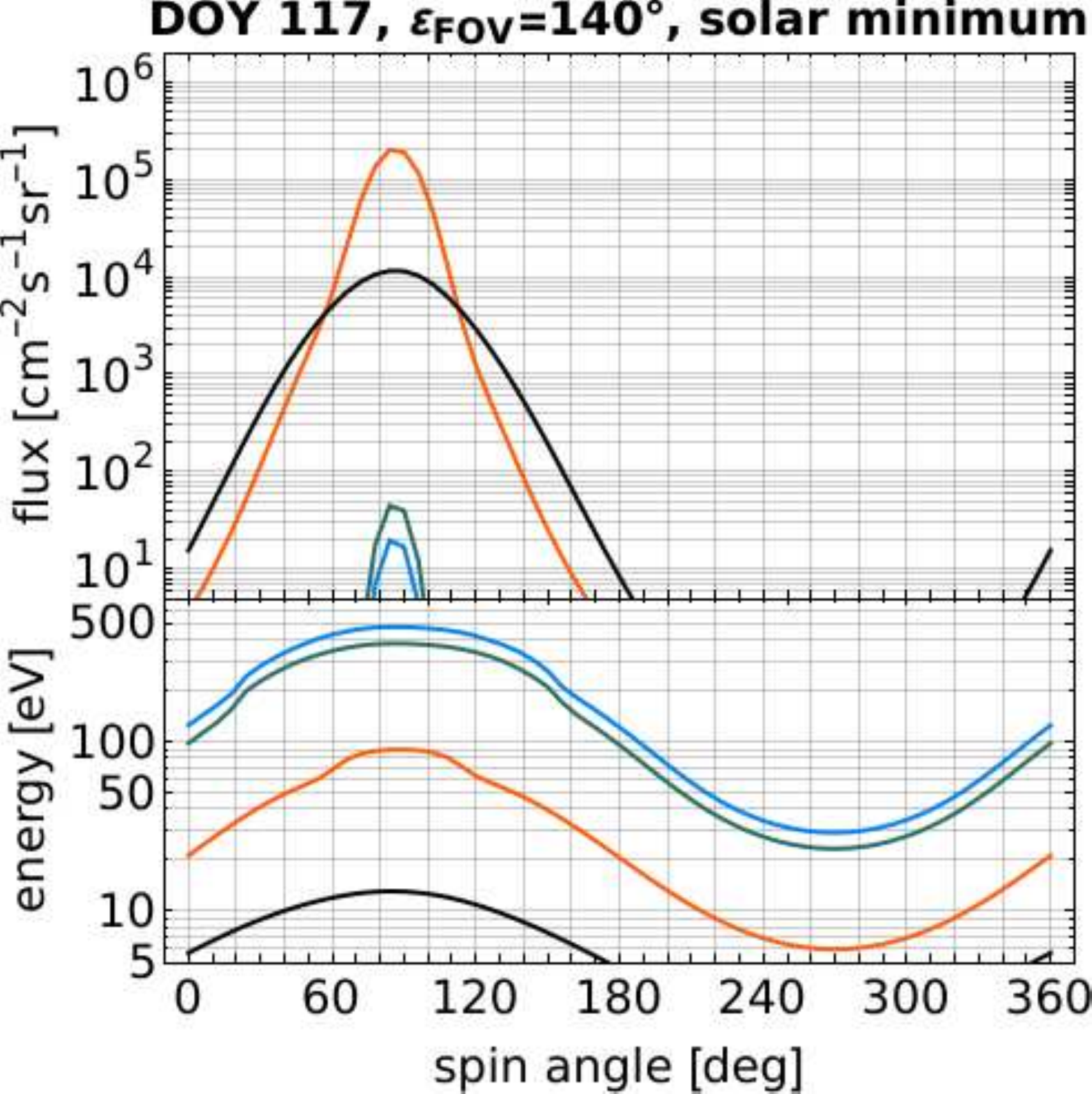} \\
\includegraphics[scale=0.3]{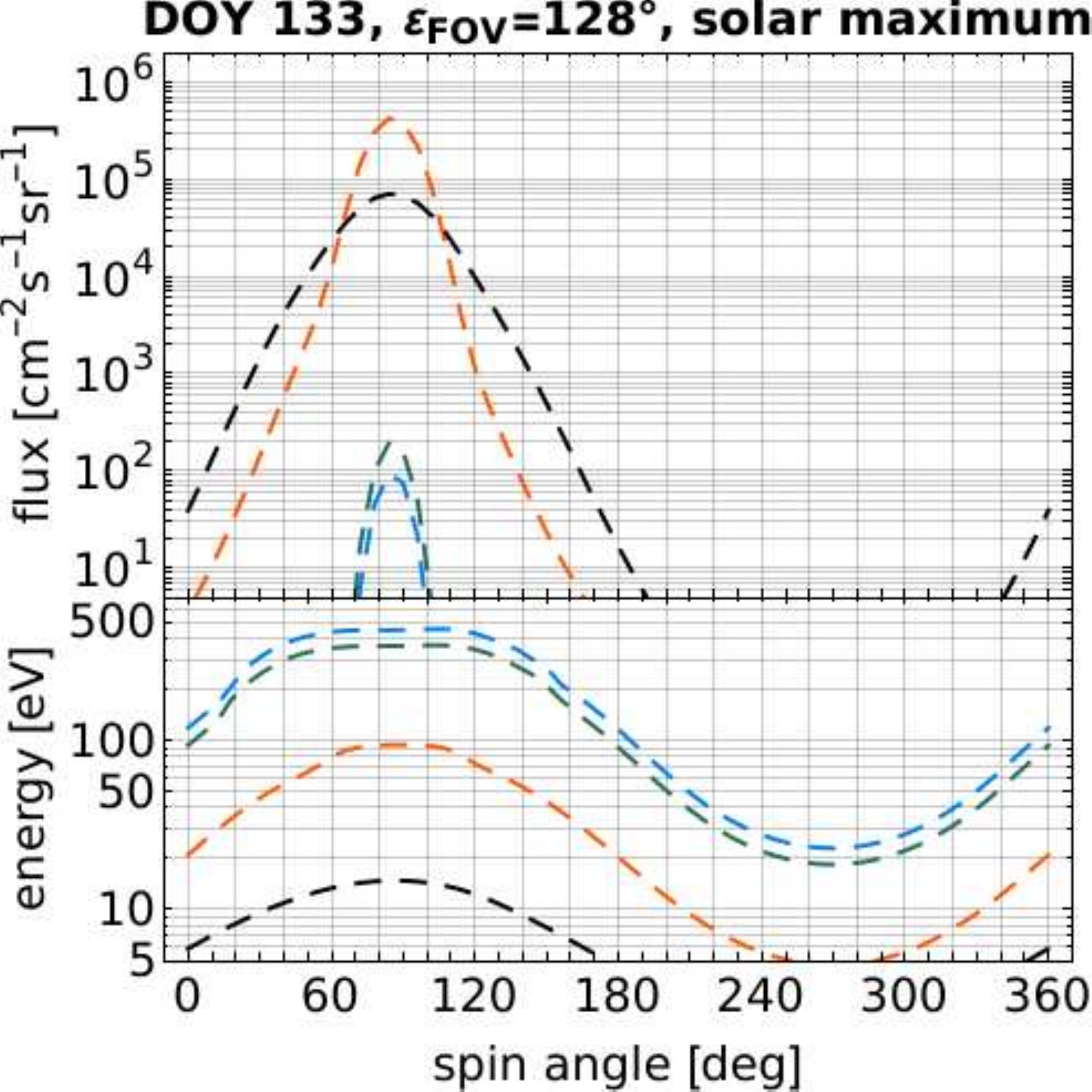} & \includegraphics[scale=0.3]{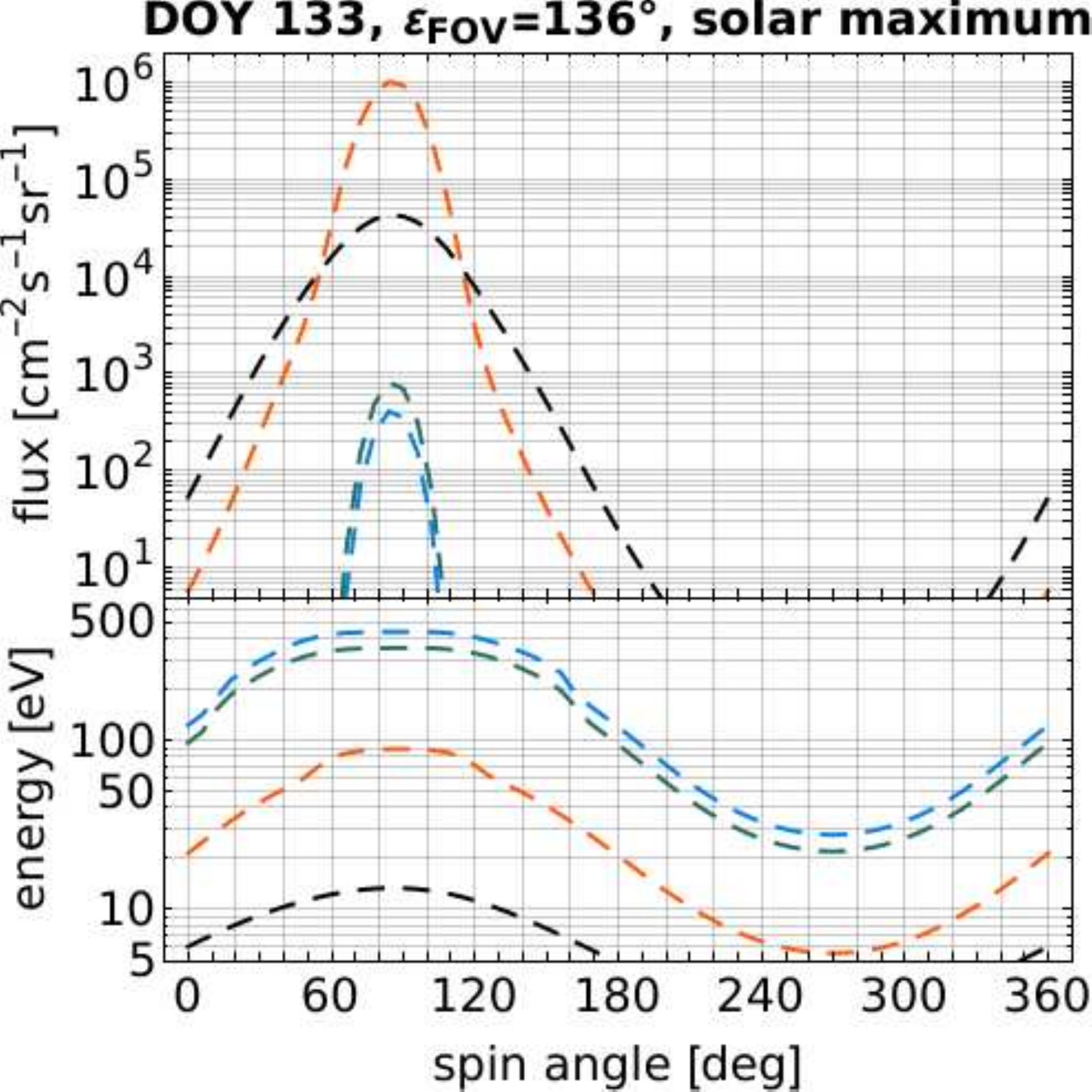} & \includegraphics[scale=0.3]{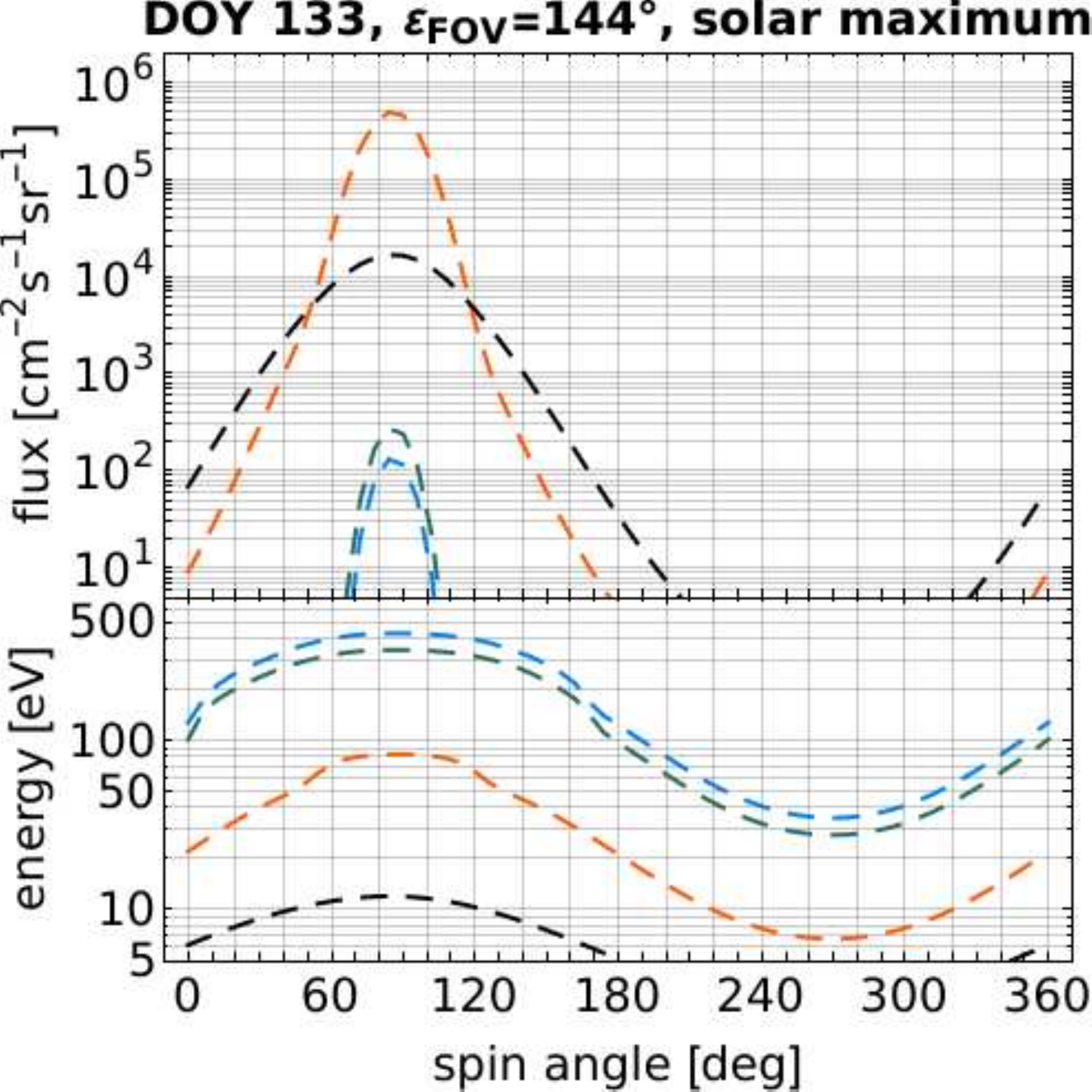} \\
\includegraphics[scale=0.3]{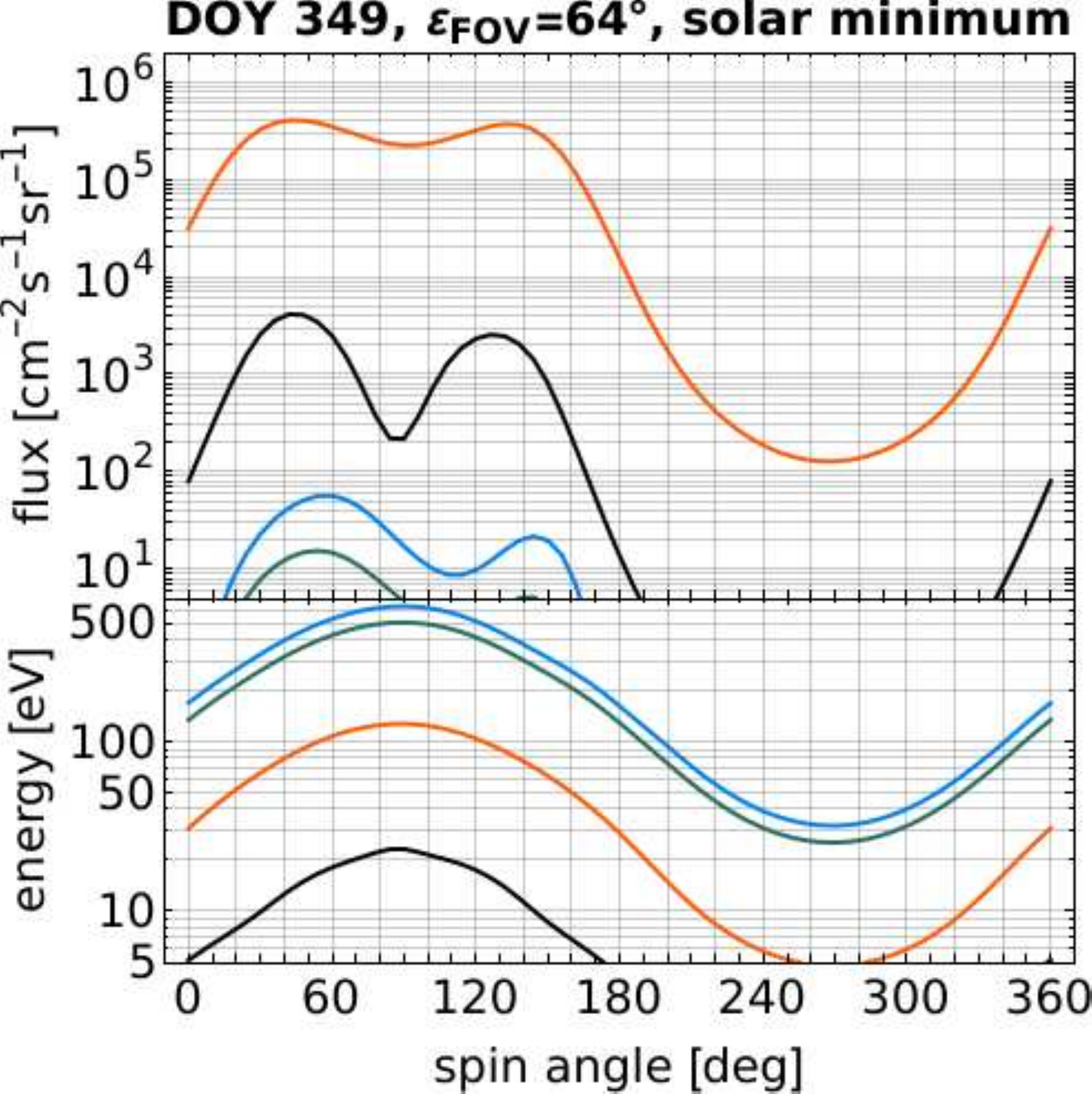} & \includegraphics[scale=0.3]{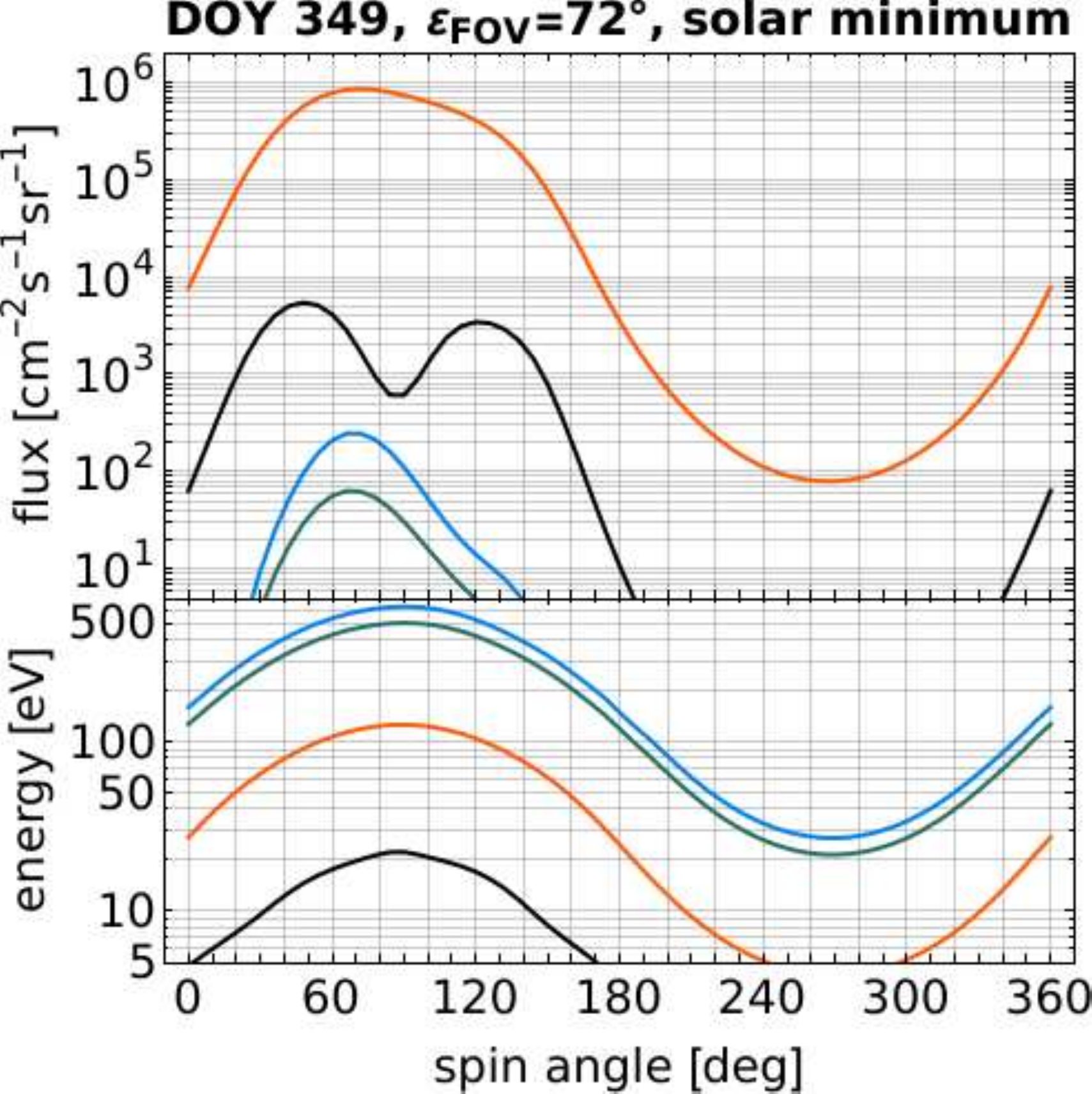} & \includegraphics[scale=0.3]{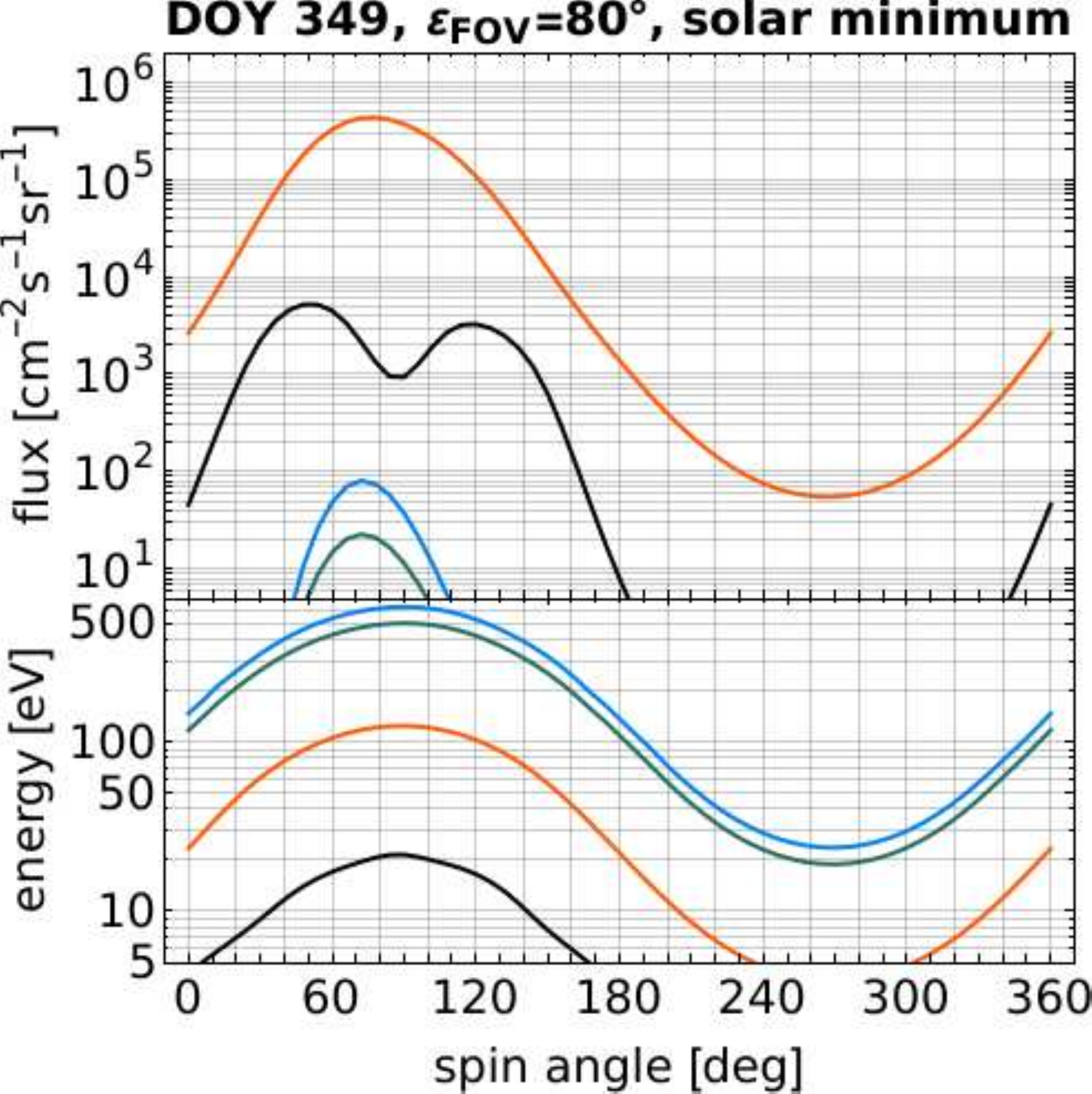} \\
\end{tabular}
\centering
\caption{ISN Ne (blue), O (green), \isnHe\, (orange), and \isnH\, (black) fluxes and the corresponding energies for selected DOY--\elong\, combinations during solar minimum (top) and solar maximum (middle). The maximum flux of ISN~Ne is observed on DOY~117 at \elong$=128\degr$ during solar minimum (shown in the middle panel of the top row) and on DOY~125 at \elong$=132\degr$ during solar maximum (not shown). The maximum flux of ISN~O is observed on DOY~133 at \elong$=136\degr$ during both solar minimum (not shown) and solar maximum (shown in the middle panel of the middle row). The bottom row presents a DOY--\elong\, combinations when the Ne flux dominates over the O flux with the maximum on DOY~349 at \elong$=72\degr$) \label{fig:NeO}}
\end{figure*}

The ISN atoms of Ne and O have been detected in the heliosphere indirectly by measurements of pickup ions \citep{geiss_etal:94a} and directly in \textit{IBEX}-Lo observations \citep{bochsler_etal:12a, park_etal:15a}. Their contribution to the ISN signal was also searched for in \textit{Ulysses} measurements \citep{wood_etal:15c}. As shown in Figure~\ref{fig:peakIbex} for the \textit{IBEX}-Lo observation geometry, measurements of ISN Ne and O atoms are limited to a few weeks between January and February. With the capability to track the ISN flow, the maximum of the Ne and O flux, expected between April and May, can be observed. Consequently, the fluxes are a factor of 1.5 (1.1) times higher for Ne and a factor of 4.2 (2.6) times higher for O during solar maximum (minimum); see Table~\ref{tab:peaks2} (Table~\ref{tab:peaks}). Due to their higher mass, Ne and O have higher energies and thus are separated in energy from the other species (see the middle panels of Figure~\ref{fig:peakImap}), which facilitates their detection and identification. However, due to lower abundance in the VLISM, the fluxes of Ne and O are small compared to \isnHe\, and \isnH\, (by three and two orders of magnitude, respectively; see Figure~\ref{fig:peakImap}). Moreover, the flow is collimated, and thus the distribution in spin angle is narrow, in contrast with those for \isnHe\, and \isnH\, (see Figure~\ref{fig:NeO}). Because the beam is narrow, the range of possible observation geometries to record sufficiently high flux of ISN Ne and O is limited to a total span of about $8\degr$, allowing for a reduction of the flux to $50\%$ of the maximum (see also Figure~\ref{fig:pxlMaps}). Thus, there are two possible scenarios to observe ISN Ne and O: first, to track the maximum flux when the energies are high and \elong\, allows for probing the spatial distribution (DOY 100--160), and second, to observe the maximum flux at \elong$>150\degr$ at small conical angle with very precise pointing, which increases the counting statistics (DOY 160--240).

Figure~\ref{fig:NeO} presents fluxes of ISN~Ne and O during the solar minimum and solar maximum. During solar minimum, the maximum flux of ISN~Ne is observed on DOY~117 at \elong$=128\degr$  (shown in the middle panel of the top row), while the maximum flux of ISN~O is observed on DOY~133 at \elong$=136\degr$ (not shown). During solar maximum, the maximum flux of ISN~Ne is observed on DOY~125 at \elong$=123\degr$  (not shown), while the maximum flux of ISN~O is observed on DOY~133 at \elong$=136\degr$ (shown in the middle panel of the middle row). For comparison purposes, the fluxes of \isnHe\, and \isnH\, for the selected DOY--\elong\, combinations are added. The Ne and O fluxes are significantly reduced when compared to He and H, but separation in energy (illustrated in the bottom panels of each plot) allows for successful detection of the species. The collimation of the Ne and O fluxes in the sky is a challenge for the detection. A shift in \elong\, by a few degrees results in a significant reduction of the measured flux (shifts by $12\degr$ during solar minimum and by $8\degr$ during solar maximum are presented in Figure~\ref{fig:NeO}). However, the observation scheme can be adjusted in flight to verify the accuracy of the flow direction and thus FOV pointing for these two species. Alternatively, keeping the boresight direction fixed for a few days allows for a transition of the boresight across the peak of the flux. Such a transition through the maximum flux of various species will give an answer about the degeneracy of the flow parameters obtained with the \textit{IBEX} observation geometry. 

\begin{figure}
\includegraphics[scale=0.4]{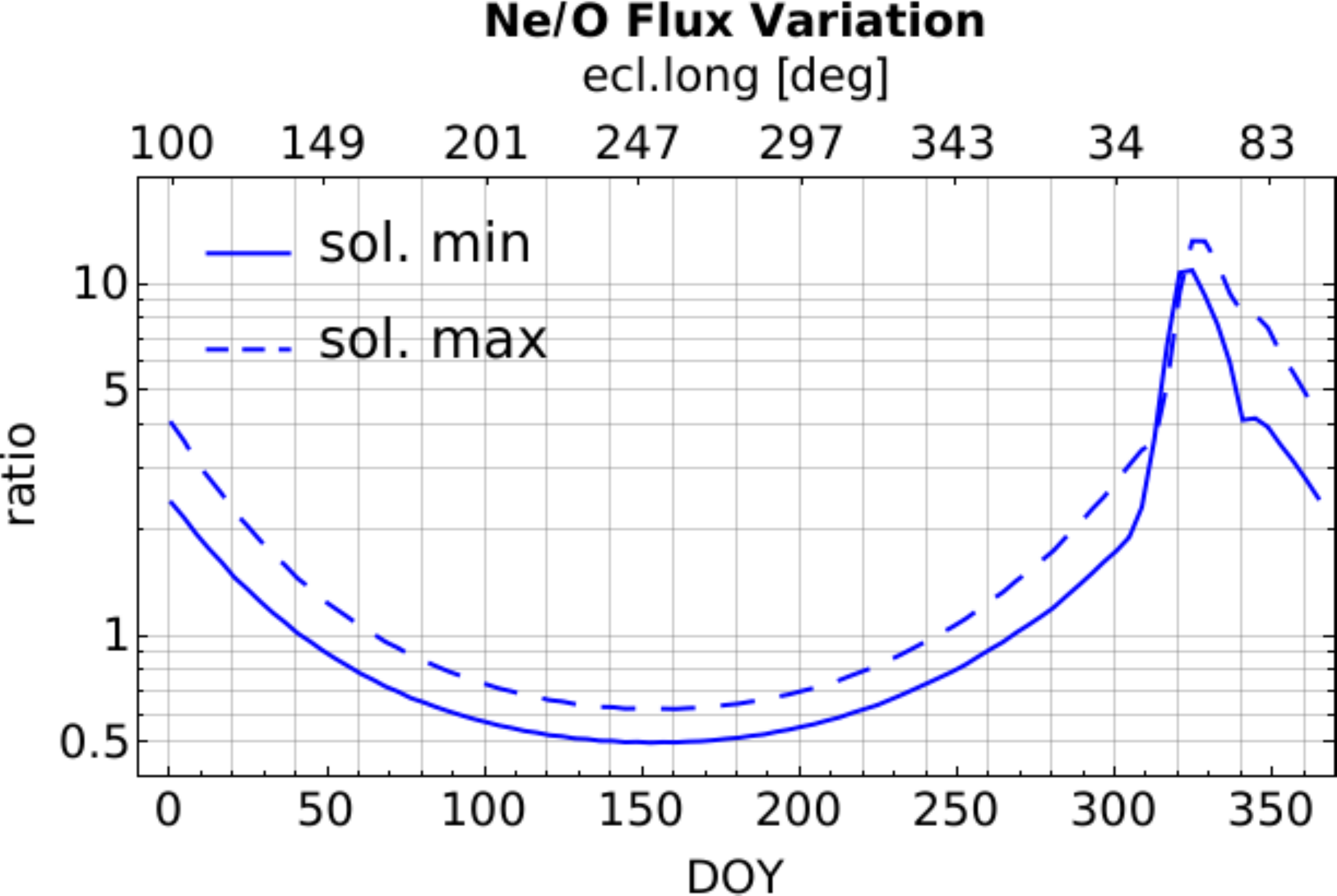}
\centering
\caption{Variation of the ratio of the maximum ISN~Ne and ISN~O flux over one year during solar minimum (solid line) and solar maximum (dashed line).
\label{fig:Ne2O}}
\end{figure}

In general, with accurate pointing, periods when the ISN~O dominates over the ISN~Ne signal and vice versa can be determined. ISN~O dominates over Ne in flux in a period DOY~$45-265$ during solar minimum and DOY~$69-241$ during solar maximum (Figure~\ref{fig:Ne2O}). However, the limitations for detection in flux and energy need to be considered, which make the last $\sim40$ days and the first $\sim40$ days in a year preferable for Ne observations, especially during solar minimum (see Figures~\ref{fig:peakImap}, \ref{fig:pxlMaps}, and the middle panel of the bottom row in Figures~\ref{fig:NeO}, and \ref{fig:Ne2O}). The observation geometry for detection of ISN~Ne solely at the end of a year needs to be set precisely, since the fluxes quickly drop when the \elong\, is shifted a few degrees off the peak, as is illustrated in the bottom row of Figure~\ref{fig:NeO}. A comparison of angular distributions from the time periods when Ne dominates in flux over O, and vice versa, can be used to search for any differences in the ISN flow parameters for these two species, which show up as a mixed distribution. Moreover, the variation of the flux ratios as a function of longitude (Figure~\ref{fig:Ne2O}) can be used to independently deduce the ratio of the total ionization rates for these two species.

%% Figure peak fluxes for Ne/O

\subsection{ISN D \label{sec:D2H}}
%% Figure peak fluxes for ISN D
\begin{figure}
\includegraphics[scale=0.4]{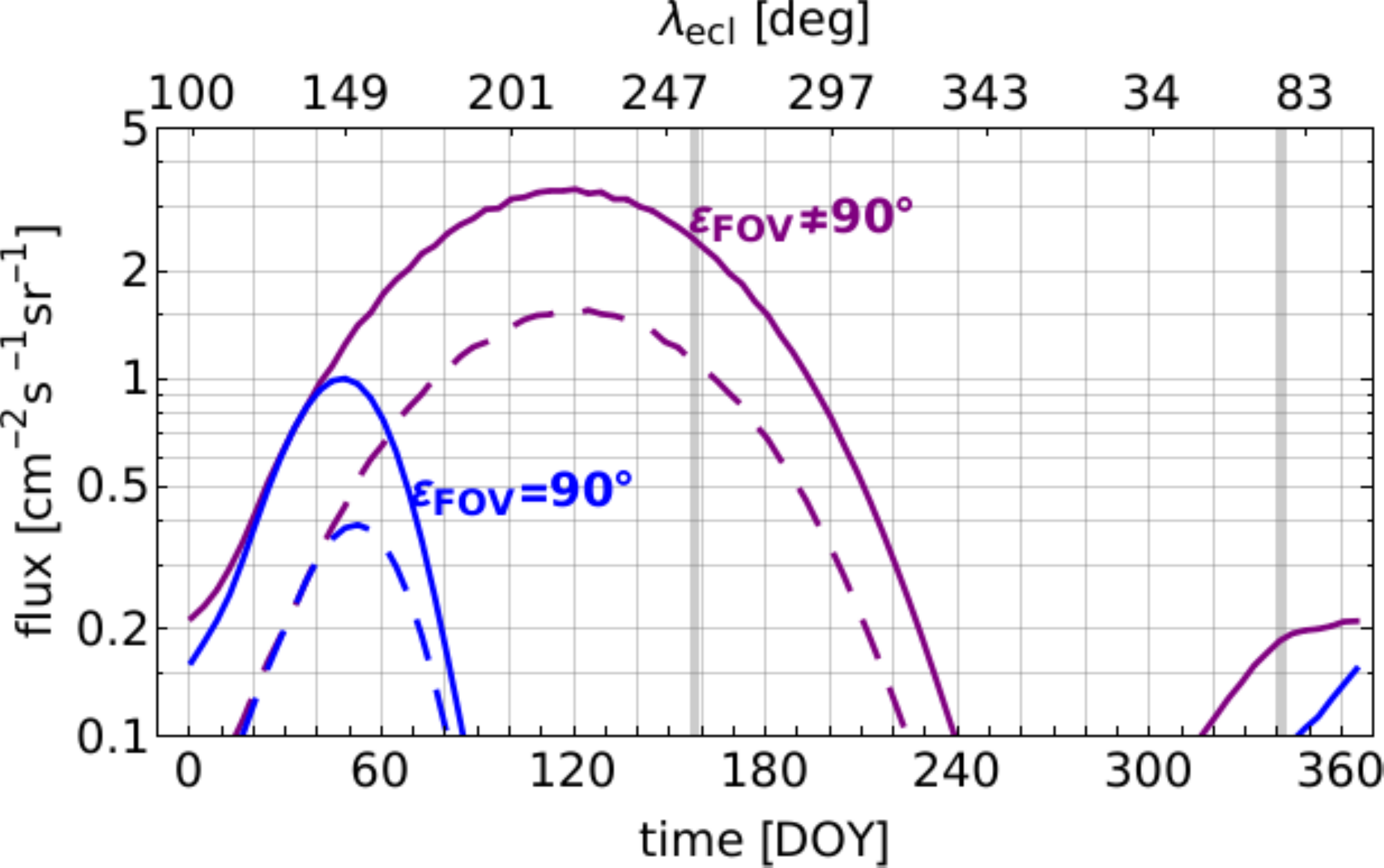}
\centering
\caption{\isnD\, (a sum of the primary and the secondary) fluxes for a boresight directed toward the maximal flux every day (\elong$\neq90\degr$; see top panel of Figure~\ref{fig:peakImap}) and kept constant (\elong$=90\degr$) during the solar minimum (solid lines) and the solar maximum (dashed lines). Please note that the fluxes are in absolute units.
\label{fig:peakFluxD}}
\end{figure}

The feasibility of \isnD\, detection was demonstrated by \textit{IBEX}-Lo during the low solar activity of 2009--2011 \citep{rodriguez_etal:13a, rodriguez_etal:14a}. With the possibility to point the ISN detector boresight toward the peak of the beam, it may be possible to increase the capability to detect and identify \isnD\, atoms. The maximum fluxes of \priD\, and \secD\, are expected in the first six months of a year (see Figure~\ref{fig:peakImap}). However, the fluxes are low, especially during solar maximum. In Figure~\ref{fig:peakFluxD} the maximum fluxes for different directions of the boresight in time are compared. The blue lines present the \isnD\, fluxes observed with the boresight directed at \elong=$90\degr$ relative to the Sun (the \textit{IBEX}-Lo observation geometry). The purple lines present the fluxes observed with the boresight adjusted to track the flux maximum within the adopted limitations (see Section~\ref{sec:methods}) with \elong\, varying with time as presented in the top panels of Figure~\ref{fig:peakImap}. As the comparison shows, the maximum flux is expected about two months later in the latter case and is about 4 times greater (Figure~\ref{fig:peakFluxD}). However, the absolute fluxes are still low compared with the fluxes of the other ISN species. Additionally, solar minimum is favorable for \isnD\, detection over solar maximum.

The observation of the \isnD\, during a period when the flux is at its maximum would be the best observation strategy in the absence of other species because of the low D abundance. Unfortunately, the maximum of the \isnD\, flux is expected when the \isnHe\, flux dominates the ISN signal ( Figure~\ref{fig:peakImap}). As discussed by \citet{rodriguez_etal:14a}, \isnD\, needs to be observed when the \isnHe\, flux is low (see their Figure~4). This is because \isnHe\, atoms sputter \ionD\, ions from the water layer on the conversion surface of the detector. Because the species are identified by the composition of the products entering the time-of-flight system (Section~\ref{sec:ibexImap}), it would be impossible to distinguish the converted ISN \ionD\, ions from the terrestrial \ionD\, ions sputtered off the conversion surface by the \isnHe. Thus, to detect \isnD\, the contribution from the \isnHe\, (and other species that sputter \ionD\, ions) needs to be as low as possible. Such conditions are found with the \textit{IBEX}-Lo geometry (\elong\, fixed at 90$\degr$), but only for very low \isnD\, fluxes.

We apply the methodology adopted by \citep{rodriguez_etal:13a, rodriguez_etal:14a} to distinguish the converted ISN \ionD\, ions from the terrestrial \ionD\, ions and to assess the \isnD\, count rates. We calculate the count rates of \ionD ions converted on the conversion surface from the simulated \isnD\, flux ($cD$) under the assumption that the conversion efficiency from neutral atom to negative ion is the same for both hydrogen isotopes. All other factors that contribute to the geometric factor adopted from \textit{IBEX}-Lo are independent of species. Because \isnD\, atoms arrive with twice the energy of \isnH\, atoms, the expected D count rate can be obtained by multiplication of the simulated D flux by the geometric factor for H for the energy step (ESA) 3 based on \textit{IBEX}-Lo calibrations and observations:
\begin{equation}
f_\mathrm{D} \cdot g_{\mathrm{H,ESA3}} = cD,
\label{eq:fD}
\end{equation}
where $f_\mathrm{D}$ is the simulated flux of \isnD, and $g_{\mathrm{H,ESA3}}$ is the geometric factor for \isnH\, in ESA~3 ($g_{\mathrm{H,ESA3}}=2.17\times10^{-5}$~$[\mathrm{cm}^{-2}\mathrm{sr}^{-1}\mathrm{eV}]$).

Similarly, the count rates of \ionD\, ions of terrestrial origin ($cD_{\mathrm{terr}}$) sputtered by the \isnHe\, flux ($f_{\mathrm{He}}$) can be assessed for each of the affected energy steps (i.e., ESA 3 and ESA 4):
\begin{equation}
f_{\mathrm{He}} \cdot g_{\mathrm{He,ESA3-4}} \cdot (D/H)_{\mathrm{terr}} = cD_{\mathrm{terr}},
\label{eq:fDterr}
\end{equation}
where the geometric factor for \isnHe\, is $g_{\mathrm{He,ESA3-4}}=5.2\times10^{-5}$~$[\mathrm{cm}^{-2}\mathrm{sr}^{-1}\mathrm{eV}]$, and $(D/H)_{\mathrm{terr}}=1.49\times 10^{-4}$ \citep{rodriguez_etal:14a}. If the ratio $cD/cD_{\mathrm{terr}}>1$, then the detection of \isnD\, is satisfied.

The bottom left panel of Figure~\ref{fig:pxlMaps} illustrates the DOY--\elong\, combinations for which the \isnD\, flux is high on one hand (purple circles) and for which the amount of converted \ionD\, ions due to \isnD\, exceeds \ionD\, ions sputtered by \isnHe\, on the other hand (blue circles). These two are clearly separated in \elong, which is a consequence of the expected variation of \isnHe\, flux in the signal and a requirement for a reduced \isnHe\, signal in the latter case.

Another observation possibility to detect \isnD\, can be a statistical analysis of the ratio of the \ionD\, ions of various origin (sputtered vs. converted) in the registered signal. Knowing the number of \ionD\, ions sputtered owing to \isnHe\, (e.g., from the instrument calibration), any statistically significant increase of the number of the \ionD\, ions will indicate a presence of the converted \ionD\, ions and thus will be an evidence of the \isnD\, in the measured signal. 

\subsection{Indirect Beam of the ISN Flow for the Study of the Ionization Rates \label{sec:twoBeams}}
The flow of the ISN atoms at every location inside the heliosphere is split into the direct and indirect beams (Section~\ref{sec:sciOpp}). Atoms of indirect beams have been exposed longer to the ionization losses, and they probe closer distances to the Sun than those of the direct beams. Ideally, if the two beams can be measured at the same time, then the differences in the measured fluxes are a direct measure of the differences in the total ionization rates that the atoms are exposed to along their trajectories.  In reality, the indirect beam is observed in the antiram direction in the spacecraft frame, i.e. when the spacecraft recedes from the ISN flow, while the direct beam is in view in the ram direction. For observations in the antiram direction, the energies of most ISN species are too low for detection, as has been demonstrated for ISN He by \citet{galli_etal:15a}. It should be noted that the indirect beam is only strong enough for observation in the downwind half of the orbit around the Sun. The further upwind the indirect trajectories cross 1~au, the lower is the survival probability for these ISN atoms, as they come very close to the Sun and their arrival direction moves very close to the Sun in the sky. While the \textit{IMAP}-Lo pivot platform provides the capability to point the boresight closer to the Sun than at $90\degr$, it is limited to $60\degr$ from the Sun to avoid exposure to the solar UV light and the solar wind. This limitation is likely to be applicable to any future detectors of ISN gas. For these reasons, the best part of the orbit to observe the indirect beam is in the downwind hemisphere and actually before the crossing of the focusing cone, when the indirect beam arrives from close to the ram direction. Concentrating on \isnHe, this condition is met on about DOY~260, when the elongation of the \isnHe\, peak drops below $90\degr$ (see the top row of Figure~\ref{fig:peakImap}), first for \secHe\, and next for \priHe.  

\begin{figure}
\includegraphics[scale=0.35]{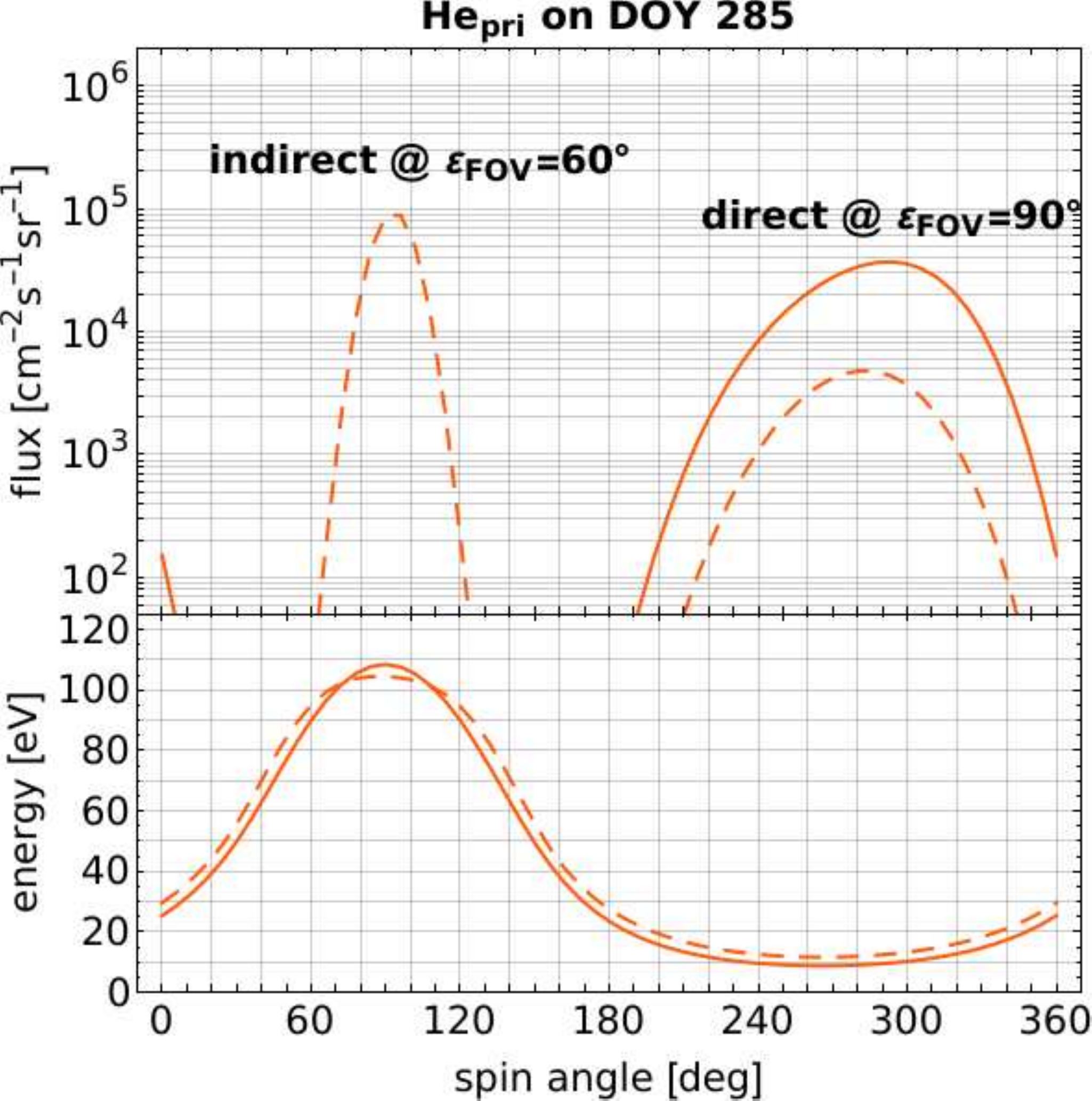} 
\centering
\caption{Flux of \priHe\, calculated on DOY~285 at \elong$=60\degr$, where the indirect beam is accessible with peak in spin angle bin 96, and at \elong$=90\degr$, where the direct beam is expected with peak in spin angle bin 294. However, the direct beam is not accessible for detection because of very low energies as illustrated in the bottom panel. This plot illustrates the spin angle variation of the flux along the FOV in the sky presented in Figure~\ref{fig:FOV}. See also similar case in the bottom panel of Figure~\ref{fig:secondariesHe}. \label{fig:indirNdir}}
\end{figure}

In the focusing cone region (November/December) direct and indirect beams cannot be untangled because of the symmetry of the flow distribution with respect to the flow direction. This is not a favorable observation geometry for separation of these two beams. The direct and indirect beams should be observed separately either at different locations around the Sun or at the same location but at different boresight directions. However, in both cases the reduced magnitude of the indirect beam flux needs to be considered. We have looked for the optimal pointing of the detector to probe the indirect beam. Within the considered pivot range ($60\degr \leq$\elong$\leq180\degr$), the indirect beam fluxes of \secHe\, start to be detectable around DOY~173 during solar minimum and around DOY~205 during solar maximum (see Table~\ref{tab:summary}). The indirect beam flux dominates the direct beam flux on about DOY~240 (DOY~260) for \secHe\, and DOY~270 (DOY~280) for \priHe\, during solar minimum (maximum) at elongation angles \elong$<90\degr$. At the same time, the direct beam is expected at elongation angles of \elong$>90\degr$. However, as illustrated in Figure~\ref{fig:indirNdir} (see also the bottom panels of Figure~\ref{fig:secondariesHe}), the energies of the direct beam atoms are of the order of about 20~eV, which might be challenging for detection, and thus only the indirect beam atoms have energies satisfying detection. On the other hand, before DOY~260, the direct beam atoms have an energy high enough for measurements at \elong$>90\degr$, but the indirect beam is visible on a elongations \elong$<60\degr$. If we had a detector with the capability to measure at elongation angles around $30\degr-40\degr$ from the Sun, the observations of both beams in the same location would then be possible.  Thus, we conclude that the simultaneous observation of both direct and indirect beams of ISN flow at the same location but with different \elong\, angles is not possible with the pivot range capabilities considered here owing to energetic reasons. The indirect beams of Ne and O could be observed during the same part of the orbit, if only the fluxes were high enough for detection.

A study of the fluxes of the indirect beams and comparison with the fluxes of the direct beams measured along the detector's orbit around the Sun can serve to investigate the variation of the total ionization rates as a function of the distance from the Sun, especially a contribution from the electron impact ionization, for which the input to the total ionization rates increases significantly inside 1~au \citep[e.g., Figure~4 of][]{sokol_etal:19a}. The locations of the perihelia of the \isnHe\, atoms on indirect trajectories are inside the Earth's orbit and reach as close to the Sun as 0.26~au. Comparison of the \isnHe\, flux of the indirect beam with that of the direct beam, which does not pass closer to the Sun than 1~au, will allow us to investigate the differences in the total ionization rate between 1 and 0.3~au. This will allow for an independent assessment of the total ionization rates for ISN gas species. 

\begin{figure}
\begin{tabular}{c}
\includegraphics[scale=0.35]{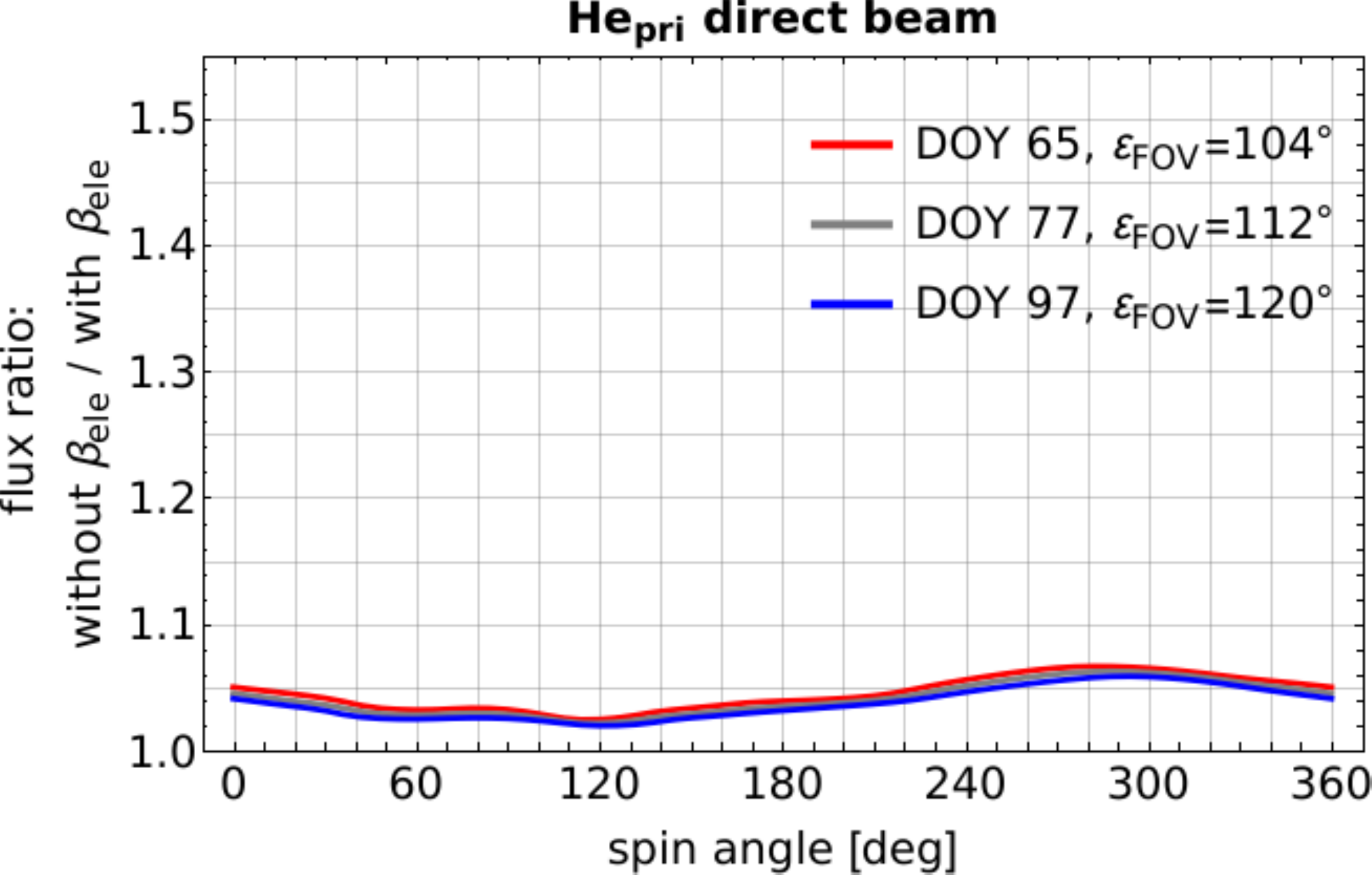}\\
 \includegraphics[scale=0.35]{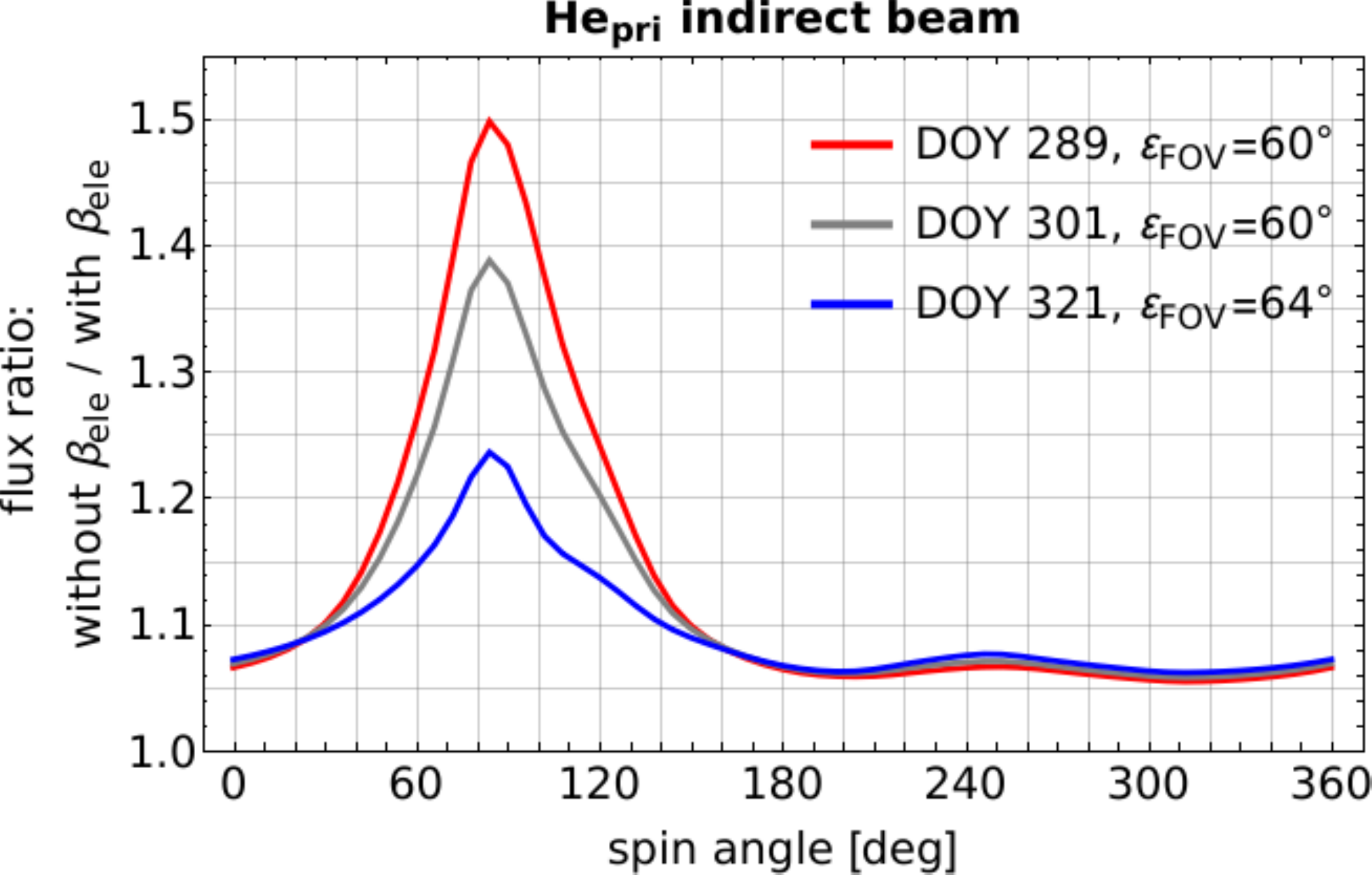}\\
\end{tabular}
\centering
\caption{Ratio of the maximum fluxes of \priHe\, calculated with the total ionization rates without the electron impact ionization ($\beta_{\mathrm{ele}}$) included  to the maximum fluxes with the electron impact ionization included in the total ionization rates for selected DOYs. Left panel: ratio for the direct beams; right panel: ratio for the indirect beams. \label{fig:El0}}
\end{figure}

Together with the knowledge about the ionization rates due to photoionization and charge exchange reactions based on simultaneous observations of the solar EUV flux and the solar wind \citep[e.g.,][]{sokol_etal:19a}, the electron impact ionization rate can be deduced for the critical region inside the Earth's orbit. Figure~\ref{fig:El0} presents ratios of the \priHe\, flux calculated with the total ionization rates without and with electron impact ionization included. The ratios are shown for maximum fluxes for selected DOYs for the direct beam (left panel) and the indirect beam (right panel). The contribution of the electron impact ionization for the direct beam atoms is less than $10\%$, while for the indirect beam atoms it may be as high as $50\%$. Additionally, the effect on the indirect beam increases with the angular distance from the downwind direction (which is approximately on DOY~342). 

%% Figure with/without electron ionization

%% Figure with/without electron ionization

In addition, the latitudinal variation of the ionization rates can be investigated through the observation of the ISN~O as pointed out by \citet{sokol_etal:19b}. The variation of the ISN~O density in the ecliptic plane reflects the latitudinal structure of the ionization rates (see Figure~14 of \citet{sokol_etal:19b}). Thus, the time series of measurements of ISN~O in the downwind hemisphere around the cone region should provide information about the anisotropy of the total ionization rates, due to the latitudinal structure of the solar wind and the solar EUV radiation. However, the fluxes of the indirect beam of ISN~O are too low to be accessible to measure.

\section{Potential Observation Schedule during Solar Maximum \label{sec:OperationPlan}}
Based on the insight discussed in the preceding sections, we present a potential observation schedule for an instrument with the capabilities similar to the planned \textit{IMAP}-Lo detector on an Earth-like orbit during solar maximum. It can be applied to any instrument with the capability to adjust the boresight direction in flight. The objective of the plan is to maximize accomplishment of the science goals presented in Section~\ref{sec:sciOpp}. The plan is sketched with the order of the science priorities, as laid out in this paper, the primary populations first, the secondary populations next, etc. The plan is flexible and can be adjusted to shift the emphasis as new information arises. It is arranged chronologically over the year, but it can start at an arbitrary moment of the year, depending on the actual launch date and the duration of the payload commissioning. 

\begin{figure}
\begin{tabular}{c}
\includegraphics[scale=0.32]{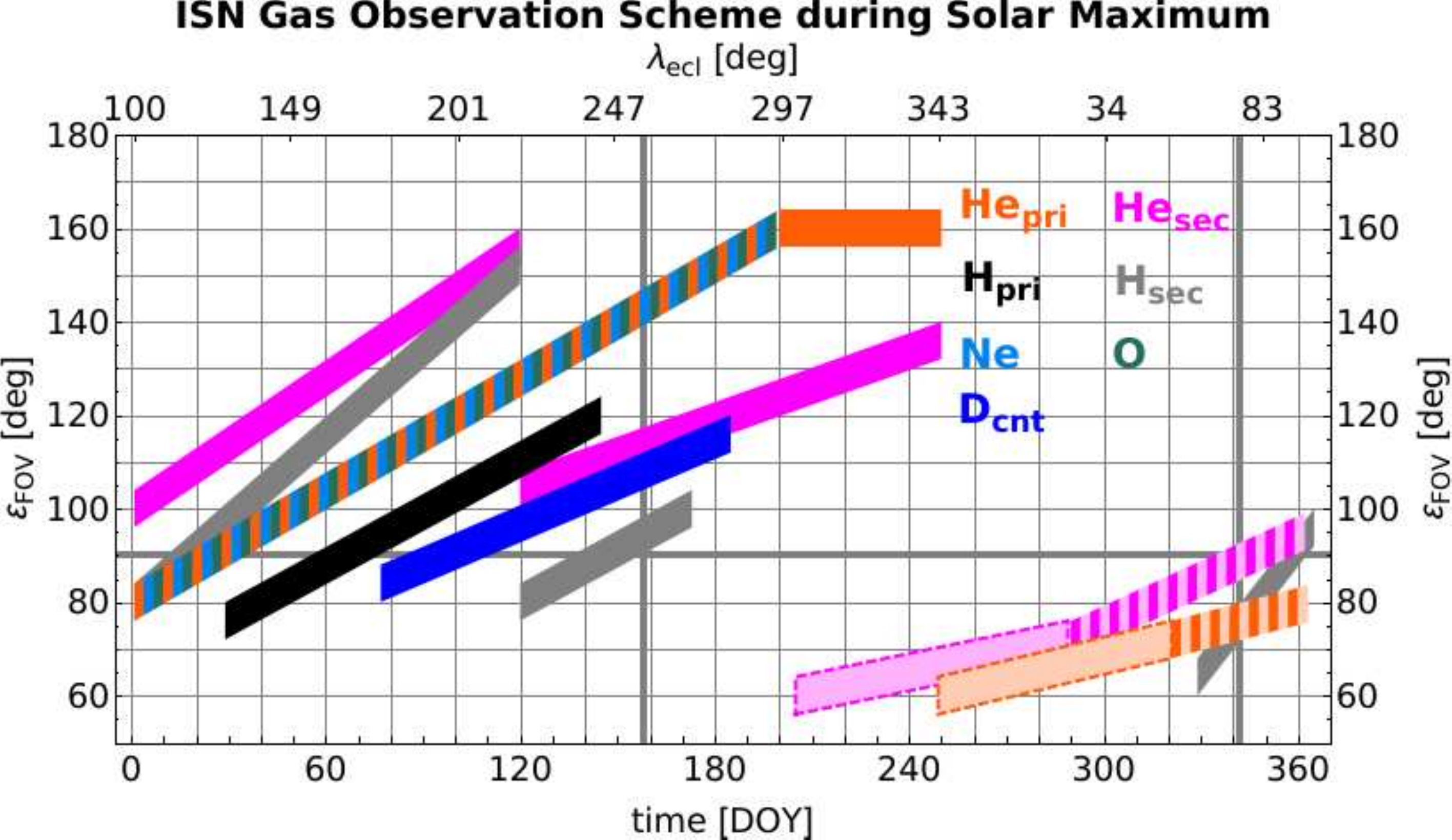}\\
\includegraphics[scale=0.3]{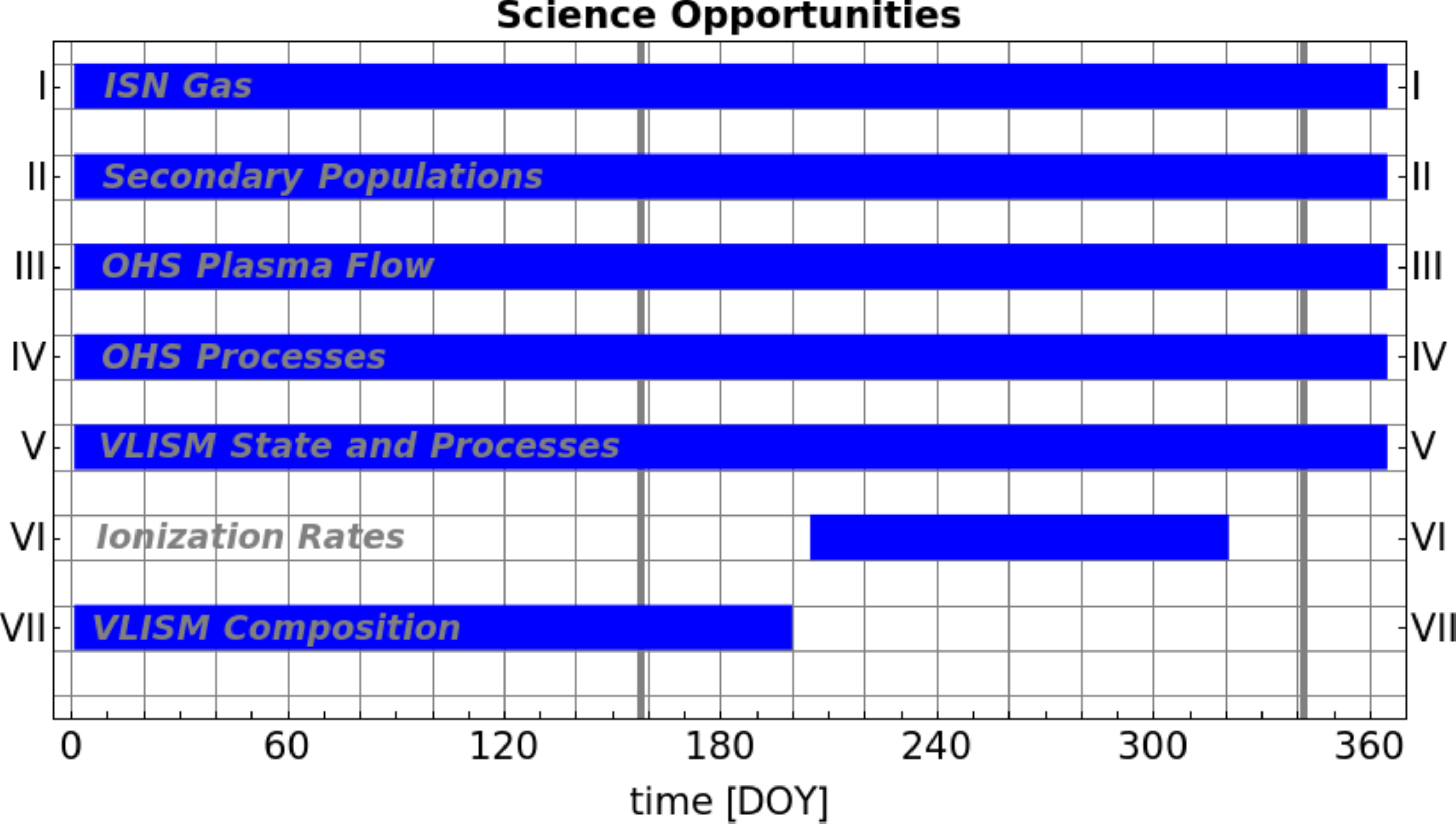}\\
\end{tabular}
\centering
\caption{Top: graphics with ISN gas observation opportunities during solar maximum as a function of DOY--\elong\, combinations.  The tetragons filled with pale color with dashed frames mark the indirect beam. The tetragons filled with alternating pale and vivid colors mark a mixture of direct and indirect beams. Bottom: diagram with the SOs for ISN gas observations discussed in Section~\ref{sec:sciOpp} and presented in Figure~\ref{fig:graph}  accessible to accomplish throughout the year in correspondence with the targets presented in the top panel.
 \label{fig:obsPlanGraph}}
\end{figure}

The plan is described in the following text sequence and supplemented by the illustrations in Figure~\ref{fig:obsPlanGraph}.  It has been compiled with the following boundary conditions and priorities in mind: (1) an upper limit for elongation is set to \elong$\approx160\degr$, to avoid emission from the Earth's geocorona in the FOV \citep{baliukin_etal:19a}; (2) we particularly include the observation geometries of the \secHe\, and \secH\, populations without contribution from \priHe\, and \priH, (Sections~\ref{sec:secondaryHe} and \ref{sec:secondaryH}); (3) the year is divided into several seasons by a range of DOYs motivated by the SOs; (4) limitations in energy, flux magnitude, and the presence of the signal in at least three spin angle bins are applied similarly to those in Figure~\ref{fig:pxlMaps} and Table~\ref{tab:summary}. When for any given range of DOYs more than one observation goal may be possible, the observational goals are ordered starting with the highest priority from the scientific perspective (starting from A, next B, next C, etc.) as argued in Section~\ref{sec:sciOpp}.

% Observation plan scheme for solar maximum
\begin{enumerate}
\item DOY 001 -- DOY 060
	\begin{enumerate}[label=\Alph*.]
	\item Track \priHe\, flow peak (incl. Ne and O), \elong$=80\degr - 100\degr$
	\item Stair-step mode with first 30 days at \elong$=84\degr$ and next 30 days at \elong$=96\degr$
	\item \secHe\, without contribution from \priHe\, at \elong$=100\degr - 136\degr$
	\item \secH\, without contribution from \priH\, at \elong$=80\degr - 124\degr$
	\item Track \priH, flow peak, DOY~29 -- 60, at \elong$=76\degr - 88\degr$
	\end{enumerate}

\item DOY 060 -- DOY 120
	\begin{enumerate}[label=\Alph*.]
	\item Track \priH\, flow peak, \elong$=88\degr - 112\degr$
	\item Track \priHe\, flow peak (incl. Ne and O), \elong$=104\degr - 128\degr$
	\item Stair-step mode with first 30 days at \elong$=110\degr$ and next 30 days at \elong$=122\degr$
	\item \secHe\, without contribution from \priHe\, at \elong$=128\degr - 156\degr$
	\item \secH\, without contribution from \priH\, at \elong$=124\degr - 152\degr$
	\item \isnD\, detection based on \ionD\, ion count rate method, DOY 077 -- DOY 120, \elong$=84\degr - 108\degr$
	\end{enumerate}
	
\item DOY 120 -- DOY 200
	\begin{enumerate}[label=\Alph*.]
	\item \elong\, at constant value from a range $142\degr - 148\degr$ to observe various species around the upwind (\isnHe, \isnH, Ne, and O should pass the FOV)
	\item Track \priHe\, flow peak (incl. Ne and O) to study parameter correlation and to increase statistics for Ne and O, \elong$=132\degr - 160\degr$
	\item \isnD\, detection based on \ionD\, ion count rate method, DOY 120 - DOY 185, \elong$=96\degr - 116\degr$
	\item \secHe\, without contribution from \priHe, \elong$=104\degr - 136\degr$
	\item \secH\, without contribution from \priH\, DOY 120 -- 173, \elong$=80\degr - 100\degr$
	\item Track \priH, flow peak, DOY~120 -- 145, \elong$=112\degr - 120\degr$
	\end{enumerate}

\item DOY 200 -- DOY 250
	\begin{enumerate}[label=\Alph*.]
	\item \priHe\, \elong$=160\degr$
	\item \secHe\, without contribution from \priHe, \elong$=128\degr - 136\degr$
	\item Indirect beam of \secHe\, DOY~205 -- 250, \elong$=60\degr - 72\degr$
	\end{enumerate}
	
\item DOY 250 -- DOY 290
	\begin{enumerate}[label=\Alph*.]
	\item Indirect beam of \secHe, \elong$=64\degr - 72\degr$
	\item Indirect beam of \priHe, \elong$=60\degr - 64\degr$
	\end{enumerate}

\item DOY 290 -- DOY 320
	\begin{enumerate}[label=\Alph*.]
	\item Indirect beam of \priHe, \elong$=60\degr - 72\degr$
	\item \secHe\, without contribution from \priHe, \elong$=76\degr - 104\degr$
	\end{enumerate}
	
\item DOY 320 -- DOY 365
	\begin{enumerate}[label=\Alph*.]
	\item \secHe\, without contribution from \priHe, \elong$=90\degr - 116\degr$
	\item Track \priHe, flow peak, \elong$=60\degr - 80\degr$ 
	\item \secH, without contribution from \priH\, DOY~329 -- 365, \elong$=64\degr - 96\degr$
	\end{enumerate}
\end{enumerate}

%% Figure peak fluxes all

The multichoice observation plan presented above is sketched in the top panel of Figure~\ref{fig:obsPlanGraph}. For any given elongation settings, most of the time several ISN species are visible, albeit at different energies. The largest number of possible observation geometries is for the first six months of a year, when all species and populations are accessible. The last six months of a year are a period when the indirect beams of \isnHe\, are accessible for detection. In the last two months both \priHe\, and \secHe\, together with \secH\, are targets for the ISN observations.

\section{Summary and Conclusions \label{sec:summary}}
The bottom panel of Figure~\ref{fig:obsPlanGraph} complements Figure~\ref{fig:graph} and, together with Figure~\ref{fig:pxlMaps} provides observational periods for the SOs discussed in Section~\ref{sec:sciOpp}. The observations of the \priHe\, flux to determine the fundamental ISN gas flow parameters are possible throughout the entire year, as well as the observations of the \secHe\, which can be observed either together with the \priHe\, flux or without its significant contribution (Section~\ref{sec:secondaryHe}). These two provide the capability to study the plasma flow and processes in the OHS. Supplemented with observations of \priH\, and \secH, which require different observation geometries, especially during solar maximum, these provide key information about the physical state of the VLISM. They will also contribute to resolving the questions about the abundance ratio of the secondary O population. Thus, the first five SOs described in Section~\ref{sec:sciOpp} are achievable with observations conducted throughout the entire year with appropriately adjusted boresight direction. 

The study of the VLISM composition through the Ne/O and D/H ratios requires successful observations of those species, which can be achieved only during a limited time period each year, when both Ne and O and both D and H satisfy the measurement limitations. With better statistics due to increased observation periods and precise adjustment of the observation directions to the peak location in the sky for heavy atoms, it will be possible to determine precisely the Ne/He, O/He, and D/H abundances at the termination shock. The determination of the abundance of ISN species in the VLISM by analysis of measurements from the Earth's orbit will additionally benefit from independent determination of the total ionization rates that act on the ISN gas at distances close to the Sun. This is achievable through the study of the indirect beams of the ISN gas, which will provide access to the electron impact ionization rate inside 1~au with carefully adjusted boresight directions in a period from July to November each year. Moreover, with the capability to track the flow maximum, measurements of the ISN gas flow in the upwind direction are possible, where the gas flow is not distorted by the Sun's gravitation contrary to the downwind hemisphere. Together with observations from different vantage points throughout the year, this should allow for breaking the parameter correlation present in the \textit{IBEX}-Lo data. 

Therefore, observations of the ISN gas with an instrument whose boresight direction is adjustable, such as \textit{IMAP}-Lo \citep{mccomas_etal:18b}, will facilitate major progress in the study of the ISN gas, the heliosphere, and the VLISM. As presented in Section~\ref{sec:OperationPlan}, many possible observation geometries may be chosen during a year, even during solar maximum, which is unfavorable for observations of ISN gas species other than He. The observation plan present here can be applied to any instrument with a similar capability to adjust the boresight direction, and it can be easily adjusted to other observation conditions. Moreover, we present when and how to observe the ISN gas flow to achieve the SO listed in Section~\ref{sec:sciOpp}. As we discussed, the observation geometry can be adjusted based on the observation target and the science goal to accomplish. The flow peak can be tracked, the wings of the distribution functions can be sampled, the secondary population flow can be separated from the primary population, the statistics for heavy species can be increased, and the observations from different vantage points can be combined to provide a three-dimensional view of the ISN gas--heliosphere interaction. The study of the ISN gas from the vicinity of the Earth's orbit is key to remotely investigating the boundary regions of the heliosphere and the very local interstellar environment (Section~\ref{sec:sciOpp} and Figure~\ref{fig:graph}), but to achieve those goals, the complete hierarchy of observation targets needs to be addressed. This paper demonstrates how this can be accomplished through planning the future ISN gas observation. As a consequence,  the boundary regions of the heliosphere and the physical state of the VLISM will become more accessible to a comprehensive investigation than ever before.  

%% Acknowledgments
\acknowledgments
We are deeply indebted to all of the outstanding people who have made the \textit{IBEX} mission possible, and we are grateful to all of the dedicated people who are making the \textit{IMAP} mission possible. Part of this research was carried out by JMS during a research visit at University of New Hampshire in Durham funded by CBK PAN award for Young Scientists KIWZ.407.02.01.2018. JMS acknowledges the support by UNH during the visit in October 2018. Part of JMS's work on the paper was supported by NAWA Bekker Program Fellowship PPN/BEK/2018/1/00049. JMS and MAK thank Marek Strumik for the help with software development. The work at CBK PAN was supported by Polish National Science Center grant 2015-18-M-ST9-00036 and partly supported by NAWA grant PPI/APM/2018/1/00032/U/001. NAS and EM were funded by the \textit{IBEX} mission as a part of the NASA Explorer Program (NNG17FC93C; NNX17AB04G) and by \textit{IMAP} as part of the Solar Terrestrial Probes Program (NNN06AA1C). EM was partially supported by NASA grant NNH15ZDA001N-HGI.

% biliography
\bibliographystyle{apj}
\bibliography{texBibFileU}{}

\end{document}